\documentclass[twocolumn]{aastex63}
\usepackage{amsmath}
\usepackage{fontawesome}
\definecolor{linkcolor}{rgb}{0.1216,0.4667,0.7059}

\newcommand{\dv}{\boldsymbol{d}}

\newcommand{\betamic}{\beta_\mathrm{mic}}

\newcommand{\muv}{\boldsymbol{\mu}}
\newcommand{\fv}{\boldsymbol{f}}
\newcommand{\Fv}{\boldsymbol{F}}
\newcommand{\rv}{\boldsymbol{r}}

\newcommand{\xv}{\boldsymbol{x}}

\newcommand{\ev}{\boldsymbol{e}}
\newcommand{\vv}{\boldsymbol{v}}
\newcommand{\uv}{\boldsymbol{u}}
\newcommand{\qv}{\boldsymbol{q}}
\newcommand{\tv}{\boldsymbol{t}}
\newcommand{\nuv}{\boldsymbol{\nu}}

\newcommand{\Tpv}{\boldsymbol{T}}
\newcommand{\thetav}{\boldsymbol{\theta}}
\newcommand{\tcv}{\boldsymbol{\phi}}
\newcommand{\Ncut}{M}
\newcommand{\mmw}{\mu} %mean molecular weight
\newcommand{\mmr}{X} %mass mixing ratio
\newcommand{\molmass}{M} %molecular mass
\newcommand{\vmr}{\xi} %volume mixing ratio

\newcommand{\kGP}{k_\mathrm{GP}} %GP kernel
\newcommand{\nulc}{\hat{\nu}} %line center
\newcommand{\nulcl}{\nulc_{m,l}} %line center
\newcommand{\numatrix}{\hat{N}} %nu-matrix
\newcommand{\ntexp}{{\mathsf{n}}} %temperature exponent
\newcommand{\gw}{{\mathsf{g}}} %statistical weight
\newcommand{\Ng}{\mathcal{N}}
\newcommand{\nbeta}{\acute{\beta}}
\newcommand{\ngamma}{\acute{\gamma}}
\newcommand{\qlc}{\hat{q}}
\newcommand{\aD}{\acute{\beta}}

\usepackage{color}
\definecolor{red}{rgb}{0.7,0.0,0.0}

\usepackage{algorithmic}
\received{May 31, 2021}
\revised{Octber 30, 2021}
\accepted{November 18, 2021}
\shorttitle{Autodiff Spectrum for Exoplanets and Brown Dwarfs}
\shortauthors{Kawahara et al.}
\begin{document}

\title{Auto-Differentiable Spectrum Model
for High-Dispersion Characterization of Exoplanets and Brown Dwarfs}

\correspondingauthor{Hajime Kawahara}

\author[0000-0003-3309-9134]{Hajime Kawahara}
\email{kawahara@eps.s.u-tokyo.ac.jp}
\affiliation{Department of Earth and Planetary Science, The University of Tokyo, 7-3-1, Hongo, Tokyo, Japan}
\affiliation{Research Center for the Early Universe, 
School of Science, The University of Tokyo, Tokyo 113-0033, Japan}

\author[0000-0003-3800-7518]{Yui Kawashima}
\affiliation{Cluster for Pioneering Research, RIKEN, 2-1 Hirosawa, Wako, Saitama 351-0198, Japan}

\author[0000-0003-1298-9699]{Kento Masuda}
\affiliation{Department of Earth and Space Science, Osaka University, Osaka 560-0043, Japan}

\author[0000-0002-1835-1891]{Ian J. M. Crossfield}
\affiliation{ Department of Physics and Astronomy, University of Kansas, Lawrence, Kansas, USA}

\author[0000-0001-6360-8062]{Erwan Pannier}
\affiliation{Laboratoire EM2C, CNRS, CentraleSup\'{e}lec, Universit\'{e} Paris-Saclay, 3 rue Joliot Curie, 91190 Gif-sur-Yvette, France}

\author[0000-0002-7554-2539]{Dirk van den Bekerom}
\affiliation{Combustion Research Facility, Sandia National Laboratories, Livermore, CA 94550, USA}

\nocollaboration{6}

\begin{abstract}
We present an auto--differentiable spectral modeling of exoplanets and brown dwarfs. This model enables a fully Bayesian inference of the high--dispersion data to fit the {\it ab initio} line--by--line spectral computation to the observed spectrum by combining it with the Hamiltonian Monte Carlo in recent probabilistic programming languages. An open source code, {\sf exojax} \href{https://github.com/HajimeKawahara/exojax}{\color{linkcolor}\faGithub}, developed in this study, was written in Python using the GPU/TPU compatible package for automatic differentiation and accelerated linear algebra, {\sf JAX} \citep{jax2018github}. We validated the model by comparing it with existing opacity calculators and a radiative transfer code and found reasonable agreements of the output. As a demonstration, we analyzed the high-dispersion spectrum of a nearby brown dwarf, Luhman 16 A and found that a model including water, carbon monoxide, and $\mathrm{H_2/He}$ collision induced absorption was well fitted to the observed spectrum ($R=10^5$ and 2.28--2.30 $\mu$m). As a result, we found that $T_0=1295_{-32}^{+35}$ K at 1 bar and C/O $=0.62 \pm 0.03$, which is slightly higher than the solar value. This work demonstrates the potential of full Bayesian analysis of brown dwarfs and exoplanets as observed by high-dispersion spectrographs and also directly-imaged exoplanets as observed by high-dispersion coronagraphy.
\end{abstract}


\section{Introduction}

High-dispersion spectroscopy is now one of the standard techniques for detecting a large number of molecular and metal lines in exoplanet spectra via  transmission, dayside emission, and direct imaging \citep{2010Natur.465.1049S,2018Natur.560..453H,2018ApJ...863L..11H,2019A&A...627A.165H,2019A&A...628A...9C,2019A&A...632A..69Y,2020MNRAS.493.2215G,2020A&A...638A..26S,2020ApJ...888L..13T,2020MNRAS.494..363C,2020A&A...641A.120H,2020ApJ...897L...5B,2020Natur.580..597E, 2012Natur.486..502B,2012ApJ...753L..25R,2013MNRAS.436L..35B,2017AJ....153..138B,2017AJ....154..221N,2019A&A...625A.107G,2020A&A...640L...5Y,2020ApJ...898L..31N,2021AJ....161..153I,2021ApJ...910L...9N,2021arXiv210510230C,
2014Natur.509...63S,2016A&A...593A..74S,2018A&A...617A.144H,2019ApJ...878L..37F}. The papers included in the long list above report the detection of various molecules such as CO, $\mathrm{H_2O, CH_4}$, TiO, HCN, OH, and a wide variety of neutral and ionized species. In addition, high-dispersion coronagraphy (HDC), which is the combination of extreme adaptive optics with coronagraphs \citep{2014ApJS..212...27K,2014arXiv1409.5740K,2015A&A...576A..59S,2017AJ....153..183W,2017ApJ...838...92M}, is currently under active development on the largest class telescopes, including REACH on Subaru \citep{2020SPIE11448E..78K}, KPIC on Keck \citep{2019arXiv190904541J}, and HiRISE on VLT \citep{2020arXiv200901841O}. High-dispersion techniques can be used not only for the detection of atoms, ions, and molecules, but also for the estimation of a variety of physical planetary parameters, including the temperature-pressure (T--P) profile \citep{2015A&A...576A.111S}, even the presence of temperature inversions \citep{2017AJ....154..221N}, planetary spin \citep{2014Natur.509...63S,2016A&A...593A..74S},   winds \citep{2010Natur.465.1049S,2020Natur.580..597E}, C/O ratio \citep[e.g][]{2021Natur.592..205G,2021arXiv210510513P}, and the signature of the night-side condensation \citep{2020Natur.580..597E}.
In addition, high-dispersion data have, in principle, the potential to constrain a planet's obliquity \citep{2012ApJ...760L..13K} and the presence of exomoons \citep{2018AJ....156..184V}, although these last goals have yet to be realized. 

In contrast, a high-dispersion study of field brown dwarfs with $R > 50,000$ remains surprisingly limited. One well-known brown dwarf dataset is the spectra of a nearby, bright brown dwarf system, the L/T transition binary Luhman 16 AB \citep[WISE J1049-5319 AB;][]{2013ApJ...767L...1L}, taken by CRIRES/VLT (2.288 -- 2.345 $\mu$m). Leveraging the high signal-to-noise ratio of CRIRES spectra over several hours, \cite{2014Natur.505..654C} applied Dopper Imaging techniques to map the surface brightness distribution of the photometrically-variable Luhman~16B. Another, earlier, such example is $\varepsilon$ Indi Ba as taken by the Phoenix spectrometer on the Gemini South telescope ( 2.308--2.317 and 1.553--1.559 $\mu$m). \cite{2003ApJ...599L.107S} derived effective temperature and surface gravity by comparing the observed spectrum with the spectrum model of brown dwarfs. The wavelength range of these previous observations was quite narrow and therefore the applicability was relatively limited. However, brown dwarfs are a promising target for recently developed high-dispersion IR spectrographs with the wide wavelength range of 8--10 m class telescopes, such as IRD on Subaru \citep{2018SPIE10702E..11K} and CRIRES+ on VLT \citep{2014SPIE.9147E..19F}. Given the many similarities between brown dwarfs and young, self-luminous exoplanets, the high-resolution spectroscopy of brown dwarfs is increasingly important as a prototype for exoplanet characterization.

Despite the wide applicability of high-dispersion spectroscopy to exoplanets, most such exoplanetary datasets have been analyzed using the cross-correlation  with a theoretical template of the planet spectrum. Recently, several efforts have been made to link the cross-correlation to the likelihood function for Bayesian inference \citep{2017ApJ...839L...2B,2019AJ....157..114B,2019AJ....158..228G,2020MNRAS.493.2215G}.  As a fully Bayesian approach, \cite{2019AJ....158..228G} proposed a combined retrieval code of low-- and high-- resolution spectra, {\sf HyDRA-H}, and demonstrated it using a dayside emission of HD209458b as observed by HST, Spitzer, and CRIRES/VLT. 
These approaches rely on a grid-based precomputation of the cross section datasets in a temperature--pressure--wavenumber grid space.

In this paper, we propose an approach that 
fits a spectral model directly from the computation of the line profile to the high-resolution data 
through a Markov Chain Monte Carlo (MCMC) method. Because the opacity and radiative transfer calculations are time-consuming at high dispersion, the key problem is how to reduce the size of the Markov chain  required to approximate a target posterior distribution. However, the traditional Metropolis--Hasting based MCMC that uses random proposal distributions tends to be inefficient as increasing the fitting parameters in particular. Such approaches exhibit a significant drop in the MCMC acceptance rate of the proposed sampling,  and/or often demonstrate a strong correlation between successive sampled states when one cannot find a proper proposal distribution in a transition kernel. In contrast, Hamiltonian Monte Carlo \citep[HMC; ][]{1987PhLB..195..216D,2011arXiv1111.4246H} has much higher acceptance rates and less correlation between samples 
than those of the conventional Metropolis--Hasting algorithm with a random walk proposal distribution \citep[see ][as a review paper]{2017arXiv170102434B}. 
{
Another advantage of HMC is that the dependence of the number of parameters $D$ on the computational time to convergence is milder ($\propto D^{5/4}$) than that of the random-walk Metropolis \citep[$\propto D^2$; see Section 4.4 in ][]{2012arXiv1206.1901N}.
}

Because of these notable features, HMC has recently begun to be used in the field of astronomy. Precise determination of a mass -- radius relation for the seven TRAPPIST--1 planets  is a recent example of leveraging the advantages of HMC in the  field of exoplanets \citep{2020arXiv201001074A}. HMC was also used in the radial velocity anomaly modeling of a star + black hole binary candidate \citep{2021ApJ...910L..17M} and an occultation mapping of Io's surface \citep{2021arXiv210303758B}. In addition, a probabilistic modeling of time series data for exoplanets, {\sf exoplanets}, provides building blocks for various fittings that can be used for HMC \citep{2021arXiv210501994F}. 

One technical difficulty of HMC is that it requires the gradient of the objective function in addition to the function itself, due to its use of the ``leap-frog algorithm.'' For complex models such as model emission spectra (which include line profile calculations and radiative transfer), it is not easy to calculate the derivative as a function of all relevant fitting parameters. To overcome this difficulty, we develop a differentiable spectral model utlizing {\it automatic differentiaion}. This powerful technique  evaluates derivatives of most numerical functions in a straightforward manner \citep{wengert1964simple,2015arXiv150205767G}. Unlike numerical or symbolic differentiation, the automatic differentiation decomposes a derivative of a target function into those of elementary functions based on the chain rule. Automatic differentiaion packages such as {\sf Autograd}, {\sf tensorflow}, and {\sf PyTorch}, have been widely used in machine learning. Among them, {\sf JAX} is an automatic differentiation package with accelerated linear algebra ({\sf XLA}) that has recently been actively developed  \citep{jax2018github}. Several recent astronomical applications utilize {\sf JAX} as a backend, {\sf JAXNS} for nested sampling \citep{2020arXiv201215286A}, and optical simulation \citep{2020arXiv201109780P}. A notable feature is that {\sf JAX} is compatible with modern probabilistic program languages (PPLs) such as {\sf NumPyro} \citep{phan2019composable}, {\sf MCX/BlackJAX}, and {\sf PyMC3} \citep{salvatier2016probabilistic}. This means that packages written in {\sf JAX} can combine the HMC of these PPLs. 

\begin{figure*}[htb]
\begin{center}
\includegraphics[width=0.85\linewidth]{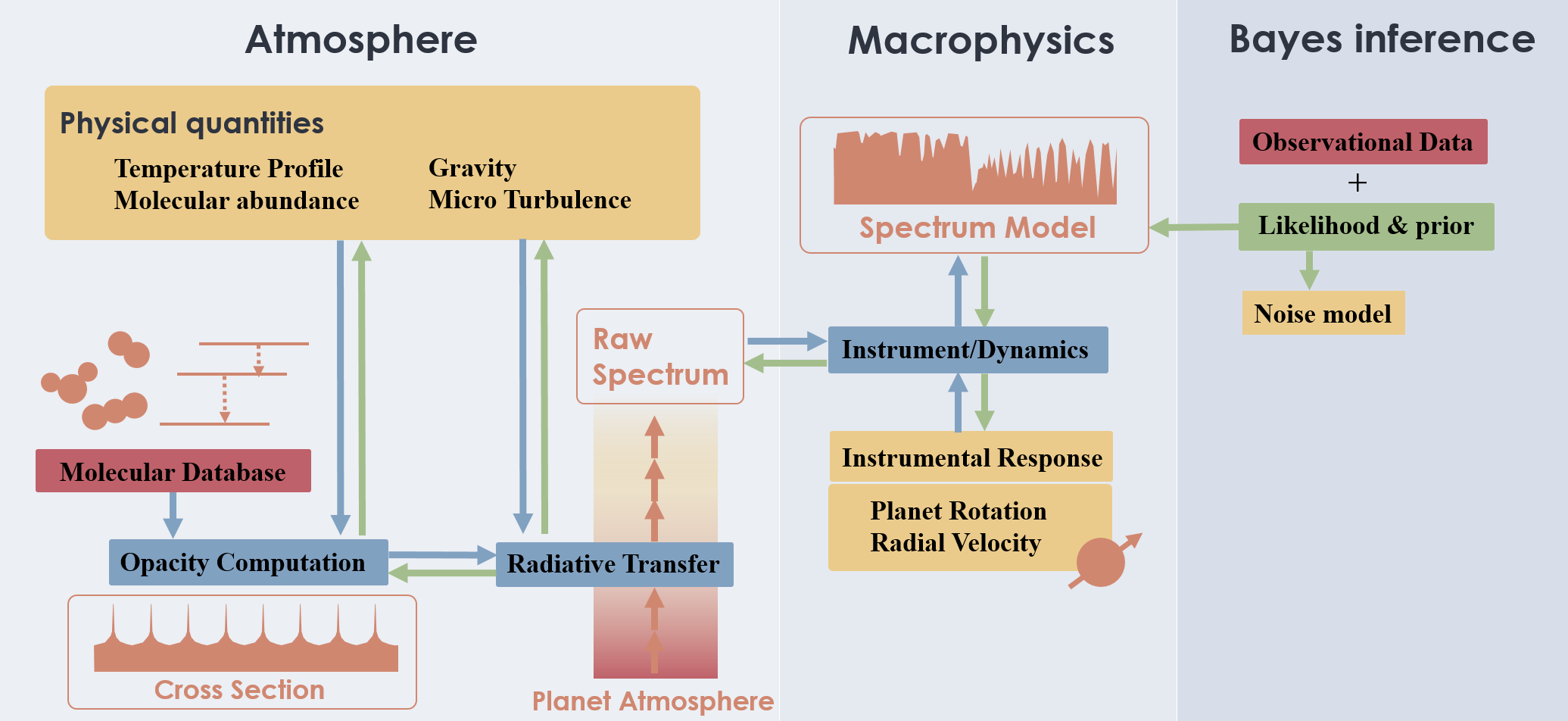}
\caption{Spectrum fitting procedure using {\sf exojax}. The blue boxes are auto-differentiable modules using {\sf JAX}.% in {\sf exojax}.  
The yellow boxes indicate the model parameters of the atmosphere. In addition, the parameters from observation and instruments are also shown in the yellow boxes. The green boxes are the statistical quantities used in HMC--NUTS by {\sf NumPyro}. The data required as external inputs are observational data and molecular databases (red boxes). { The blue arrows indicate the direction of the forward modeling of the spectrum.  HMC--NUTS uses the derivatives of the posterior ( $\propto$ prior $\times$ likelihood) by fitting parameters. The green arrows shows the propagation of auto--gradient (Appendix D). } \label{fig:exojax}}
\end{center}
\end{figure*}

Figure \ref{fig:exojax} shows a schematic of our auto-differentiable spectral model, {\sf exojax}. We require only observational data and a molecular database as external inputs. The main modules in {\sf exojax} are written in {\sf JAX}, and therefore they are differentiable for parameters such as the T--P profile, volume mixing ratios, gravity, and $V \sin{i}$ and so on. The derivatives of these parameters are used in PPLs such as {\sf NumPyro} to infer posterior distributions.  {\sf exojax} is an open-source package and its code is publicly available  \href{https://github.com/HajimeKawahara/exojax}{\color{linkcolor}\faGithub}.

The remainder of this paper is organized as follows. Section \ref{sec:cs} provides a formulation of differentiable cross sections, including an auto-differentiable Voigt profile. Differentiable radiative transfer is introduced in Section \ref{sec:rt}.
The astronomical and instrumental response functions are described in Section \ref{sec:resp}. We demonstrate {\sf exojax} by applying it to the high-dispersion spectra of a nearby brown dwarf, Luhman 16A in Section \ref{sec:ap}. We summarize the achievements of this study in Section \ref{sec:sum}.

\section{Auto-differentiable Cross Section}\label{sec:cs}

\subsection{Line Profile}
{
We consider two different methods for computing auto--differentiable line profiles. The first method is to calculate the Voigt function directly on a line-by-line basis, which is advantageous when the number of lines ($N_l$) is small, typically less than $10^3$. We call the first method the direct LPF as the direct computation of the line profile function. The other method is to synthesize the cross section of a large number of lines together, which is beneficial when computing the cross section for $N_l \gtrsim 10^3$. The second method is based on a modification of the recently proposed method, that is, the discrete integral transform \citep{van2021discrete}. We call the second method MODIT, a modified version of the discrete integral transform.

\subsubsection{Direct LPF}\label{ss:directopa}}
The cross section of a single molecular line is expressed as 
\begin{eqnarray}
\label{eq:crosssec}
  \sigma(\nu) &=& S \, V(\nu - \nulc, \beta, \gamma_L) \\
  &=& \frac{S}{\sqrt{2 \pi} \beta} H\left( \frac{\nu - \nulc}{\sqrt{2} \beta},\frac{\gamma_L}{\sqrt{2} \beta} \right),
\end{eqnarray}
 where $S$ is the line strength, { $V(\nu, \beta, \gamma_L)$ is the Voigt profile, $\nulc$ is the line center, $\gamma_L$ is the Lorentzian half--width (gamma factor), and the standard deviation of a Gaussian broadening is denoted by $\beta$ in this paper, to avoid confusion with the notation of the cross--section of $\sigma$. Considering the thermal broadening as the source of Gaussian broadening, we obtain
\begin{eqnarray}
\label{eq:thermal}
\beta = \beta_T = \sqrt{\frac{k_B T}{M m_u}} \frac{\nulc}{c}
\end{eqnarray}
where $m_u$ is the atomic mass constant, $M$ is the molecular/atomic weight, $k_B$ is the Boltzmann constant, and $c$ is the speed of light}\footnote{In addition to the thermal broadening (\ref{eq:thermal}), the Doppler broadening due to microturbulence can be included (layer by layer) as a Gaussian broadening with a standard deviation of $\betamic$. In this case, the standard deviation of the core broadening is expressed as 
$\beta^2 = {\beta_T^2 + \betamic^2}$. We can also model the microturbulence by the post process after the spectrum computation at the top of atmosphere when we assume a common microturbulence velocity in the entire atmosphere, as explained in Section \ref{sec:resp}.}.

The Voigt--Hjerting function \citep{1938ApJ....88..508H} is expressed as 
\begin{eqnarray}
H(x,a) = \frac{a}{\pi} \int_{-\infty}^{\infty} \frac{e^{-y^2}}{(x-y)^2 + a^2} dy.
\end{eqnarray}
The Voigt-Hjerting function is a real part of the Faddeeva function $w(z) = \exp{(-z^2)} \, \mathrm{erfc}(-i z)$ for $z = x + i a \in \mathbb{C}$,
\begin{eqnarray}
H(x,a) = \mathrm{Re}[w(x +i a)]
\end{eqnarray}
The Faddeeva function $w(z)$ is also known as {\sf wofz} (w--of--z) following the implementation of Algorithm 680 in the Transactions on Mathematical Software \citep{poppe1990algorithm}. Although {\sf wofz} in {\sf scipy.special} is widely used as $w(z)$ in Python, it is not compatible with {\sf JAX}\footnote{{\sf scipy.special.wofz} consists of a combination of Algorithms 680 and 916 (\href{http://ab-initio.mit.edu/wiki/index.php/Faddeeva_Package}{http://ab-initio.mit.edu/wiki/index.php/Faddeeva\_Package}).}. Therefore, we implemented the Voigt-Hjerting function compatible with automatic differentiation based on a combination of Algorithm 916 and an asymptotic representation, as explained below. 

To compute the Voigt--Hjerting function, \cite{2011arXiv1106.0151Z} proposed using its expansion,
\begin{eqnarray}
\label{eq:algo916}
\hat{H}_{\Ncut}(x,a) &=& e^{-x^2} \mathrm{erfcx} (a) \cos{(2 x a)} \nonumber \\
&+& \frac{2 \eta x \sin{(a x)}}{\pi}  e^{-x^2} \mathrm{sinc} (a x/\pi) \nonumber \\
&+& \frac{2 \eta}{\pi} \left\{  - a \cos{(2 a x)} \Sigma_1 + \frac{a}{2} \Sigma_2 + \frac{a}{2} \Sigma_3 \right\},
\end{eqnarray}
where
\begin{eqnarray}
\Sigma_1 &=& \sum_{n=1}^{\Ncut} \left( \frac{1}{\eta^2 n^2 + a^2}\right) e^{-(\eta^2 n^2 + x^2)}  \\
\Sigma_2 &=& \sum_{n=1}^{\Ncut} \left( \frac{1}{\eta^2 n^2 + a^2}\right) e^{-(\eta n + x)^2}  \\
\Sigma_3 &=& \sum_{n=1}^{\Ncut} \left( \frac{1}{\eta^2 n^2 + a^2}\right) e^{-(\eta n - x)^2}, 
\end{eqnarray}
{ $\mathrm{erfcx} (x) = 2 \pi^{-1/2} e^{x^2} \int_x^\infty e^{-t^2} dt $ is the scaled complementary error function }$, \mathrm{sinc}(x) = \sin{(\pi x)}/{(\pi x)}$ is the normalized sinc function, and $\eta \le 1$ is a scalar control parameter (we adopted $\eta = 0.5$). The implementation using Equation (\ref{eq:algo916}) is known as ``Algorithm 916'' \citep{2011arXiv1106.0151Z}. The original implementation of Algorithm 916 uses a recursive evaluation of the convergence of the series of $\Sigma_1$,$\Sigma_2$,and $\Sigma_3$. To make the function compatible with automatic differentiation, we fix {\Ncut} and vectorize each term as $\Sigma_1 = \vv_0 \cdot \vv_1$, $\Sigma_2 = \vv_0 \cdot \vv_2$, and $\Sigma_3 = \vv_0 \cdot \vv_3$, where $(v_0)_n = (\eta^2 n^2 + a^2)^{-1}$, $(v_1)_n = \exp{[-(\eta^2 n^2 + x^2)]}$, $(v_2)_n = \exp{[-(\eta n + x)^2]}$, and $(v_3)_n = \exp{[-(\eta n - x)^2]}$. 

A scaled complementary error function,
$\mathrm{erfcx} (a) = e^{a^2} \mathrm{erfc} (a)$,
suffers from an overflow for a large value of $a$ when a naive implementation of a combination $e^{a^2}$ and a complementary error function, $\mathrm{erfc} (a)$, is applied. Therefore, we implemented {\sf erfcx} using the Chebyshev approximation \citep{shepherd1981chebyshev}.  Appendix \ref{ap:scipy} provides a comparison with the implementation of {\sf scipy}. 

Equation (\ref{eq:algo916}) is accurate for $|x| \lesssim \Ncut/2$, however, the value rapidly drops to zero for $|x| > \Ncut/2$. This means that we need to increase $\Ncut$ to compute for a large $|x|$, and Equation (\ref{eq:algo916}) requires more computational time. Instead, we use an asymptotic representation for a larger $|z|$, which is expressed as
\begin{eqnarray}
\label{eq:asywofz}
  w_n^\mathrm{asy}(z) &=& \frac{i}{z \sqrt{\pi}} ( 1 + \tilde{\alpha} ( \mathsf{s}_0 + \tilde{\alpha} ( \mathsf{s}_1 +  \cdots \tilde{\alpha}  ( \mathsf{s}_n )\cdots) \\
  \tilde{\alpha}  &\equiv& \frac{1}{2 z^2}
\end{eqnarray}
where $\mathsf{s}_0=1, \mathsf{s}_1=3, \mathsf{s}_2=15, \cdots, \mathsf{s}_k= (2 k + 1)!!$. This asymptotic representation provides an accurate approximation of the Feddeeva function for $|z| \gtrsim 10$ \citep{2018arXiv180601656Z}\footnote{The other asymptotic representation in \cite{2018arXiv180601656Z}, $w(z) = i z/\sqrt{\pi} (z^2 - 2.5)/(z^2(z^2-3) + 0.75$) is slightly faster than $\mathrm{Re} [w_2^\mathrm{asy}(x+i a)]$.  However, the residual from {\sf scipy.wofz} is one order of magnitude larger than $\mathrm{Re} [w_2^\mathrm{asy}(x+i a)]$. }. As the {\sf exojax} implementation of the Voigt--Hjerting function, we decided to use Equation (\ref{eq:algo916}) with $\Ncut=27$ for $|z|^2<111$ and the real part of $w_2^\mathrm{asy} (z)$ for $|z|^2 \ge 111$: 
\begin{eqnarray}
H(x,a) = 
  \begin{cases}
    \hat{H}_{27}(x,a) & \mbox{for $x^2 + a^2 < 111$} \\
    \mathrm{Re} [ \,w_2^\mathrm{asy}(x+i a)\,] & \mbox{otherwise.} 
  \end{cases}
\end{eqnarray}
This implementation is quite accurate: the absolute difference from {\sf scipy.special.wofz} is smaller than $10^{-6}$, at least for $10^{-3}<a<10^{5}$ and  $10^{-3}<x<10^{5}$, as shown in Appendix \ref{ap:scipy}.

\begin{figure*}[htb]
\begin{center}
\includegraphics[width=\linewidth]{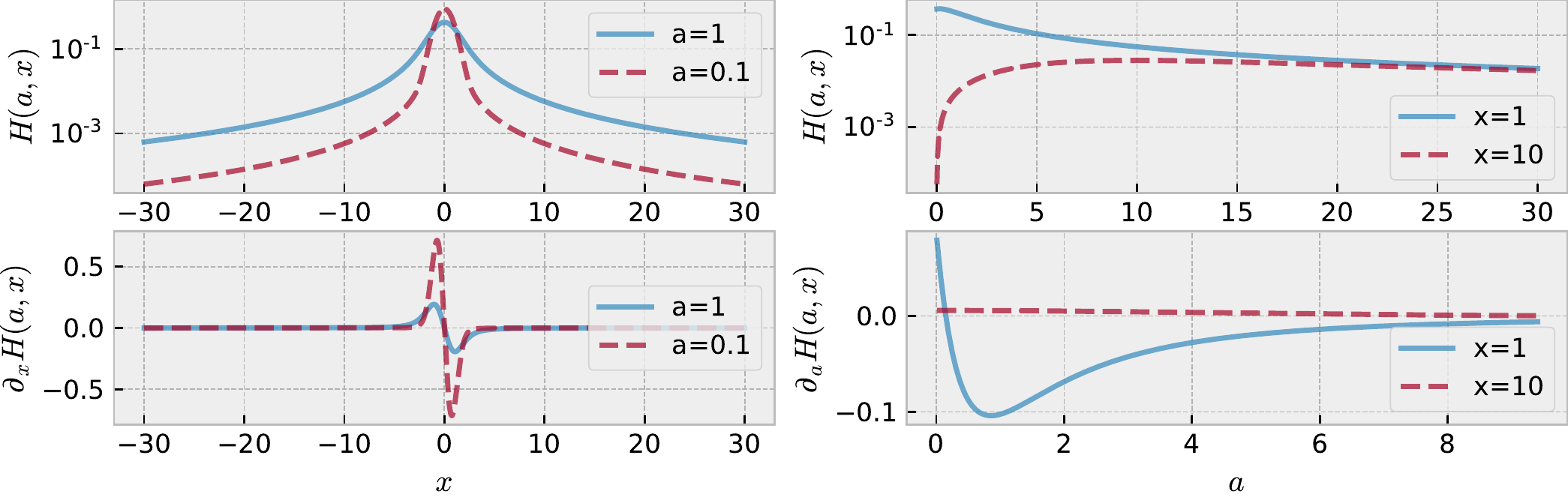}
\caption{Examples of $H(x,a)$ and its automatic differentiations computed by {\sf hjert} in {\sf exojax}.\label{fig:exwofz_ej2}}
\end{center}
\end{figure*}

Because {\sf hjert} is implemented using only {\sf JAX} functions, it is auto-differentiable, as shown in Figure \ref{fig:exwofz_ej2}. However, custom-defined derivatives of $H(x,a)$ are useful for speeding up the computation.  We implemented custom derivative rules for differentiation using the following analytic relation: 
\begin{eqnarray}
\partial_x H(x,a) &=&  2 a L(x,a) - 2 x H(x,a) \\
\partial_a H(x,a) &=&  2 x L(x,a) + 2 a H(x,a) - \frac{2}{\sqrt{\pi}},
\end{eqnarray}
where we defined the imaginary part of $w(x+i a)$ as
\begin{eqnarray}
L(x,a) = \mathrm{Im}[w(x +i a)].
\end{eqnarray}
For $L(x,a)$, we used an expression similar to $H(x,a)$ that uses the expansion of Algorithm 916 for $x^2 + a^2 < 111$ and the asymptotic representation otherwise (see Appendix \ref{ap:Lxa} for more details). We define a custom Jacobian-vector product \citep[JVP;][]{hirsch1974differential,2015arXiv150205767G} for the Voigt-Hjerting function, which enables us to compute the forward automatic differentiation (Appendix \ref{ap:JVP})\footnote{{\sf NumPyro} supports the JVP since version 0.6.0. There is another type of automatic differentiation: a vector-Jacobian product (VJP). Currently, {\sf exojax} does not support VJP.}.  

%%%MODIT
{\subsubsection{MODIT}\label{ss:modit}

With an increase in the number of lines of $N_l$, the direct LPF tends to be intractable even when using GPUs, in particular for $N_l \gtrsim 10^3$ (see the benchmark test in \S \ref{ss:benchmark}).
\cite{van2021discrete} proposed an efficient method for rapid opacity computation for large $N_l \gtrsim 10^3$, the discrete integral transform. In the framework of the discrete integral transform, the synthesis of the line opacities is expressed as the integral of the lineshape distribution $\mathfrak{S}(\nulc,\beta,\gamma_L)$, as follows:
\begin{eqnarray}
\sigma (\nu) &=& \sum_l S_l V(\nu - \nulc_l, \beta_l, \gamma_{L,l}) \\
&=& \int d \nulc \int d \beta \int d \gamma_L  \mathfrak{S}(\nulc,\beta,\gamma_L) V(\nu - \nulc, \beta, \gamma_{L}), \nonumber \\
\end{eqnarray}
where $S_l, \nulc_l, \beta_l,$ and $ \gamma_{L,l}$ are the line strength, line center, standard deviation of the Doppler broadening, and Lorentzian width of the $l$-th molecular line. By discretizing  $\nulc,\beta,\gamma_L$, and $\mathfrak{S}(\nulc,\beta,\gamma_L)$, the synthesized cross section can be computed as follows:
\begin{eqnarray}
\sigma (\nu_i) &=& \sum_{jhk} \mathfrak{S}_{jhk} V(\nu_i - \nulc_j, \beta_h, \gamma_{L,k}) \\
\label{eq:FTFT}
&=& \sum_{hk} \mathrm{FT}^{-1} [ \mathrm{FT} (\mathfrak{S}_{jhk})  \mathrm{FT} (V (\nulc_j, \beta_h, \gamma_{L,k})) ], \nonumber \\
\end{eqnarray}
where  $\mathfrak{S}_{jhk} = \mathfrak{S}(\nulc_j, \beta_h, \gamma_{L,k})$ denotes the lineshape distribution matrix. In this framework, the number of convolutions is reduced from one per line to one per point of the lineshape distribution matrix grid. The lineshape distribution matrix ${\mathfrak{S}}_{jk}$ is constructed using a simple weight explained in \S 3.2 of \cite{van2021discrete}, where each line contributes to the grids of ${\mathfrak{S}}_{jk}$ adjacent to the exact position for $q$ and $\ngamma_L$ with a linear weight based on the distance between the exact position and the grids. In Equation (\ref{eq:FTFT}), $\mathrm{FT}$ is the discrete Fourier transform for $j$  and $\mathrm{FT}^{-1}$ is its inverse discrete Fourier transform. The fast Fourier transform is available for the numerical operations of $\mathrm{FT}$ and $\mathrm{FT}^{-1}$.

In the original version of the discrete integral transform, the wavenumber grid should be evenly spaced on a linear scale. For the spectral modeling of exoplanets/stars, the grid evenly spaced on a logarithmic scale is useful because $\log{\nu}$ is proportional to the velocity. Another advantage of the evenly spaced--log grid is that one can use the common Doppler width for a given temperature and isotope. To do so, we slightly modify the original formulation of the discrete integral transform, tagged by MODIT. In MODIT, we define a new variable $q= R_0 \log{\nu}$, where $R_0$ is the spectral resolution of the wavenumber grid\footnote{For the dimensional consistency, $\nu$ should be divided by a reference wavenumber, $\nu_0=1 \mathrm{cm}^{-1}$.}. The discretization of $q$ provides an evenly spaced grid on the logarithm scale of the wavenumber. Then, the Gaussian profile is expressed as:
\begin{eqnarray}
&f_G&(\nu; \nulc; \beta_T)d \nu = \frac{1}{\sqrt{2 \pi} \beta_T} e^{-(\nu - \nulc)^2/2 \beta_T^2} d \nu \\
&=& \frac{e^{\frac{q - \qlc}{R}}}{\sqrt{2 \pi} \aD} \exp\left[{- \frac{R_0^2}{2 \aD^2} \left(e^{\frac{q - \qlc}{R_0}} -1\right)^2 }\right] d q 
%&\sim& \frac{1 + \frac{\delta q}{R}}{\sqrt{2 \pi} \aD} e^{- (\delta q)^2/(2 \aD^2)} d q 
\label{eq:fgfg}
%&\approx& \frac{1}{\sqrt{2 \pi} \aD} e^{- (q - \qlc)^2/2 \aD^2} d q 
\approx  f_G(q; \qlc; \aD) d q \nonumber \\
\end{eqnarray}
where $\qlc \equiv R_0 \log{\nulc}$, and
\begin{eqnarray}
\aD \equiv \frac{R_0 \beta_T}{\nulc} = \sqrt{\frac{k_B T}{M m_u}} \frac{R_0}{c}
\end{eqnarray}
is the standard deviation of the Gaussian profile in the $q$ space, which does not depend on the line center $\nulc$. Therefore, the Doppler broadening parameter in the evenly spaced logarithmic grid is common for a given temperature $T$ and molecular mass $M$. We also define $\gamma_L$ normalized by $\nulc/R_0 $ as
\begin{eqnarray}
\ngamma_L &\equiv& \frac{R_0 \gamma_L}{\nulc}. 
\end{eqnarray}
This quantity depends on the line properties, but is demensionless. MODIT defines the lineshape distribution matrix in two dimensions: $q$ and $\ngamma$.

For a given temperature and molecule, the synthesis of the cross--section can be expressed using the 2D lineshape distribution matrix $\mathfrak{S}_{jk} = \mathfrak{S} (\nulc_j,\ngamma_k) $ as follows:
\begin{eqnarray}
\sigma (q_i) &=&  \sum_{jk} \mathfrak{S}_{jk} \acute{V}(q_i - q_j;\nbeta, \ngamma_k) \\
\label{eq:ftmodit}
&=& \sum_{k} \mathrm{FT}^{-1} [ \mathrm{FT} (\mathfrak{S}_{jk})  \mathrm{FT} (\acute{V} (\nulc_j, \nbeta, \ngamma_{L,k})) ] 
\end{eqnarray}
where $\acute{V}(q_i;\nbeta,\ngamma_j)$ is the profile in the $q$ space that satisfies $V (\nulc, \beta, \gamma_{L}) d \nu = \acute{V}(q;\nbeta;\gamma_L) dq$ . Similar to Equation (\ref{eq:fgfg}), we can approximate the Lorentz profile of the $q$ space as $f_L(\nu;\nulc;\gamma_L) d \nu \approx f_L(q;\qlc;\ngamma_L) d q$ for $|\log{\nu} - \log{\nulc}| \ll 1$. In this case, we can regard $\acute{V}$ as a Voigt profile itself, that is, $\acute{V}(q;\nbeta;\ngamma_L) \approx V(q;\nbeta;\ngamma_L) $. Equation (\ref{eq:ftmodit}) can be rewritten as:
\begin{eqnarray}
\label{eq:MODIT}
%\sigma (q_i) &=& \displaystyle{\sum_{k} \mathrm{FT}^{-1} \left[ \tilde{\mathfrak{S}}_{jk} \exp{(-2 \mathfrak{K}_{jk} )} \right]} \\
%\mathfrak{K}_{jk} &=&  \pi^2 \nbeta^2 (\tilde{q}_j)^2 + \pi \ngamma_{L,k} |\tilde{q}_j| 
\sigma (q_i) &=& \displaystyle{\sum_{k} \mathrm{FT}^{-1} \left[ \tilde{\mathfrak{S}}_{jk} \exp{(-2  \pi^2 \nbeta^2 \tilde{q}_j^2 - 2 \pi \ngamma_{L,k} |\tilde{q}_j|  )} \right]}, \nonumber \\
\end{eqnarray}
where $\tilde{\mathfrak{S}}_{jk}$ is the discrete Fourier conjugate of ${\mathfrak{S}}_{jk}$ for the $q$ direction (index of $j$), and $\tilde{q}_j$ is the conjugate of $q_i$. In addition to Equation (\ref{eq:MODIT}), we include the correction of the aliasing effect of the Lorentz profile (van den Bekerom et al. in preparation). Because the molecular mass and temperature should be common in the above formulation, we use Equation (\ref{eq:MODIT}) to compute the synthesized cross section for each isotope and atmospheric layer.

}
%%%
\subsection{Molecular Database}
\subsubsection{HITRAN/HITEMP}\label{ss:hit}

\begin{table}[]
  \caption{Parameters for molecular lines}
  \begin{center}
  \begin{tabular}{lccc} 
  \hline\hline
qunatity & symbol & HITRAN & ExoMol \\
 &  & HITEMP & \\
\hline
line center & $\nulc$ & \checkmark & \checkmark \\
line strength  & $S(T)$  & ref & \\
broadening coefficient & $\alpha$ & air,self & $\alpha (J_\mathrm{low})$ \\
temperature exponent & $\ntexp$ & air & $\ntexp (J_\mathrm{low})$  \\
state energy & $E$ & lower & \checkmark \\
statistical weight & $\gw$ & upper & \checkmark \\
Einstein coefficient & $A$ & \checkmark & \checkmark \\
partition function & $Q(T)$ & & \checkmark \\
$J$--quantum number & $J$ & & \checkmark \\ 
\hline
  \end{tabular}
  \end{center}
  \tablecomments{air = value for ``air'' on Earth, self = self pressure (partial pressure), ref = at the reference $T$ and/or $P$, lower = lower state, upper = upper state.}
\end{table}

To avoid complex notations, we omit the molecular and line indices of $m,l$ in this section \S \ref{ss:hit} and \ref{ss:exomol}.

For HITRAN, a line's strength at a temperature of $T$ is computed by
\begin{eqnarray}
\label{eq:linestT}
S (T) &=& S_0 \frac{Q(T_\mathrm{ref})}{Q(T)} \frac{e^{- h c E_\mathrm{low} /k_B T}}{e^{- h c E_\mathrm{low}  /k_B T_\mathrm{ref}}} \frac{1- e^{- h c \nulc /k_B T}}{1-e^{- h c \nulc /k_B T_\mathrm{ref}}} \\
\label{eq:Slog}
 &=& (q_r(T))^{-1} e^{  s_0 - c_2 E_\mathrm{low}  (T^{-1} - T_\mathrm{ref}^{-1}) }  \frac{1- e^{- c_2 \nulc/ T}}{1-e^{- c_2 \nulc/T_\mathrm{ref}}}, \nonumber \\
\end{eqnarray}
where $S_0 = S (T_\mathrm{ref})$,  $T_\mathrm{ref}=296$ K, $c$ is the speed of light, $h$ is the Planck constant, $k_B$ is the Boltzmann constant, $\nulc$ is the line center wavenumber, $E_\mathrm{low} $ is the lower-state energy of the transition in the unit of wavenumber, $Q(T)$ is the partition function, $s_0 = \log_{e}{S_0}$, $q_r(T) = Q(T)/Q(T_\mathrm{ref})$, and $c_2 = h c/k_B = 1.4387773 \, \mathrm{cm \, K}$. The expression of Equation (\ref{eq:Slog}) can avoid an overflow of the value when we use the single precision floating point number (float 32 bit) used in {\sf JAX}.

The gamma factor is the summation of the pressure broadening\footnote{In HITRAN, the broadening parameters are provided in the form of $\gamma_\mathrm{air} = \alpha_\mathrm{air}/P_\mathrm{ref}$ and $\gamma_\mathrm{self} = \alpha_\mathrm{self}/P_\mathrm{ref}$.},
\begin{eqnarray}
\label{eq:gammahitran}
  \gamma_L^p = \left( \frac{T}{T_\mathrm{ref}} \right)^{-\ntexp_\mathrm{air}} \left[ \alpha_\mathrm{air} \left( \frac{P - P_\mathrm{self}}{P_\mathrm{ref}}\right) + \alpha_\mathrm{self} \frac{P_\mathrm{self}}{P_\mathrm{ref}} \right] \nonumber \\
\end{eqnarray}
and the natural broadening\footnote{The natural profile obeys the Lorentzian with the gamma factor given by Equation (\ref{eq:naturalgamma}). The natural profile is derived from the fact that the probability that an electron remains in an excited state is proportional to $e^{-A t}$  \citep[for example, ][]{2003adu..book.....D}.}
\begin{eqnarray}
\label{eq:naturalgamma}
  \gamma_L^n = \frac{A}{4 \pi c},
\end{eqnarray}
that is, $\gamma_L=\gamma_L^p+\gamma_L^n$,
where $\alpha_\mathrm{air}$  and  $\alpha_\mathrm{self}$ are the air-broadened and self-broadened half width at half maximum at $T_\mathrm{ref}$ and $P_\mathrm{ref}=1$ atm, $\ntexp_\mathrm{air}$ is the temperature exponent for air (or the coefficient of the temperature dependence of the air-broadened half width), $A$ is the Einstein A coefficient, and $P_\mathrm{self} = \xi P$ is the partial pressure of the molecules ($\xi$ is the { volume mixing ratio }). In HITRAN/HITEMP, ``air'' means the $\mathrm{N}_2$ atmosphere. In the case of brown dwarfs/gaseous planets, we usually assume the $\mathrm{H}_2$ atmospheres; therefore, we need to be careful to take the differences into account. For instance, Table 1 in \cite{2019ApJ...872...27G} provides the difference in $\gamma_L$ between ``air'' and the $\mathrm{H}_2$ atmospheres for several molecules, which are approximately 1 to 1.5 times different. 

In summary, we require the parameter set of $S_0, A, E_\mathrm{low}, \nulc, \alpha_\mathrm{self},\alpha_\mathrm{air}$, and $\ntexp_\mathrm{air}$ for each line. As the interface of the HITRAN/HITEMP database, we imported the official HITRAN API, {\sf HAPI} \citep{2016isms.confETG12K}, into {\sf exojax} under an MIT license\footnote{\url{https://opensource.org/licenses/MIT}}. 

\subsubsection{ExoMol}\label{ss:exomol}

ExoMol \citep{2016JMoSp.327...73T} provides $\nulc$, $A$, the state energy $E$, and the statistical weight $\gw$. Using these quantities, the line strength was computed as follows: 
\begin{eqnarray}
\label{eq:STexomol}
S(T) =\displaystyle{ \frac{\gw_\mathrm{up}}{Q(T)} \frac{ A}{8 \pi c \nulc^2} e^{-  c_2 E_\mathrm{low}/T} {\left(1- e^{-c_2 \nulc/T}\right)}},
\end{eqnarray}
where $E_\mathrm{low}$ is the lower energy state of the transition and $\gw_\mathrm{up}$ is the statistical weight of the upper state  \citep[e.g.][]{2009JQSRT.110..533R}.
In practice, we compute $s_0 = \log_{e} S(T_\mathrm{ref})$ using Equation (\ref{eq:STexomol}) and use it to compute $S(T)$ using Equation (\ref{eq:Slog}). 

The pressure broadening is modeled as
\begin{eqnarray}
\label{eq:gammaexomol}
  \gamma_L^p =  \alpha_\mathrm{ref}  \left( \frac{T}{T_\mathrm{ref}} \right)^{-\ntexp} \left( \frac{P}{1 \,\mathrm{bar}} \right),
\end{eqnarray}
where $\ntexp$ is the temperature exponent, $\alpha_\mathrm{ref}$ is the broadening parameter, and $T_\mathrm{ref}=296$K. In ExoMol, the broadening parameters $\alpha_\mathrm{ref}$ and $\ntexp$ are not provided for each line. Instead, these parameters for the $\mathrm{H_2}$--atmosphere are a function of, for instance, the lower $J$-quantum number of the transition, $J_\mathrm{low}$, or a set of $J_\mathrm{low}$ and the upper $J$-quantum number of the transition $J_\mathrm{up}$ \citep{2016JMoSp.327...73T,2017JQSRT.203..490B}. 

The natural and thermal broadenings are computed using Equations (\ref{eq:naturalgamma}) and (\ref{eq:thermal}), respectively.

\subsection{Partition Function}

The partition function $Q(T)$ is a function of temperature. For instance, ExoMol provides a grid model of $Q(T)$. To be differentiable, we apply a one-dimensional interpolation function, {\sf jax.numpy.interp}, to the grid data as a function of $T$ for each grid of $\nu$. { In this study, we used the interpolation of the grid model provided by {\sf ExoMol}. }

\subsection{Collision-Induced Absorption}
The collision-induced absorption (CIA) of two molecules is an important continuum in the near-infrared region for a clear-sky atmosphere. We used the HITRAN CIA database \citep{2012JQSRT.113.1276R,2019Icar..328..160K}, which provides the absorption coefficient $\alpha_\mathrm{CIA} (T,\nu)$ as a grid table.  The optical depth was computed as follows: 
\begin{eqnarray}
\Delta \tau^{\mathrm{cont, CIA(m,m^\prime)}} &=& n_m n_{m^\prime} \alpha_\mathrm{CIA} (T,\nu) \Delta r \\
\label{eq:taucont}
&=&  n_m n_{m^\prime} \alpha_\mathrm{CIA} (T,\nu) H \frac{\Delta P}{P}  
\end{eqnarray}
where $ n_m = \xi_m P/k_B T $ and $n_{m^\prime} $ are the number densities of two molecules that collide with each other, $\Delta r$ is the physical height of the layer, and $H$ is the pressure scale height expressed as 
\begin{eqnarray}
\label{eq:scaleheight}
H = \frac{k_B T}{\mu m_u g}
\end{eqnarray}
where  $\mu$ is the mean molecular weight, $g$ is local gravitational acceleration (hereafter gravity), and $m_u$ is the atomic mass unit.

The absorption coefficient is smooth in the wavenumber direction, and the change is quite negligible for the Doppler shift. Hence, we only require the derivative of $\alpha_\mathrm{CIA} (T,\nu)$ by $T$. We adopted a one-dimensional interpolation using {\sf jax.numpy.interp}.

\subsection{Comparison with other opacity calculators}\label{ss:comp}
\begin{figure*}[htb]
\begin{center}
\includegraphics[width=\linewidth]{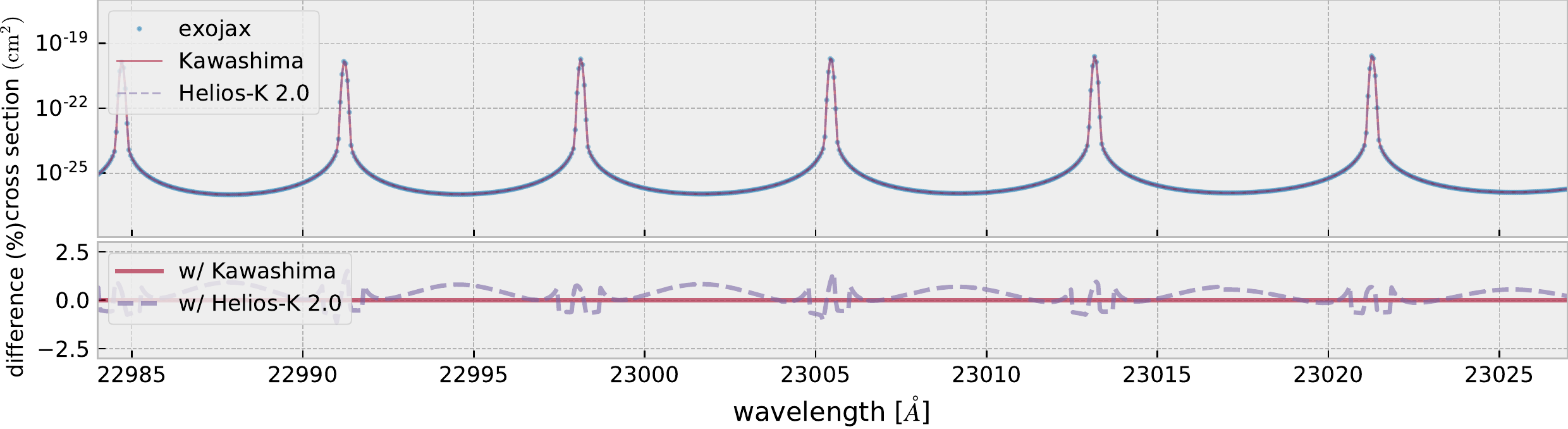}
\caption{Comparison of the HITRAN CO cross section with \cite{2018ApJ...853....7K} (Kawashima) and {\sf HELIOS-K } 2.0. $\Delta \nu = 0.01 \mathrm{cm}^{-1}$, $T=1000$K, $P=10^{-3}$ bar. The bottom panel shows the fractional differences between {\sf exojax} and both \cite{2018ApJ...853....7K} and {\sf HELIOS-K}. \label{fig:comp}}
\end{center}
\end{figure*}
To verify the accuracy of our approach, we compared cross sections calculated using the techniques described above with the results obtained using the existing opacity calculators. One is the cross section made for the study of exoplanet spectra \citep{2018ApJ...853....7K}, which uses a polynomial expansion of the Voigt profile by \cite{1997JQSRT..57..819K} \citep[see also][]{2004JQSRT..86..231R}. The other is a fast GPU-based opacity calculator, {\sf HELIOS-K} 2.0 \citep{2021arXiv210102005G}.
In this test, we compute the cross section of a 100 \% CO atmosphere using HITRAN assuming $T=1000$K and $P=10^{-3}$ bar. The partition function was computed using the Total Internal Partition Sums \citep[TIPS;][]{2011Icar..215..391L}. The pressure shift is included in both {\sf exojax} and \cite{2018ApJ...853....7K}. Figure \ref{fig:comp} shows a comparison of the cross sections of the three opacity calculators. The difference between {\sf exojax} and {\sf HELIOS--K} version 2.0, is within $\sim$ 1 \%. \cite{2021arXiv210102005G} reported that the difference of the Voigt--Hjerting function between HELIOS-K 2.0 and {\sf scipy.special.wofz} is within about 1 \% (as shown in their Figure 4), which is consistent with the 1 \% deviation we found. The difference between {\sf exojax} and \citet{2018ApJ...853....7K} is much smaller, despite the different algorithms for the Voigt profile computation. {In addition, the difference between the direct LPF and MODIT is $<$ 1 \% except for the same setting as Figure \ref{fig:comp} when using a double--precision floating point (F64). When using a single--point floating precision (F32), the error becomes larger at values lower than $10^{-25} \mathrm{cm^2}$, as shown in Figure \ref{fig:comp} (see Appendix \ref{sec:moditf64} for more details). MODIT has an advantage in computation time for a large number of lines, as we will see \S \ref{ss:benchmark}. In this case, a mild dynamic range of the cross section is naturally expected. However, we can compare the spectra using F32 and F64 for several cases when using MODIT+F32 for an HMC-NUTS.}

{
\subsection{Benchmark of opacity computation}\label{ss:benchmark}

We investigate the dependence of the computation time on the number of lines ($N_\mathrm{line}$) and wavenumber bins ($N_\nu$) as shown in Figure \ref{fig:bklpf}. Because the computation time depends on both the actual computation on the GPU device and data transfer from the main memory to the memory on the GPU, we show two different cases with and without data transfer to the GPU device for the direct LPF (\S \ref{ss:directopa}). The HMC-NUTS fitting corresponds to the latter case because it reuses the values in the GPU memory many times. The computation time with data transfer was approximately ten times slower than that without transfer. For the direct LPF, the computation time is approximately proportional to the number of lines and the wavenumber. The mean computation time without transfer was $\sim 0.1$ ns per line per wavenumber bin using NVIDIA/DGX A100. 

The MODIT algorithm (\S \ref{ss:modit}) exhibits almost no dependence on the number of lines until $N_\mathrm{line} \sim 10^5$ and converges to a linear dependence for larger $N_\mathrm{lines}$. This trend is consistent with the results of \cite{van2021discrete} (See Figure 3 and 11 in their paper). Notably, MODIT does not depend significantly on the number of wavenumber bins $N_\nu$. For a large number of lines, the calculation of the lineshape density $\mathfrak{S}_{jk}$ takes so much longer than the convolution step that it dominates computation time. For a small number of lines, this is probably because batch computation tends to be advantageous for FFT in GPU computations. 

\begin{figure}[htb]
\begin{center}
\includegraphics[width=\linewidth]{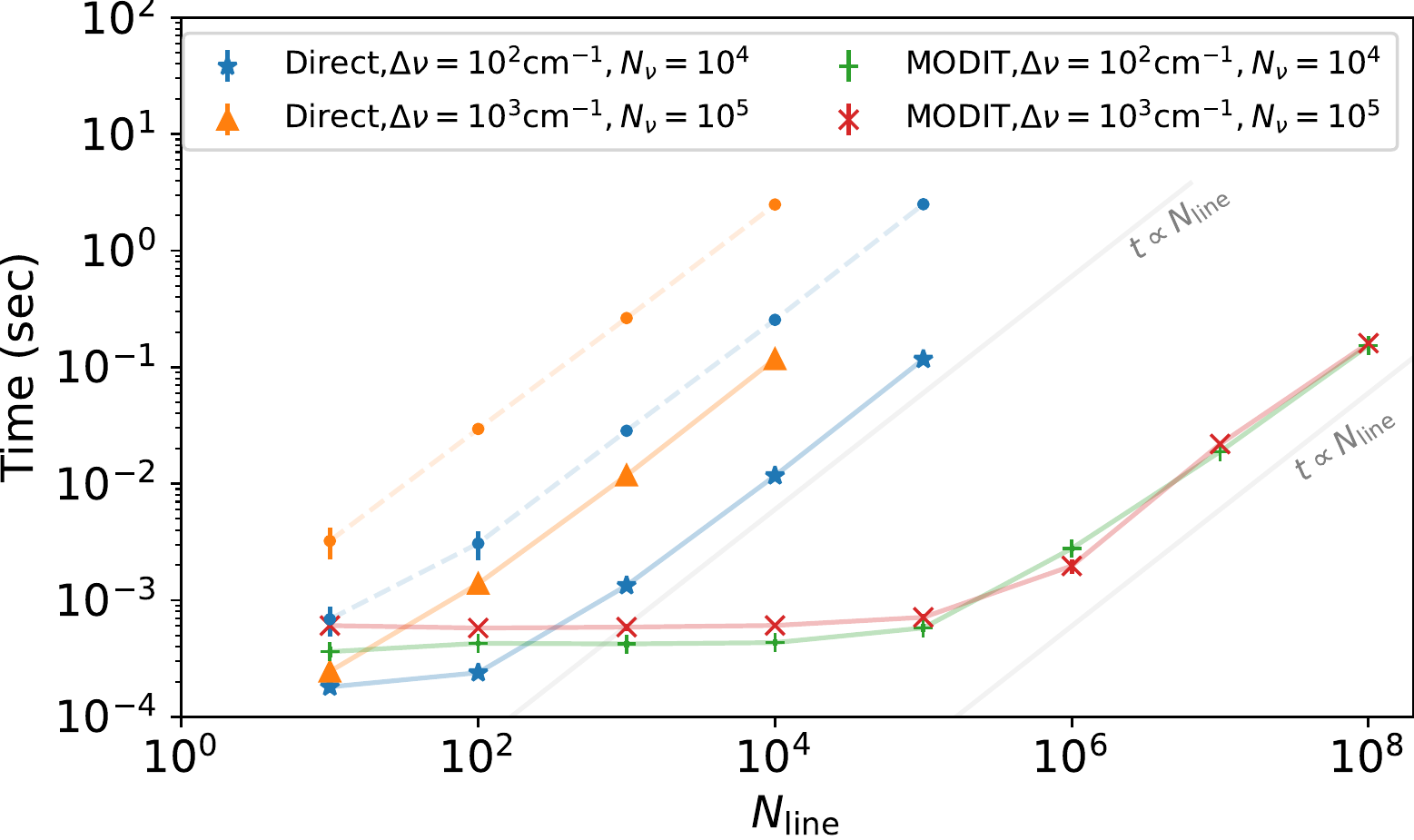}
\caption{Computation time as a function of the number of the molecular lines. We used NVIDIA/DGX A100 for this benchmark test. The star and triangle symbols correspond to the direct LPF without data transfer to the GPU memory. Dashed lines with the same color indicate the cases with data transfer. The plus and cross are MODIT without data transfer.  \label{fig:bklpf}}
\end{center}
\end{figure}
}

\section{Auto-differentiable Radiative Transfer}\label{sec:rt}

The differential optical depth in the $n$-th layer consists of lines and continuum opacities.
\begin{eqnarray}
\Delta \tau_n (\nu) &=& \sum_m^{N_m} \Delta \tau_n^{\mathrm{mol}, m} (\nu) + \sum_{c}  \Delta \tau_n^{\mathrm{cont}, c} (\nu). 
\end{eqnarray}
The contribution of the $m$-th molecule/atom is given by 
\begin{eqnarray}
\label{eq:dtau_line}
 \Delta \tau_n^{\mathrm{mol}, m} (\nu) &=& \frac{\Delta P_n }{\mmw m_u g}  \vmr_{m,n} \sigma_m(\nu)
  \\
  \label{eq:dtau_line2}
&=& \frac{\Delta P_n }{m_u g} \frac{\mmr_{m,n}}{\molmass_m}  \sigma_m(\nu)
\end{eqnarray}
where { $ \sigma_m(\nu)$ is the cross--section of the $m$-th molecule/atom}, $\vmr_{m,n}$ and  $\mmr_{m,n}$ are the volume mixing ratio (or equivalently the mole fraction) and mass mixing ratio of the $m$-th molecule/atom in the $n$-th layer, $\molmass_m$ is the molecular mass of the $m$-th molecule/atom normalized by the atomic mass constant.

The radiative transfer { with no scattering } is usually formulated as an iterative computation process of an upward flux at the $n$-th atmospheric layer, which consists of the source $\mathcal{S}$ from the layer and the upward flux from one layer below, considering its transmission,
\begin{eqnarray}
\label{eq:rerr}
F_{n} = \mathcal{T}_n F_{n+1} + (1-\mathcal{T}_n) \, \mathcal{S}_n,
\end{eqnarray}
where $F_{n}$ is the upward flux of the $n$--th layer, $\mathcal{T}_n $ is transmission within the $n$--th layer, and $\mathcal{S}_n$ is the source function of the $n$-th layer. Here, indexing is made in an ascending order for pressure, that is, $P_{n-1} < P_{n}$, and $F_0$ is the outgoing flux to space.
{ In Appendix \ref{ap:rt}, we describe the derivation of Equation (\ref{eq:rerr}) from a general expression of the two-stream approximation \citep{1989JGR....9416287T,heng2017exoplanetary}.}
Although {\sf JAX} supports the automatic differentiation of an iterative procedure, its implementation of {\sf jax.lax.scan} tends to result in a high computational cost. Instead, we reformulate the radiative transfer using a linear algebraic expression to leverage the power of {\sf XLA}. 

Denoting the transmitted source term by $ \mathcal{Q}_{n} \equiv (1-\mathcal{T}_n) \mathcal{S}_n$, the recurrence formula (\ref{eq:rerr}) can be expanded as 
\begin{eqnarray}
\label{eq:rerr2}
F_0 &=&  \mathcal{T}_0 ( \mathcal{T}_1 ( \mathcal{T}_2 (\cdots\mathcal{T}_{N-2} (\mathcal{T}_{N-1} F_B + \mathcal{Q}_{N-1} ) + \mathcal{Q}_{N-2})  \nonumber \\
&+& \cdots + \mathcal{Q}_2) + \mathcal{Q}_1) + \mathcal{Q}_0.
\end{eqnarray}
We set the upward flux at the bottom layer $F_B = \mathcal{Q}_{N}$ and define the transmitted source vector by $\qv \equiv (\mathcal{Q}_0, \mathcal{Q}_1, \cdots , \mathcal{Q}_{N} )^\top$, the transmission vector $\tv \equiv (1,\mathcal{T}_0,\mathcal{T}_1,\cdots,\mathcal{T}_{N-1})^\top $, and the cumulative product vector of the transmission vector, $\tcv = \mathcal{P} (\tv) \equiv (1, \mathcal{T}_0,  \mathcal{T}_0 \mathcal{T}_1,\cdots,\mathcal{T}_0 \mathcal{T}_1 \cdots \mathcal{T}_{N-2}, \mathcal{T}_0 \mathcal{T}_1 \cdots \mathcal{T}_{N-1})^\top$. We obtain the upward flux from the top layer to the space as
\begin{eqnarray}
\label{eq:rerr_tq}
F_0 = \tcv \cdot \qv. 
\end{eqnarray}

For the emission spectrum, the source function in the $n$-th layer is given by 
\begin{eqnarray}
\label{eq:sourcepla}
\mathcal{S}_n (\nu) = \pi B_x (T_n),
\end{eqnarray}
where $ B_x (T_n)$ is the Planck function\footnote{Here, we do not specify the unit of the $x$-axis, in the Planck function for generality, and when adopting $x=\nu$, the unit of the Planck function in cgs becomes $\mathrm{erg/s/cm^2/cm^{-1}}$.  } and $T_n$ is the representative temperature in the $n$-th layer.  
The transmission matrix for pure absorption under the two-stream approximation is expressed as
\begin{eqnarray}
\label{eq:trans}
\mathcal{T}_n  &=& 2 E_3(\Delta \tau_n ) \\
&=&( 1 - \Delta \tau_n) \exp{(- \Delta \tau_n)} + (\Delta \tau_n )^2 E_1(\Delta \tau_n ),
\end{eqnarray}
where $\Delta \tau_n$ is the optical depth of the $n$-th layer, and $E_N(x)$ is the exponential integral of the $N$-th order
\citep{2014ApJS..215....4H}. We adopt the analytic fitting formula of $E_1(x)$ by \cite{1970hmfw.book.....A},
\begin{eqnarray}
\label{eq:transco}
E_1(x)=
  \begin{cases}
    - \ln{x} + \sum_{k=0}^{5} A_k x^k \mbox{\,\,\, for $x \le 1,$}\\
    \displaystyle{\frac{e^{-x}}{x} \frac{\sum_{k=0}^{4} B_k x^{4-k}}{\sum_{k=0}^{4} C_k x^{4-k}}} \mbox{\,\,\, otherwise.}
  \end{cases}
\end{eqnarray}
 as used in {\sf HELIOS-R} (version 1) with the coefficients of $A_k, B_k,$ and $C_k$ are given in \cite{2017AJ....154...91L}. It should be noted that this approximation is accurate for $\mathcal{T}(x)$ and $1-\mathcal{T}(x)$ within a difference of $<2 \times 10^{-7}$ from {\sf scipy.special.expn}, as described in Appendix \ref{ap:scipy}.

\subsection{Matricization for Parallel Computing}

So far, we have considered $\nu$ as a continuous variable.
In practice, we need to discretize $\nu$ to a sufficient resolution to capture the line features, that is, $\nu \to \nu_j$ for $j=0,1,\cdots, N_\nu -1$. Because Equation (\ref{eq:rerr_tq}) provides the upward flux in a single wavenumber, we extend it to a vectorized version for the set of $\nu_j$. We define the transmission matrix and the transmitted source matrix as follows:
\begin{eqnarray}
\label{eq:rerr_TM}
\mathcal{T}_{nj} &=& \mathcal{T}_n (\nu_j) 
\end{eqnarray}
and
\begin{eqnarray}
\label{eq:rerr_QM}
\mathcal{Q}_{nj} &=& [1-\mathcal{T}_n (\nu_j)] \mathcal{S}_n (\nu_j).
\end{eqnarray}
We also define the cumulative product of the transmission matrix for the axis of the atmospheric layer by $\Phi$. By denoting $\tv_j \equiv (1,\mathcal{T}_0  (\nu_j) ,\mathcal{T}_1  (\nu_j) ,\cdots,\mathcal{T}_{N-1}  (\nu_j) )^\top $ and  $\tcv_j = \mathcal{P} (\tv_j)$, we can express $\Phi_{nj} = (\tcv_j)_n$.

Then, we obtain the upward flux vector $(F_0)_j = F_0 (\nu_j)$ as 
\begin{eqnarray}
\label{eq:lart}
\Fv_0  =  (\Phi \odot \mathcal{Q})^\top \uv, 
\end{eqnarray}
where $\odot$ is the Hadamard product, and $\uv \in \mathbb{R}^{N+1}$ is a 1-vector, as $u_i = 1$. This linear algebraic formulation of radiative transfer is simply implemented by a combination of  summation and cumulative product operator\footnote{Using {\sf JAX}, Equation (\ref{eq:lart}) can be expressed as {\sf  jax.numpy.sum( Q*jax.numpy.cumprod(T,axis=0),axis=0))}, where {\sf Q} is $\mathcal{Q}$ and {\sf T} is $\mathcal{T}$.}.

Using Equation (\ref{eq:trans}), the upward flux becomes a function of $\Delta \tau_n ({\nu_j})$ and $\mathcal{S}_n(\nu_j)$,
\begin{eqnarray}
\label{eq:lart2E3}
\Fv_0  =  \Fv_0 (\mathcal{D}, \mathcal{S}),
\end{eqnarray}
where  $\mathcal{S}_{nj} = \mathcal{S}_n(\nu_j)$ is the source function matrix, and $\mathcal{D}_{nj} = \Delta \tau_n({\nu_j})$ is the optical depth matrix, which requires the computation of the cross section of the entire line at every $\nu_j$. These implementations are given in {\sf exojax} as {\sf rtrun}. 

{ For the direct LPF, the cross--section of the $m$-th molecule/atom can be computed by
\begin{eqnarray}
\sigma_m(\nu) = \sum_l \sigma_{m,l} (\nu - \nulcl),
\end{eqnarray}
where $\sigma_{m,l}$ and  $\nulcl$ are the cross--section and line center of the $l$-th line of the $m$-th molecule/atom. } To leverage the parallel computation of GPU or TPU for the direct LPF, we introduce the $\nu$--matrix $\numatrix^{(m)}$ of the $m^{th}$ absorbing species (molecule or atom), defined by
\begin{eqnarray}
\numatrix_{jl}^{(m)} \equiv \nu_j - \nulcl \in \mathbb{R}^{N_\nu \times N_l},
\end{eqnarray}
where $\nu_j$ denotes the discretized wavenumber. The $\nu$-matrix needs to be precomputed for each absorber  used in a parallel calculation of the optical depth. 

\subsection{Comparison with petitRADTRANS}\label{ss:comprt}

\begin{figure*}[htb]
\begin{center}
\includegraphics[width=\linewidth]{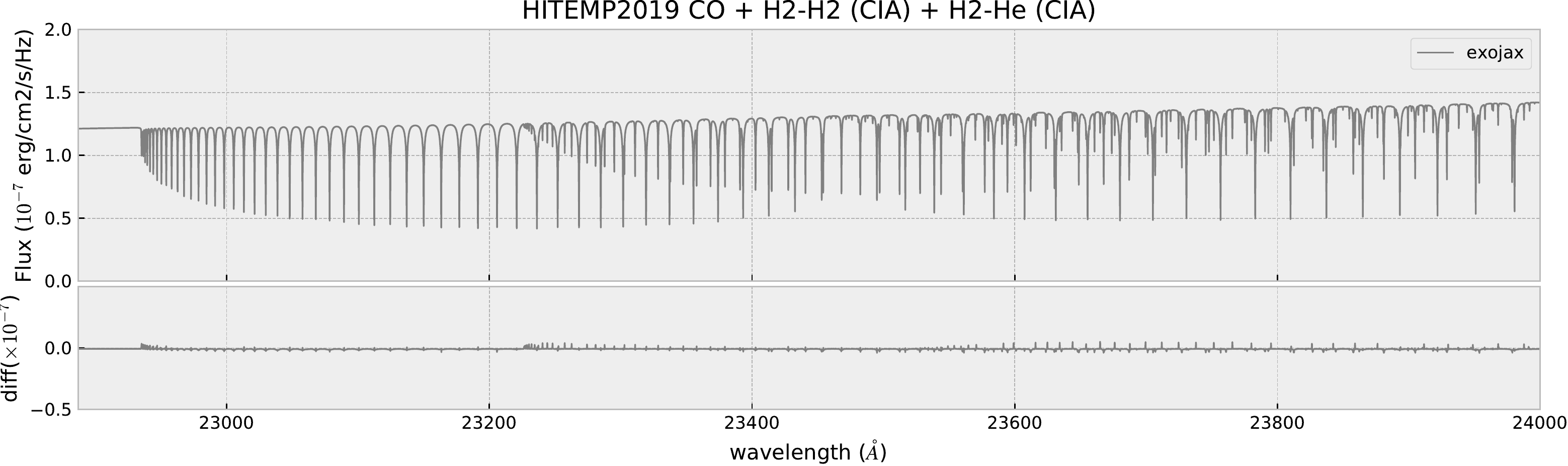}
\caption{Outward spectrum computed by {\sf exojax} (top panel) and its comparison with {\sf petitRADTRANS} (bottom panel). The bottom panel shows the difference between {\sf exojax} and {\sf petitRADTRANS} (i.e. {\sf exojax} - {\sf petitRADTRANS}).  \label{fig:comprt}}
\end{center}
\end{figure*}

We compared the output spectrum of our code with an existing radiative transfer code, {\sf petitRADTRANS} \citep{2019A&A...627A..67M}. {\sf petitRADTRANS} uses a precomputed grid model of the cross section in a pressure--temperature plane. First, we computed the cross sections of HITEMP2019/CO using {\sf exojax} and created a grid model for {\sf petitRADTRANS}. Therefore, in this comparison, both use the same cross section of CO. To avoid the error caused by the interpolation of the temperature and pressure grids used in {\sf petitRADTRANS}, we assume a linear $T$--$\log_{10}{P}$ profile with $T=1000$K at $\log_{10}{(P \mathrm{\,bar})} =2$ and $T=500$ K at $\log_{10}{(P \mathrm{\,bar})} =-10$. We assume a constant mass mixing ratio of 0.01 for CO, 0.75 for $\mathrm{H_2}$, 0.24 for $\mathrm{He}$, and $g=10^5 \,\mathrm{cm/s^2}$. We also include the $\mathrm{H_2-H_2}$ and $\mathrm{H_2-He}$ CIA as a continuum. We set 130 atmospheric layers from 100 bar to $10^{-10}$ bar. Figure \ref{fig:comprt} shows the output spectrum of {\sf exojax} and the difference from {\sf petitRADTRANS}. They are in excellent agreement with the linear T--$\log_{10}{P}$ profile.

For general T-P profiles, we noticed that there could be an order of 10 \% deviation between {\sf exojax} and {\sf petitRADTRANS}. This is likely due to the interpolation used in {\sf petitRADTRANS} considering the excellent agreement for the linear T--$\log_{10}{P}$ profile. The nominal temperature grid in {\sf petitRADTRANS} has a 30 \% increase, such as 666 K, 900 K, 1215 K as grid points. If the interpolation fails to recover the real value by 3 \%  at approximately 1000 K, (i.e., a 30 K error), for instance, it induces the error of the flux approximately as $(1.03)^4 - 1 = 13$ \%. In general, a precomputed grid model has such a systematic error in radiative transfer. To reduce this interpolation error, the model must be prepared in a finer grid. 

\section{Astronomical and Instrumental Responses}\label{sec:resp}

So far, we have described the procedure for computing the emission outward spectra from the top of atmosphere $\Fv_0$. Several processes remain to connect the calculated outward spectra with the observed one ($\Fv$).  These processes include  Doppler broadening by planet/stellar rotation, a Doppler shift by radial velocity, and an instrumental profile (IP) response. Finally, one must also  convert the calculated wavenumber grid, $\nu_j$, to that of the observed spectrum (denoted by $\nu_i^\prime$). { We note that one should use a fine grid for $\Fv_0$ enough to resolve the line profile (typically $R_0 \gtrsim 500,000$), which is usually finer than the grid of the observed spectrum $\Fv$.} In this section, we present our approach for accounting for all these effects. 

These astronomical and instrumental responses can be modeled by the convolution of the kernel and an outward spectrum. We define a response operator using a kernel $k(v)$ as
\begin{eqnarray}
\label{eq:response}
(k \ast F_0) (v_i^\prime) &\equiv& \displaystyle{\frac{\int F_0 (v_j) k(v_i^\prime - v_j; \rv) d v_j}{\int k(v; \rv) d v}}, \\  
\label{eq:response2}
&\approx&\displaystyle{ \frac{\sum_j k(\Delta v_{ij}; \rv) F_0 (v_j)}{\sum_j k(\Delta v_{ij}; \rv) }}
\end{eqnarray}
where $F_0(v_j) = (\Fv_0)_j$, 
$v_j$ is the velocity equivalence of $\nu_j$, $v_i^\prime$ is that of $\nu_i^\prime$, and $\Delta v_{ij} \equiv v_i^\prime - v_j = c \ln (\nu_j/\nu_i^\prime)$ is a velocity difference of the two grids. In Equation (\ref{eq:response2}), we assume that $\Delta v_j = \mathrm{constant.}$  to avoid the computation of $\Delta v_i$; that is, the grid of $v_i$ is evenly spaced in the linear space. This means that we use the grid of $\nu_i$  (or $\lambda_i$) with logarithmically-equal spacing. Including the radial velocity $\mathrm{RV}$, the wavenumber form of Equation (\ref{eq:response2}) is simply given by  
\begin{eqnarray}
\label{eq:responsenu}
F(\nu_i^\prime) = k \ast F_0  \approx \displaystyle{\frac{\sum_j k(c \ln{(\nu_i^\prime/\nu_j) + \mathrm{RV}}; \rv) F_0 (\nu_j)}{\sum_j k(c \ln{(\nu_i^\prime/\nu_j)}; \rv) }}, \nonumber\\ 
\end{eqnarray}
where $\nu^\prime_j$ is the wavenumber grid of the observational data, which is not necessarily evenly spaced in the log space\footnote{Because Equation (\ref{eq:responsenu}) is the matrix product, the computation cost and required memory size is proportional to the product of the numbers of the wavenumber grids $\nu_i$ and $\nu_j^\prime$. When this number exceeds approximately $10^4$--$10^5$, {\sf jax.numpy.convolve} can be used instead of the direct computation of Equation (\ref{eq:responsenu}). In this case, the linear interpolation is required to connect the output to the wavenumber grid of the observational data. In this paper, we do not use this extension. }.  

\subsection{Rigid Rotation Kernel}

The rigid (solid-body) rotation kernel \citep[e.g.][]{Gray} is given by 
\begin{eqnarray}
&\,&k_R(v) = \zeta(v/V\sin{i}) \\
&\,&\zeta(x) \equiv 
\displaystyle{\frac{\int_{-\sqrt{1-x^2}}^{\sqrt{1-x^2}} dy \, I(x,y)}{
\int_{-1}^{1} dx \int_{-\sqrt{1-x^2}}^{\sqrt{1-x^2}} dy  \, I(x,y) }} = \nonumber \\
&\,&
\displaystyle{\frac{-\frac{2}{3} \sqrt{1-x^2} \left(3 {u_1}+2 {u_2}
   x^2+{u_2}-3\right)+\frac{1}{2} \pi  u_1 
   \left(1-x^2\right)}{\pi (1 -u_1/3 - u_2/6) } },\nonumber \\
\end{eqnarray}
where we take the stellar disk coordinates $(x,y)$ such that the distance from the center to the edge of the star is unity and the axis of rotation corresponds to the y-axis and assumes a quadratic form of the limb darkening \citep{1950HarCi.454....1K} as
\begin{eqnarray}
I(x,y) &=& 1 - u_1 (1- \mu(x,y)) - u_2 (1-\mu(x,y))^2 \\
\mu(x,y) &\equiv& \sqrt{1 - (x^2 + y^2)}.
\end{eqnarray}
Here, $u_1$ and $u_2$ are the limb darkening coefficients.

\subsection{Instrumental Response Kernel}

A simple choise of the IP response is a Gaussian broadening
\begin{eqnarray}
k_G(v) = \frac{1}{\sqrt{2 \pi \beta^2}} \exp{\left( - \frac{v^2}{2 \beta^2}\right)}
\end{eqnarray}
 with a standard deviation of $\beta_\mathrm{IP} = c /(2\sqrt{2\ln{2}} R )$, %(km/s), 
 where $R$ is the resolving power. 
We can also include a microturbulence term $\beta_\mathrm{mic}$ in this form,  although we can directly include the microturbulence in each layer.

\section{Application to Luhman 16A}\label{sec:ap}

As a demonstration of {\sf exojax}, we fit our auto-differentiable spectrum model to a high-resolution spectrum of a nearby brown dwarf Luhman 16 A as taken by CRIRES/VLT \citep{2014Natur.505..654C}. Luhman 16 A has the spectral type of L7.5 and is located near the L-T transition. In contrast to its companion brown dwarf, Luhman 16 B, there is no strong signature of photometric variability in Luhman 16 A. The spectral features of this dataset have the advantage of being a simple system consisting mainly of CO and $\mathrm{H_2O}$ as dominant lines and CIA as continuous opacity.  Clouds are considered to be located below the CIA photosphere in the observing band  (22876--23453 \AA) for the following reason. The absorption coefficient of CIA  ($\mathrm{H_2}-\mathrm{H_2}$) has a maximum value near 2.4 $\mu$m, close to the observing band. This is approximately 10 times higher than the value in the $H$ band. In contrast, the cloud opacity should be flat or larger on the blue side considering that most of the cloud species that can form in this temperature range do not have their absorption features in the observing band \citep[e.g.][]{2015A&A...573A.122W}. The fact that the $H$-band flux is $\sim$1.5 times larger than that in the observing band \citep{2013ApJ...772..129B} suggests that the continuous absorption of the dataset is dominated by CIA, not clouds. { In this demonstration, we use both the direct LPF and MODIT for opacity calculation and compare the results with each other. } 

%\subsection{Reducing the number of molecular lines}
\subsection{Opacity calculation}

In this demonstration, Li2015 was used for $\mathrm{\,^{12}C^{16}O}$ \citep{2015ApJS..216...15L} and POKAZATEL for $\mathrm{\,^{1}H_2^{16}O}$ \citep{2018MNRAS.480.2597P} as obtained from the ExoMol database. In addition, CO in HITEMP2019 was used to observe the effect of different broadening parameters on the line shape.

Modern molecular databases contain a large number of transitions. A large number of weak lines exist below the photosphere. To reduce the computational complexity { for the direct LPF}, we exclude weak lines that do not contribute to the emission spectrum. { Such weak lines are identified by computing the layer where the line contributes the most to the emission. When this layer is below the photosphere determined by the CIA, we exclude the line from the opacity computation. A more detailed explanation is provided in Appendix \ref{ap:weakline}. For MODIT, we used all the CO and water lines for opacity computation. 

The number of wavenumber bins of the opacity computation (i.e., the dimension of $\Fv_0$) is $N_\nu=$4500 for both methods, which corresponds to  $R_0=7.2 \times 10^5$ as the wavenumber resolution. This is sufficient to resolve the original line profile.

}
\subsection{Model and HMC-NUTS}

%penchan:~/exojax/examples/LUH16A/postfig(feature/doc5)>python resfig.py
\begin{figure*}[htb]
\begin{center}
\includegraphics[width=\linewidth]{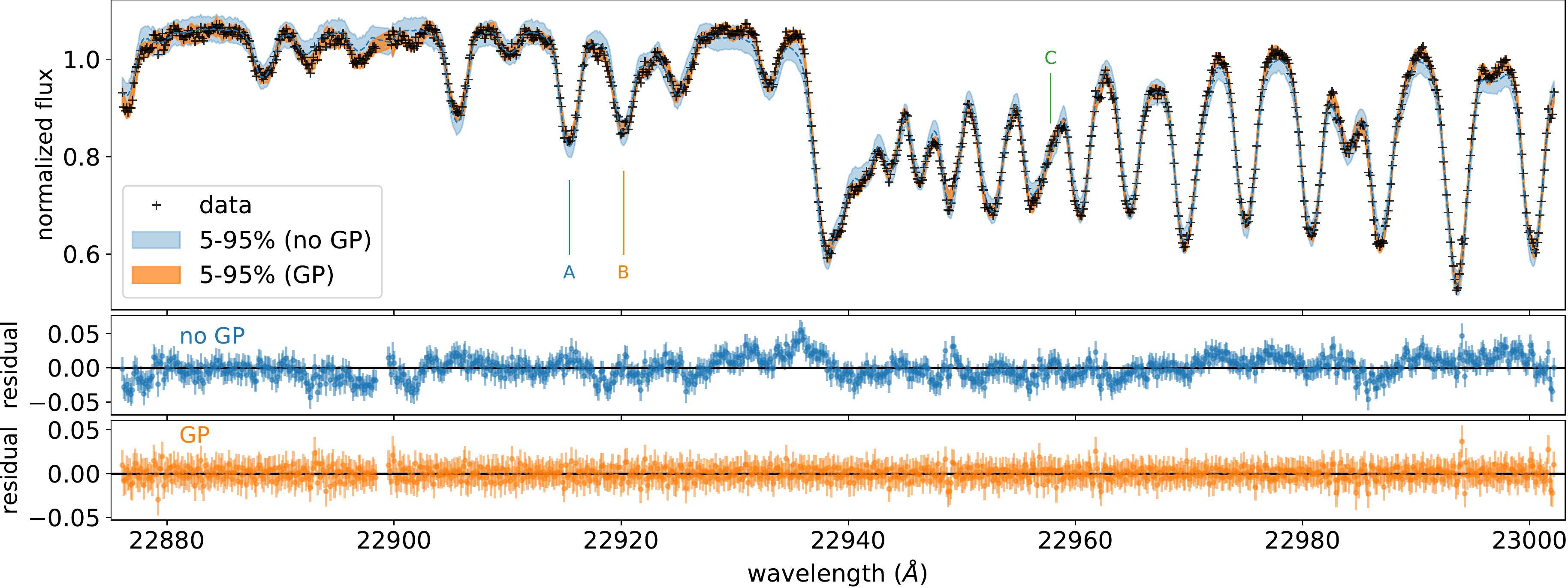}
\caption{ Spectrum of Luhman--16 A (detector 1) and the 90\% credible interval by {\sf exojax} (top). The blue and orange areas correspond to the case of the independent Gaussian noise (Case I) and that of the correlated noise using the Gaussian process (Case II). The bottom two panels show the residuals of the median of the predictions from the data for Cases I and II.  The lines labeled A, B, and C are described in the text. A direct LPF was used to create this figure. \label{fig:results}}
\end{center}
\end{figure*}

The dataset consists of spectra from four CRIRES detectors (S1--S4). First, we fit the model to the spectrum from the first detector (S1; 22876--23002 \AA).  S1 contains a CO band head at 2.29 $\mu$m. Therefore, the baseline level shortward of the CO band head wavelength can be considered to be radiation from the CIA photosphere.

The original dataset is normalized by the blaze function. Therefore, we renormalized the model spectrum by the flux density at 2.29 $\mu$m, $F_{\nu,\mathrm{ref}}$, where the emission is likely dominated by the CIA. We evaluated $F_{\nu,\mathrm{ref}}$ using the mid-resolution spectrum from the Magellan/FIRE Prism Spectrograph ($R \approx 8,000$) \citep{2013ApJ...772..129B} that was normalized to match the $K$-band magnitude of $9.44 \pm 0.07$. 

We take the absolute flux density of $F^{(A)}_\lambda =3 \times 10^{-12} \mathrm{\,\, erg/cm^2/s/\mu m}$ at $\lambda_\mathrm{ref}=$ 2.29 $\mu$m from Figure~2 of \cite{2013ApJ...772..129B}. Then, the reference flux is estimated as
\begin{eqnarray}
F_{\nu ,\mathrm{ref}} (R_p) &=&  \lambda_\mathrm{ref}^2 \left( \frac{R_p}{10 \mathrm{\, pc}} \right)^{-2} F^{(A)}_\lambda \\
&=&  3 \times 10^4 (R_p/R_J)^{-2} \, \mathrm{\,\,erg/cm^{2}/s/cm^{-1}} \nonumber \\
\end{eqnarray}
where $R_p$ is the radius of the brown dwarf, and $R_J$ is the Jovian radius. The spectrum at the top of atmosphere is modeled as
\begin{eqnarray}
\label{eq:model}
F(\nu^\prime) = k_G \ast k_R \ast F_0 (\nu)/F_{\nu ,\mathrm{ref}} (R_p).
\end{eqnarray}
For the IP profile $k_G$, we assume a Gaussian with $R=100,000$, which is much narrower than the broadening due to brown dwarf's rotation. 
%For the stellar rotation $k_R$, we fixed $u_1=0.6$ and $u_2=0.1$ as the limb darkening coefficient based on a theoretical prediction for 1500K and logg=5 for $K$ band \citep{2012A&A...546A..14C}. 
{ For the stellar rotation, we adopted the non-informative prior suggested by \cite{2013MNRAS.435.2152K} for limb darkening, $u_1=2\sqrt{q_1}q_2, u_2=\sqrt{q_1}(1-2 q_2)$ and $q_1,q_2 \sim \mathcal{U}(0,1)$.}
%
%    q1 = numpyro.sample('q1', dist.Uniform(0.0,1.0))
%    q2 = numpyro.sample('q2', dist.Uniform(0.0,1.0))
%    sqrtq1=jnp.sqrt(q1)
%    u1=2.0*sqrtq1*q2
%    u2=sqrtq1*(1.0-2.0*q2)
%    
We do not include the microturbulence in this demonstration. We set 100 atmospheric layers and modeled a T--P profile by a powerlaw with an index of $\alpha$ and $T_0$ as parameters, 
\begin{eqnarray}
T(P) = T_0 \left(\frac{P}{1 \, \mathrm{bar}}\right)^{\alpha},
\end{eqnarray}
where $T_0$ is the temperature at 1 bar.

An emission spectrum contains information on gravity through optical depths of both molecular lines in Equation (\ref{eq:dtau_line2}) and continuous opacity in Equations (\ref{eq:taucont}) and (\ref{eq:scaleheight}). Therefore, the pressure of the photosphere depends on gravity. While the temperature near the photosphere is well constrained from line strength ratios, the pressure is determined from the line pressure broadening and is more difficult to constrain. In addition, the accuracy of the broadening parameters is still limited.
Instead of making gravity a free parameter, we therefore incorporate the dynamical mass constraint from astrometric observations as the prior.  \cite{2017ApJ...846...97G} derived the dynamical mass of Luhman 16 A as $M_p=34.2^{+1.3}_{-1.1} \, M_J$ from a 31 year astrometric dataset. \cite{2018A&A...618A.111L} updated the astrometric mass $M_p = 33.5 \pm 0.3 M_J$. We adopt $\mathcal{N} (33.5,0.3)$ based on the latter constraint as a prior of $M_p$. Under the strong constraint of mass, gravity is simpliy connected to the radius, in other words, a combination of the temperature of the photosphere and the absolute flux.
In section \ref{ss:logg}, we will relax this tight prior on mass and adopt a uniform distribution in order to test whether we can infer spectroscopic mass only using a high dispersion spectrum and absolute flux. 

 \begin{table}[]
  \caption{Model Parameters and Priors}
  \begin{center}
  \begin{tabular}{lcc} 
  \hline\hline
symbol & parameter & prior  \\
 \hline
$M_p$ & mass ($M_J$) & $\mathcal{N} (33.5,0.3)$ \\
$R_p$ & radius ($R_J$) & $\mathcal{U} (0.5,1.5)$ \\
$V\sin{i}$ & projected rotation (km/s) & $\mathcal{U} (10,20)$\\
$\mathrm{RV}$ & radial velocity (km/s) & $\mathcal{U} (26,30)$\\
$\alpha$ & power law index& $\mathcal{U} (0.05,0.15)$ \\
$T_0$ & temperature $T$ at 1 bar & $\mathcal{U} (1000,1700)$\\
$\mathrm{MMR_{CO}}$ & mass mixing ratio of CO & $\mathcal{U} (0,0.01)$\\
$\mathrm{MMR_{H_2O}}$ & mass mixing ratio of water & $\mathcal{U} (0,0.005)$\\
$q_1$ & limb darkening parameter & $\mathcal{U} (0,1)$\\
$q_2$ & limb darkening parameter & $\mathcal{U} (0,1)$\\
\hline
\multicolumn{3}{l}{\it Case I: Independent Gaussian noise} \\
$\sigma_s$ & jitter noise & $\mathcal{E} (10)$ \\ 
\hline
\multicolumn{3}{l}{\it Case II: Correlated noise (GP+RBF kernel)} \\
$\log_{10}{\tau}$ &  correlation length ($\mathrm{cm^{-1}}$) & $\mathcal{U} (-1.5,0.5)$ \\
$\log_{10}{a}$ &  
correlation amplitude 
& $\mathcal{U} (-4,-2)$ \\
  \hline
  \hline
  \end{tabular}
  \end{center}
    \tablecomments{$\mathcal{U}(x_l, x_h)$: uniform distribution in the range between $x_l$ and $x_h$, 
    $\mathcal{E}(\lambda)$: exponential distribution with a rate parameter $\lambda$, 
    $\mathcal{N}(\mu,\sigma)$: normal distribution with a mean of $\mu$ and a standard deviation of $\sigma$.  }
    \label{tab:prior}
\end{table}

\subsubsection{Case I: Independent observational noise}

{ Assuming that the observational noise obeys an independent normal distribution, } the likelihood can be modeled as 
\begin{eqnarray}
\label{eq:likelihood}
\mathcal{L} = \prod_i \frac{1}{\sqrt{{\sigma}_{e,i}^2 + \sigma_s^2}} \exp{\left( - \frac{(d(\nu_i) - F(\nu_i))^2}{2 ({\sigma}_{e,i}^2 + \sigma_s^2) }\right)},
\end{eqnarray}
where $\sigma_{e,i}$ is the observation error for each $\nu_i$ and $\sigma_s$ is an additional error (jitter) to be modeled. The priors are listed in Table \ref{tab:prior}.

\subsubsection{Case II: Including a correlated noise to the model}\label{ss:gpobs}
{

As we will see, there remains a small correlated noise in the residual of Case I. A Gaussian process (GP) is straightforward for modeling correlated noise because {\sf exojax} is compatible with PPLs. We can replace the likelihood in Equation (\ref{eq:likelihood}) with a multivariate normal distribution as 
\begin{eqnarray}
\mathcal{L} = \mathcal{N} (\dv - \Fv(\nuv), \Sigma(\nuv)),
\end{eqnarray}
where $\mathcal{N} (\muv, \Sigma)$ is a multivariate normal distribution with a mean of $\muv$ and a covariance of $\Sigma$, $\boldsymbol{\nu} = (\nu_0,\nu_1,\cdots,\nu_{N-1})^\top$, and $\Fv(\nuv) = (F(\nu_0),F(\nu_1),\cdots,F(\nu_{N-1}))^\top$, In PPLs, this can be simply formulated as 
\begin{eqnarray}
\dv &\sim& \mathcal{N} (\Fv, \Sigma(\nuv)).
\end{eqnarray}
 The covariance is modeled by a GP kernel $K$ as
\begin{eqnarray}
\Sigma_{ij} = K_{ij} + \sigma_{e,i}^2 I.
\end{eqnarray}
The GP kernel takes a functional form as 
\begin{eqnarray}
K_{ij} = a \, \kGP (\nu_i-\nu_j; \tau),
\end{eqnarray}
where $\tau$ expresses the coherent length in the unit of wavenumber and $a$ is the normalization. In this demonstration, we use the radial basis function (RBF) kernel, $\kGP(\nu,\tau)=e^{-\nu^2/2\tau^2}$. In Case II, we infer $\tau$ and $a$ on the priors given in Table \ref{tab:prior}.  The way to calculate the credible interval of the predicted spectrum is described in Appendix \ref{ap:ci}. 

}

\subsubsection{HMC-NUTS setting and computational time}

We performed an HMC \citep[e.g.][]{2012arXiv1206.1901N,2017arXiv170102434B} --  No-U-Turn Sampler \citep[NUTS;][]{2011arXiv1111.4246H} implemented in a PPL, {\sf NumPyro} \citep{2019arXiv191211554P}, with 500 warm-up steps for adaptation phase of HMC and 1000 samples. In this demonstration, we used {\sf Python} 3.8.5, {\sf NumPyro} 0.7.2, and {\sf JAX} 0.2.12 (with {\sf jaxlib} 0.1.64 + {\sf cuda 11.2}) for the external modules and NVIDIA/DGX A100 as a GPU facility and the full analysis took { 29.5 hours (Case II; Direct LPF) and 16.5 hours (Case II; MODIT). We set the maximum number of leapfrog steps at each iteration to be 1023\footnote{{\sf max\_tree\_depth}=10.}. The median number of the steps per sample was 511; most of the iterations did not reach 1023 steps, except for the warm-up phase. The total number of the evaluated steps was $6 \times 10^5$. Hence, the mean computational time of one leapfrog step
was approximately 0.2 s, corresponding to $T_\mathrm{layer} = 2 \times 10^{-3}$ s per layer. This value is a factor larger than the corresponding computational time investigated in \S \ref{ss:benchmark}, which is presumably due to gradient computation and/or the instrumental response. Nevertheless, the order of $T_\mathrm{layer}$ is close to that required to compute the line profile. The additional cost of a gradient--based MCMC {\it per spectrum} is only a factor of several to that of non--gradient based methods. 

}

\subsection{Results}\label{ss:tp}

The observed spectrum and the prediction (mean) with 5--95 \% area from the fitting are shown in Figure \ref{fig:results}. In this fitting, we excluded the region around 22889 \AA, which is significantly contaminated by telluric lines.

{
For Case I, the residual remains below $\sim 5$ \%, which is smaller than that of the forward modeling approach using the BT-Settl library \citep{2013MSAIS..24..128A} by \cite{2014Natur.505..654C} in the original analysis of the data. We find the correlated pattern in the residuals for Case I, including the largest deviation around the left (bluer) side of the band head of CO. 
In the GP model (Case II), this deviation is modeled as a realization of the correlated Gaussian noise, which is taken into account in estimating uncertainties of the other model parameters.
}

\begin{figure*}[htb]
\begin{center}
\includegraphics[width=\linewidth]{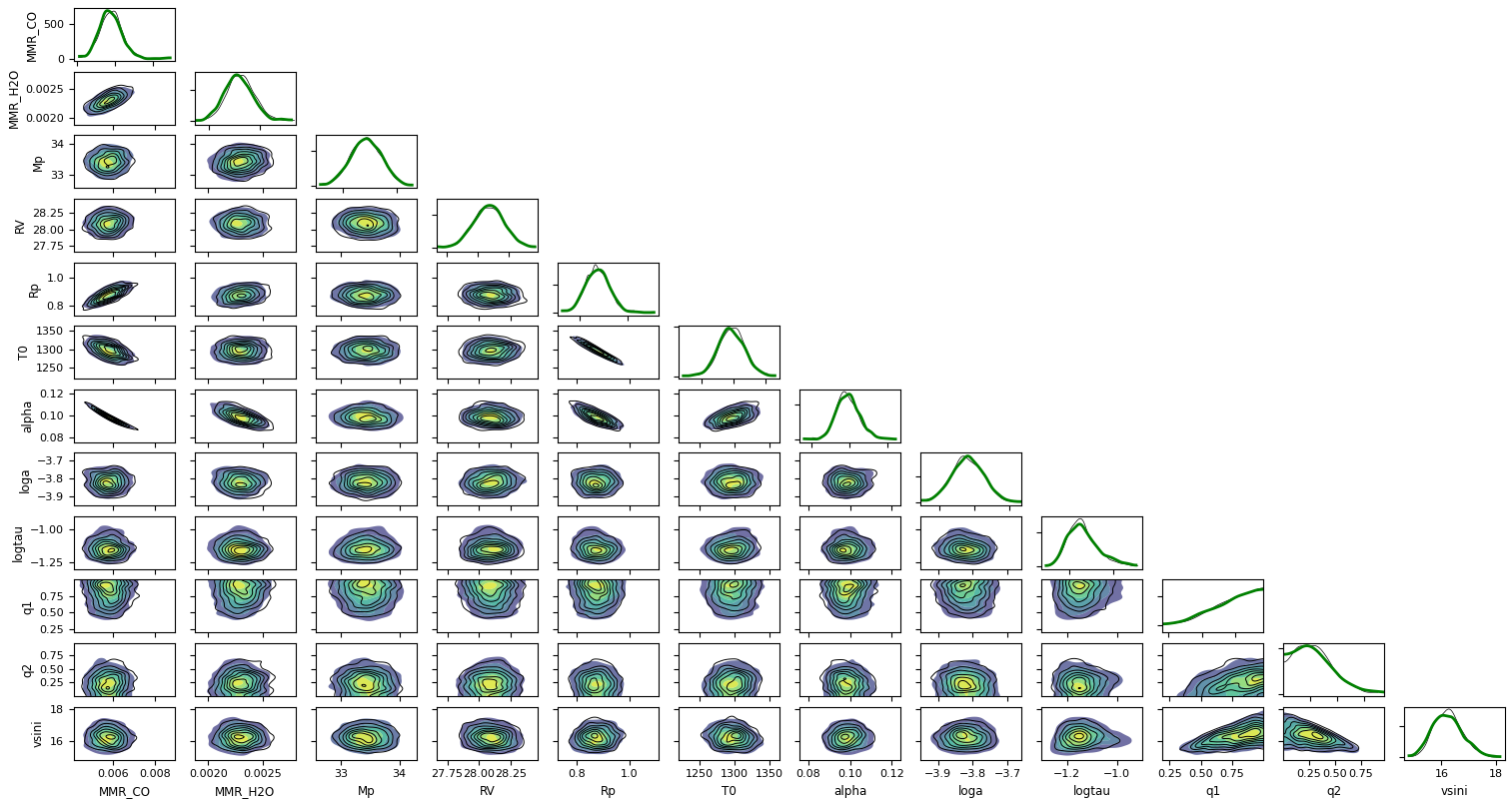}
\caption{Corner plot of the posterior sampling. { The posterior and marginal probability are shown by the colored contours and green lines (Direct LPF), or the thin black contours/lines (MODIT), respectively. }\label{fig:posterior}}
\end{center}
\end{figure*}

 { The posterior samples for Case II are shown in Figure \ref{fig:posterior}. The direct LPF and MODIT provide almost the same results as expected.} There were several clear positive and negative correlations between parameters. The radius is negatively correlated with $T_0$ and $\alpha$ because of the constraint on the absolute flux. The mass mixing ratios of CO and water are also negatively correlated with $\alpha$ to match the observed line depths, that is, less abundance requires a larger gradient of the temperature profile to keep the same observed line depth. These two negative correlations may explain the positive correlations between the mass mixing ratio of CO and $R_p$ and the mass mixing ratios of CO and water.

%\begin{landscape}
%\end{landscape}

Table \ref{tab:prioir} provides the medians of the marginal posteriors and 
their credible intervals (5\%--95\%). The estimated temperature at 1 bar is $T_0=1298_{-35}^{+40}$ K. This value is between the two previous estimates from forward modelings; $T_\mathrm{eff} = $1200 K \citep{2015ApJ...798..127B} and $T_\mathrm{eff} = $1500 K \citep{2014Natur.505..654C}. The estimated $V \sin{i}$ is $16.1_{-0.6}^{+0.5}$ km/s (Case II), which is slightly smaller than the previous estimate of $17.6 \pm 0.1$ km/s using the same data \citep{2014Natur.505..654C}. 
{ We found that the error of $V \sin{i}$ is significantly smaller ($\sim 0.1$ km/s) when fixing the limb darkening coefficient than our fiducial case. This means that incorporating the uncertainty of the limb darkening coefficient into the model is important for inferring  $V \sin{i}$.}

We also computed the carbon-to-oxygen ratio (C/O) as
\begin{eqnarray}
\label{eq:corat}
\mathrm{C/O} &=& \displaystyle{\frac{{\mathrm{MMR}_\mathrm{CO}}/{m_\mathrm{CO}}} {{\mathrm{MMR}_\mathrm{CO}}/{m_\mathrm{CO}} + {\mathrm{MMR}_\mathrm{H2O}}/{m_\mathrm{H2O}}} },
%&=& \left( 1 + \frac{m_\mathrm{CO} \mathrm{MMR}_\mathrm{H2O} }{m_\mathrm{H2O} \mathrm{MMR}_\mathrm{CO}} \right)^{-1}, 
\end{eqnarray}
assuming that all of the carbon and oxygen atoms are distributed to CO or $\mathrm{H_2O}$. Here $m_\mathrm{CO}$ and $m_\mathrm{H2O}$ are the molecular masses of CO and $\mathrm{H_2O}$, $\mathrm{MMR}_\mathrm{CO}$, and $\mathrm{MMR}_\mathrm{H2O}$ are the mass mixing ratios of CO and $\mathrm{H_2O}$. The resultant value of C/O $=0.62_{-0.04}^{+0.03}$ is close to, but slightly higher than the solar value of 0.59 \citep{2021arXiv210501661A} or 0.55 \citep{2009ARA&A..47..481A}.

%penchan:~/fig>python posterior.py
\begin{table*}[]
  \caption{ Median and 5--95\% interval of inferred parameters. \label{tab:prioir}}
\begin{center}
\begin{tabular}{lccc} 
\hline\hline
noise modeling & Case I (no GP) & \multicolumn{2}{c}{Case II (GP)} \\
opacity calculator & direct LPF & direct LPF & MODIT \\
\# of CO lines & \multicolumn{2}{c}{40}  & 204 \\
\# of H2O lines & \multicolumn{2}{c}{334}  & 13874 \\
%computation time & 25.0 hr & 29.5 hr & 24.6 hr \\
\hline
$\mathrm{MMR_{CO}}$ & $0.0059_{-0.0005}^{+0.0004}$ & $0.0057_{-0.0008}^{+0.0011}$& $0.0058 \pm {0.001}$\\
$\mathrm{MMR_{H2O}}$ & $0.0023 \pm {0.0001}$  & $0.0023 \pm {0.0002}$&  $0.0023 \pm 0.0002$\\
$M_p (M_J)$ & $33.2 \pm 0.5$ & $33.2 \pm 0.5$ & $33.5_{-0.4}^{+0.5}$\\
RV (km/s)$\,^\dagger$ &$28.1 \pm {0.1}$ & $28.1 \pm {0.2}$  & $28.1 \pm {0.2}$ \\
$R_p (R_J)$ & $0.89_{-0.04}^{+0.03}$ & $0.88 \pm {0.08}$ & $0.87 \pm {0.08}$ \\
$T_0$ (K) & $1293 \pm 15$ &$1295_{-32}^{+35}$ & $1298 \pm 31$ \\
$\alpha$ & $0.098 \pm {0.005}$ &$0.099_{-0.009}^{+0.011}$ & $0.098_{-0.009}^{+0.012}$ \\
$q_1$ & $0.6 \pm {0.4}$ & $0.8_{-0.3}^{+0.2}$ & $0.8_{-0.3}^{+0.2}$ \\
$q_2$ & $0.5_{-0.4}^{+0.3}$ & $0.3 \pm 0.3$ & $0.3 \pm 0.3$\\
$V \sin{i}$ (km/s) & $15.5_{-0.6}^{+0.8}$ & $16.2_{-0.8}^{+1.0}$ & $16.2_{-0.7}^{+0.9}$ \\
$\sigma$ & $0.0134_{-0.0006}^{+0.0006}$ & -- & -- \\
$\log_{10} a$ & -- & $-3.8 \pm 0.1$ & $-3.8 \pm 0.1$ \\
$\log_{10} \tau $ & -- & $-1.1 \pm 0.1$ &  $-1.1 \pm 0.1$\\
\hline
C/O$\,^\ddagger$ & $0.62_{-0.02}^{+0.01}$ &$0.62 \pm 0.03$ & $0.62_{-0.04}^{+0.03}$ \\
  \end{tabular}
  \end{center}
 \tablecomments{$\dagger$: relative to Earth. $\ddagger$: derived using Equation (\ref{eq:corat})}
\end{table*}

\begin{figure*}[htb]
\begin{center}
\includegraphics[width=0.53\linewidth]{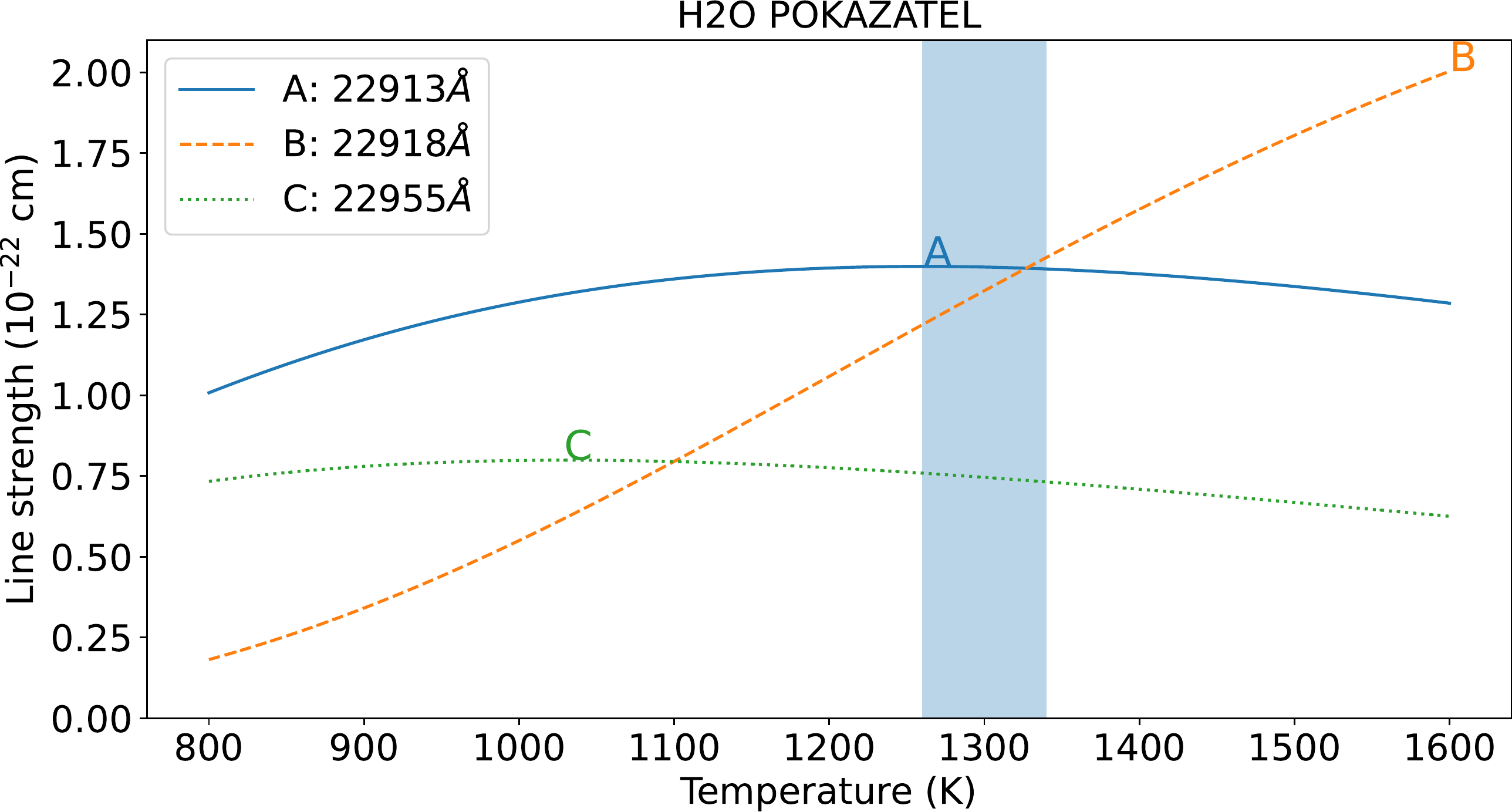}
\includegraphics[width=0.46\linewidth]{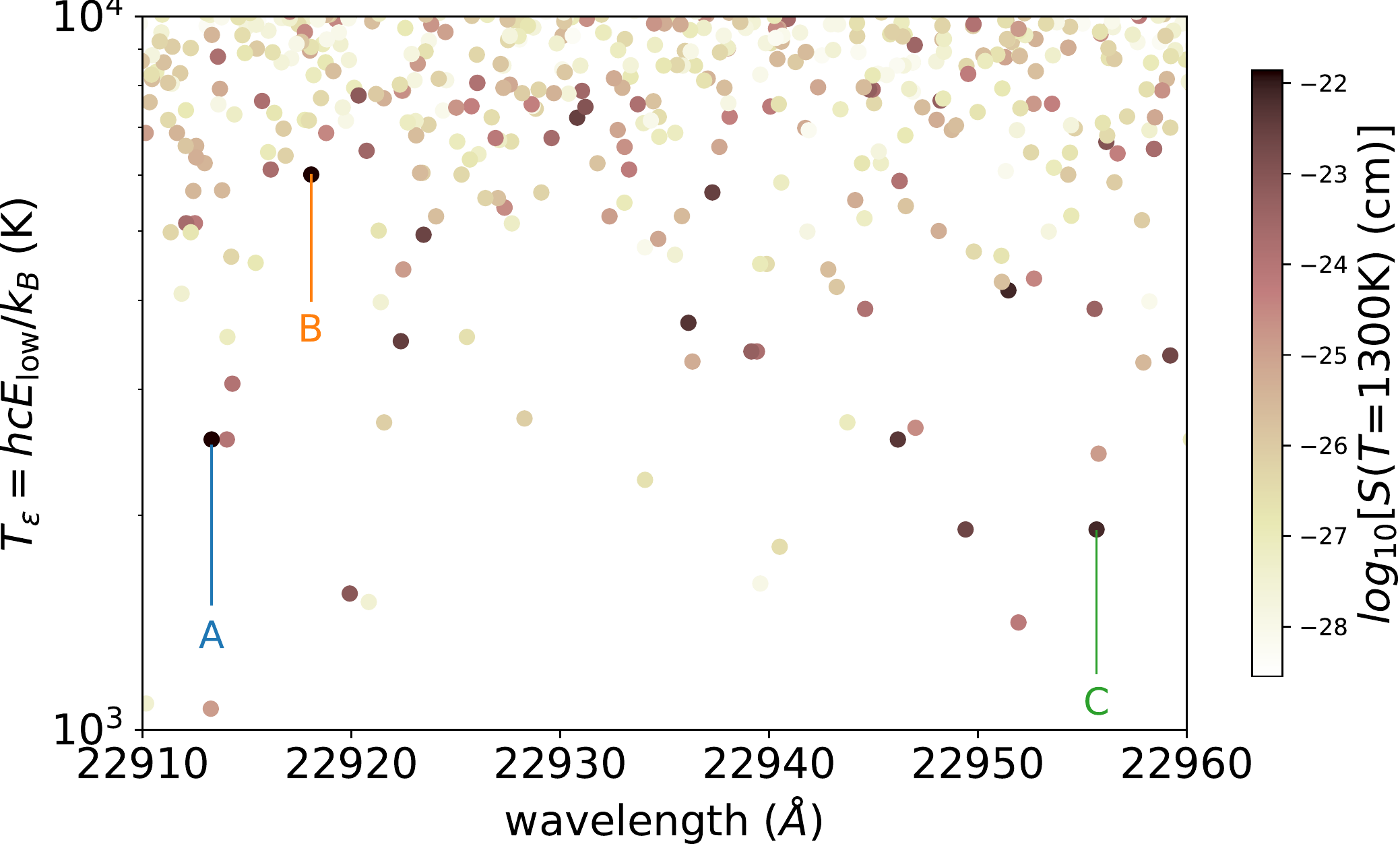}
\caption{Line diagnostics using water (ExoMol POKAZATEL). Left: Temperature dependence of line strength for three representative lines (A, B, and C). The shaded area corresponds to 1260--1340 K, which is approximately the range of posterior of $T_0$. Right: Temperature equivalence of the lower state of water lines. The color indicates the logarithm of the line strength at 1300 K. The lines of A, B, and C are indicated by vertical lines. \label{fig:ls}}
\end{center}
\end{figure*}

Although the HMC-NUTS directly provides the posterior sampling of the parameters, let us look at how the high-dispersion spectrum contains information on these parameters. The ratios of some lines can be regarded as a temperature probe. Figure \ref{fig:ls} shows the temperature dependence of the selected water lines. The strengths of the lines at 22913 and 22918 \AA\ have different temperature dependences, as shown in the left panel. In the case of Luhman 16 A, these two lines exhibit approximately the same level of line depth. This fact directly indicates that the temperature in the layer where the optical depth of these lines is unity is approximately 1300 K, which is consistent with the posterior sampling of $T_0$ at 1 bar.   

How can we deterimine which line pairs are most useful for temperature measurements? The derivative of the line strength ratio with respect to temperature can be regarded as temperature sensitivity of the pair of two lines. From Equation (\ref{eq:linestT}), the sensitivity for the pair of lines whose line centers are close to each other is given by 
\begin{eqnarray}
\label{eq:LDsensitivity}
\eta_{\mathrm{LD}}(T) \equiv \frac{d}{d T} \left( \frac{S_1(T)}{S_2(T)} \right) \approx \frac{T_{\varepsilon,1}-T_{\varepsilon,2}}{T} e^{-\frac{T_{\varepsilon,1} + T_{\varepsilon,2}}{T} }
\end{eqnarray}
where 
\begin{eqnarray}
T_{\varepsilon,l} \equiv h c E_\mathrm{low}/ k_B 
%=  1.44 \, E_\mathrm{low} \mathrm{\,(K)}
= 1.44\,\mathrm{K}\,\left(E_\mathrm{low} \over \mathrm{cm^{-1}}\right)
\end{eqnarray}
is the temperature equivalence of the lower state of the $l$-th line. Equation (\ref{eq:LDsensitivity}) indicates that the temperature can be well constrained if there is a pair with a strong line intensity, a large difference in the value of $T_\epsilon$, and $T_\epsilon$ that is not significantly larger than the temperature range to be measured. 
As shown in the right panel of Figure \ref{fig:ls}, the pair of 22913 and 22918 \AA\ satisfy such conditions. These lines are exhibited in the spectrum in a form that is easy to separate from the other lines. In contrast, the line at 22955 $\mathrm{\AA}$, labeled by ``C'', is more difficult to separate from the CO lines; however, it still contains the temperature information, as shown in Figure \ref{fig:ls}. The full modeling of the high-dispersion spectrum by {\sf exojax} has the advantage of being able to fully use such contaminated lines.

\subsection{Spectroscopic mass}\label{ss:logg}

In Section \ref{ss:tp}, we used a prior of mass from the astrometric results, which was well constrained within 1 \%.  Considering that the temperature is well determined by the line ratio and that we used the absolute flux from the mid-resolution spectrum, the radius was well determined. In this case, gravity was well constrained to $\log{g} = 5.03 \pm 0.04$. 

\begin{figure}[htb]
\begin{center}
\includegraphics[width=\linewidth]{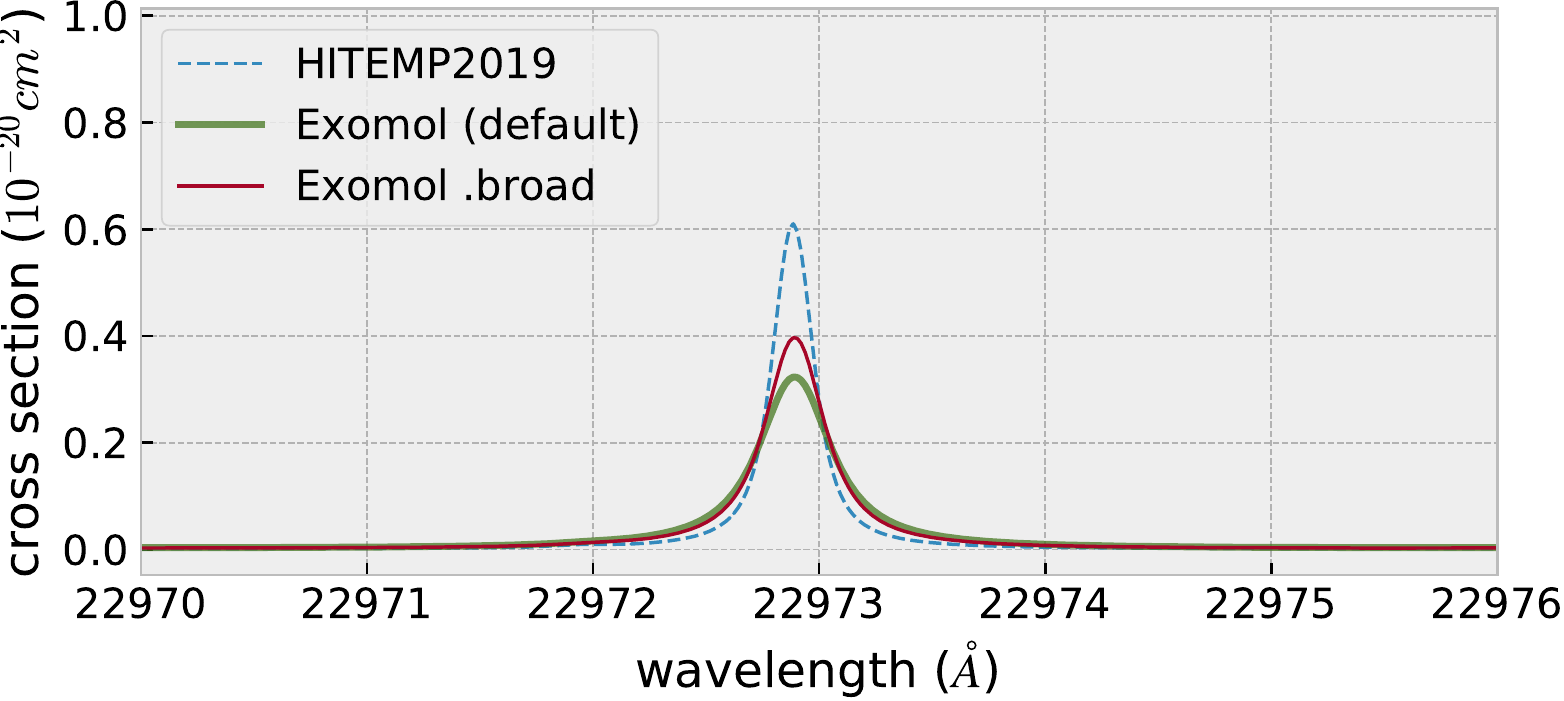}
\caption{Comparison of the CO cross section between HITEMP2019 and ExoMol/Li2015 (default: constant default values of $\alpha_\mathrm{ref}=0.07$ and $n_\mathrm{texp}=0.5$ in the definition file, .broad: broadening parameters in the ``.broad'' file). We assume $T=1300$ K, $P=1$ bar ($P_\mathrm{air}=0.99$ bar and $ P_\mathrm{self}=0.1$ bar for HITEMP2019). \label{fig:compdb}}

\end{center}
\end{figure}

%In principle, 
The high-dispersion spectrum contains information on $\log{g}$ through the line profile, which is constrained by the pressure of the region from which the molecular lines come. This means that the combination of a high-dispersion spectrum and absolute flux, in principle, can constrain the planet/brown dwarf mass without other information. However, this inference should be extremely sensitive to the accuracy of the molecular properties, in particular, $\gamma_L$ and $\ntexp$ in Equations (\ref{eq:gammahitran}) and (\ref{eq:gammaexomol}). In fact, there are significant differences between molecular databases in the broadening parameters of CO, as shown in Figure \ref{fig:compdb}. The difference in the gamma parameter between HITEMP2019 and ExoMol are as follows: $\gamma_L=$ 0.0146-0.025 for HITEMP2019, $\gamma_L=$ 0.0326 for ExoMol (default values in the definition file), and $\gamma_L=$ 0.026-0.033 in the observing band. We note that the broadening parameters in HITEMP2019 are for ``air'' and not for the $\mathrm{H}_2$ atmosphere as explained in Section \ref{ss:hit} and that the broadening parameters given in the broadening file of ExoMol (labeled as ``.broad'' in the caption) were produced by averaging the values of other molecules \citep{2016JQSRT.168..193W}.  

\begin{figure}[htb]
\begin{center}
\includegraphics[width=\linewidth]{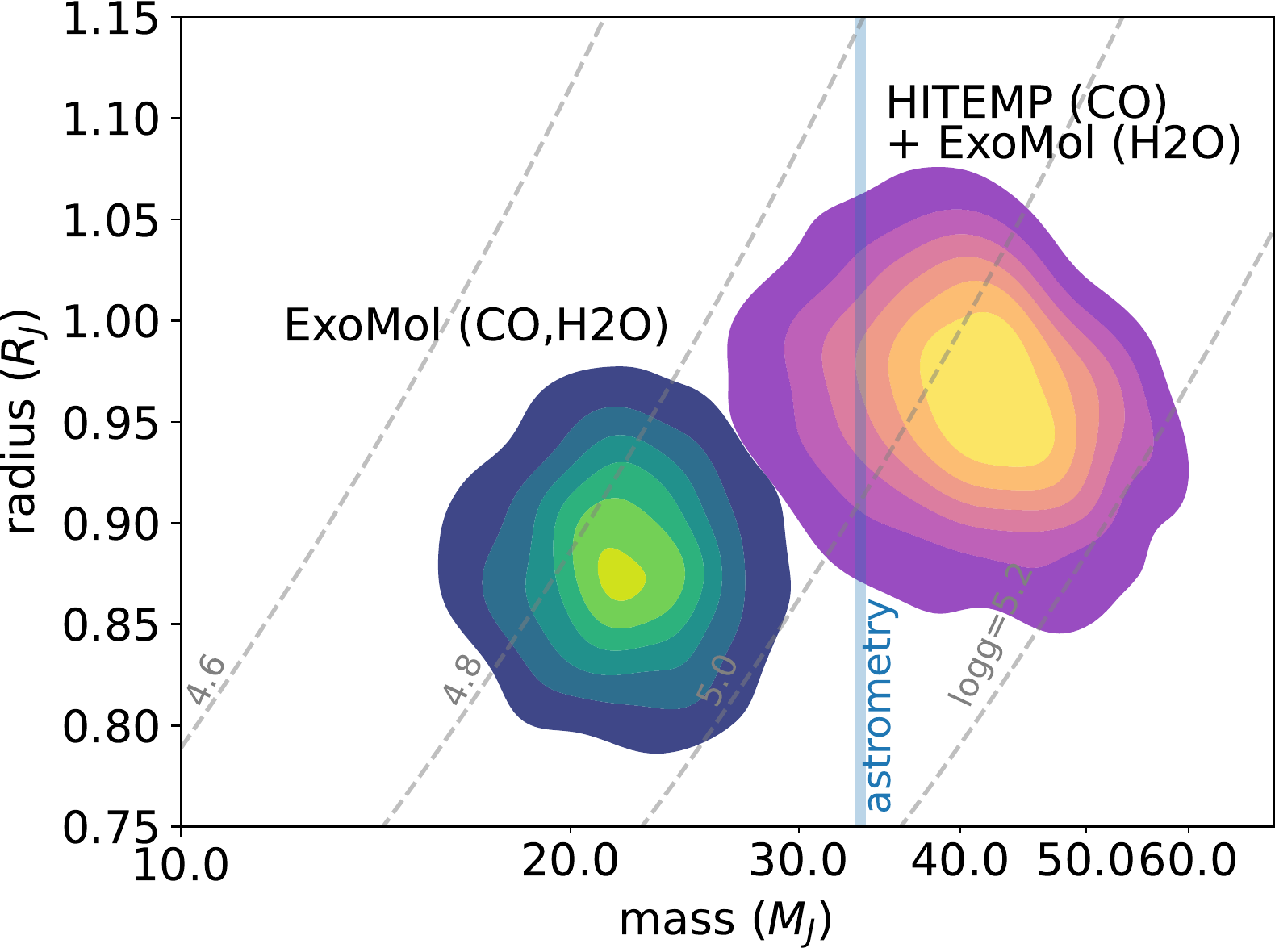}
\caption{%Inferred mass and radius. 
The mass and radius inferred using two different models for CO (ExoMol using the ``.broad'' file and HITEMP2019).
The astrometric mass is shown in the blue shaded region \citep{2018A&A...618A.111L}. The dashed lines show the iso--gravity lines. %The two models for CO (ExoMol using the ``.broad'' file and HITEMP2019) are shown.  
\label{fig:massinf}}
\end{center}
\end{figure}

Despite the incompleteness of the broadening parameters, we attempted to estimate the brown dwarf's mass using ExoMol broadening parameters (.broad) and HITEMP2019 for CO, adopting a uniform prior on the mass. { We adopted the same model as Case II (with GP) except for the mass prior. The results are shown in Figure \ref{fig:massinf}. Although the two estimates does not agree with each other, 
this result indicates that spectroscopic mass estimation is promising if the accuracy of the braodening paramaters is sufficient to model the line profile to determine gravity.} 
Conversely, it is possible to validate the broadening parameter from the high-dispersion analysis of the binary whose astrometric mass is available. 

{

\section{Future Expansion}\label{sec:sum}

We constructed an automatic differentiation model of line-by-line opacity and radiative transfer calculations. We showed that our emission spectrum model works well for a real high-dispersion spectrum of a nearby bright brown dwarf near the L-T transition, Luhman 16 A. Here, we discuss the potential applicability and expansion of our model.

\subsection{Retrieval with Higher Dimensionality}

As mentioned in the Introduction, one of the advantages of HMC over a random-walk Metropolis is rapid convergence for large dimensions of parameters to be inferred, $D$\footnote{See the \href{https://statmodeling.stat.columbia.edu/2017/03/15/ensemble-methods-doomed-fail-high-dimensions/}{blogpost} by Bob Carpenter  for the geometrical limitation of a random walk metropolis in high dimension.}. Thus far, we have 11 or 12 parameters. These values are not significantly larger than those of recent retrievals, for instance, $D=10$ for CHIMERA  \citep{2020AJ....159..117T}, $D=10$--13 for HYDRA-H \citep{2019AJ....158..228G}. 

One of the applications for large dimensionality is the model flexibility of the quantities in layers. Layer-by-layer modeling  \citep{2013ApJ...775..137L} provides more flexibility in the temperature and VMR profiles. Here, we demonstrate a layer-by-layer approach to the temperature profile using {\sf exojax}. This model takes 100 temperature parameters, denoted by vector $\Tpv$. To avoid model instability, we use a multivariate Gaussian type of the prior for $\Tpv$ with an RBF kernel as
\begin{eqnarray}
p(\Tpv) = \mathcal{N}( T_0 {\boldsymbol{u}}, K)
\end{eqnarray}
and
\begin{eqnarray}
 K_{ij} = a_T \, \kGP (\log_{10} {P_i} - \log_{10} {P_j}; \tau_P) + \epsilon a_T \delta_{ij},
\end{eqnarray}
where the second term of the righthand side indicates a small diagonal matrix ($\delta_{ij}$ is the Kronecker delta) that stabilizes the computation of the multivariate normal distribution. We adopted $\epsilon=10^{-5}$. In principle, we can infer $T_0$, a GP amplitude of $a_T$, and a coherent log pressure scale of $\tau_P$ as free (hyper)parameters, as we did in \S \ref{ss:gpobs}. However, it took a long time to converge when we considered $a_T$ and $\tau_P$ as hyperparameters. Therefore, in this test, we take $T_0 \sim \mathcal{U} (1000,1600)$ K and fixed $a_T=10^6$ and $\tau_P=10^{0.5}$. These values ensure both smoothness of the profile and sufficient variance of tempreature, as shown by a random sampling of the prior distribution (the left panel in Figure \ref{fig:hund}).  
 
 Figure \ref{fig:hund} (right) shows a random sampling of the posterior distribution of $\Tpv$. The posterior profile exhibits a tight constraint on temperature for $P = 0.1-10$ bar while the posterior takes wide range of temperature for the other pressure range. For this test, it took about 28 hours even though we reduced the wavenumber range to 22910 -- 22950 \AA  (22 CO lines and 128 water lines), which corresponds to about one third of the data used in Section 5, and adopted Case I (no GP) to the noise model and LPF for the opacity calculator. We adopted 2047 to the maximum number of leap-frog steps. With current GPU performance, the layer-by-layer approach is marginally applicable to real high-resolution data. However, given the current speed of improvement in GPU performance, we expect the layer-by-layer approach to be fully practical within several years even for a large number of lines using MODIT.  

\begin{figure}[htb]
\begin{center}
\includegraphics[width=0.49\linewidth]{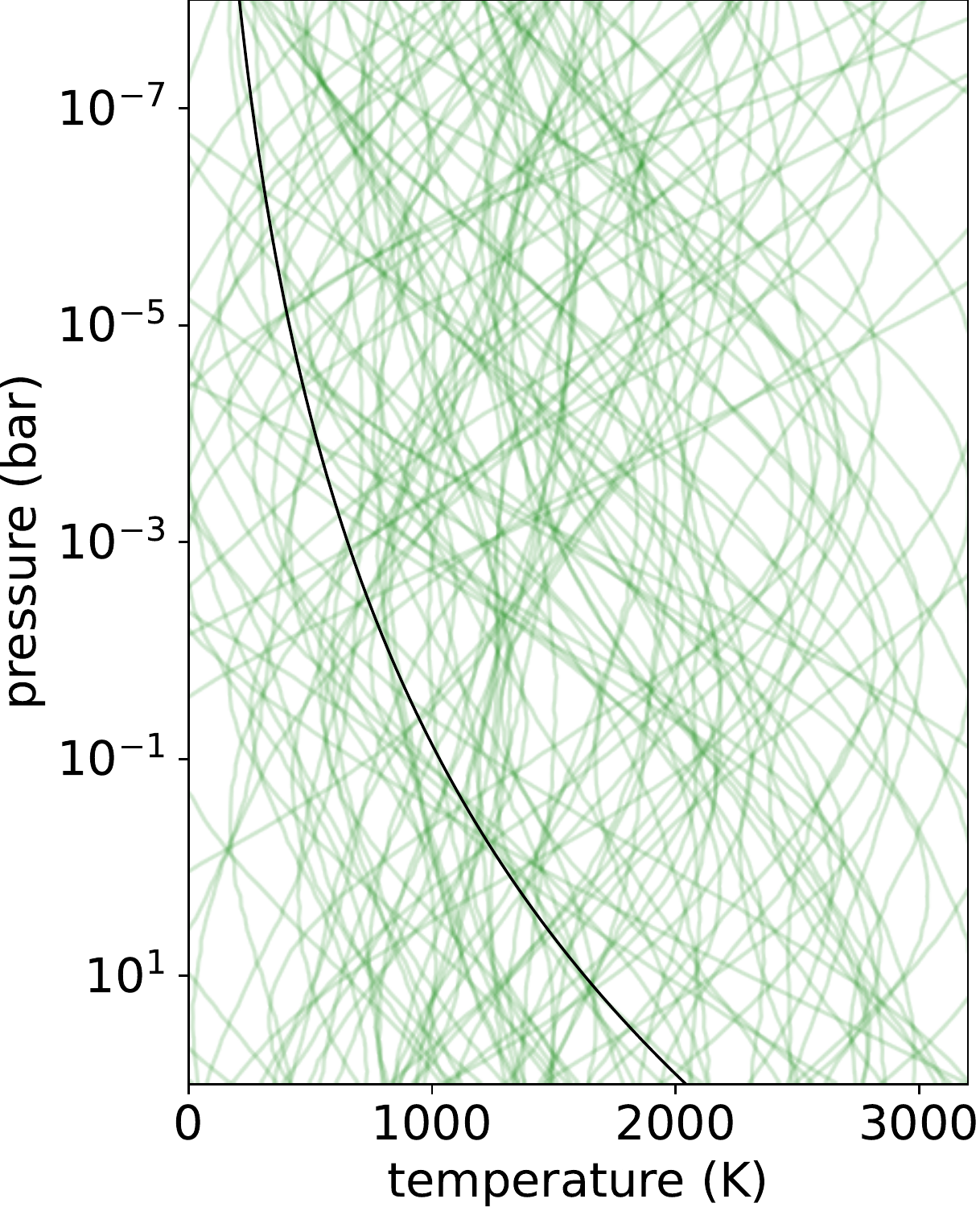}
\includegraphics[width=0.49\linewidth]{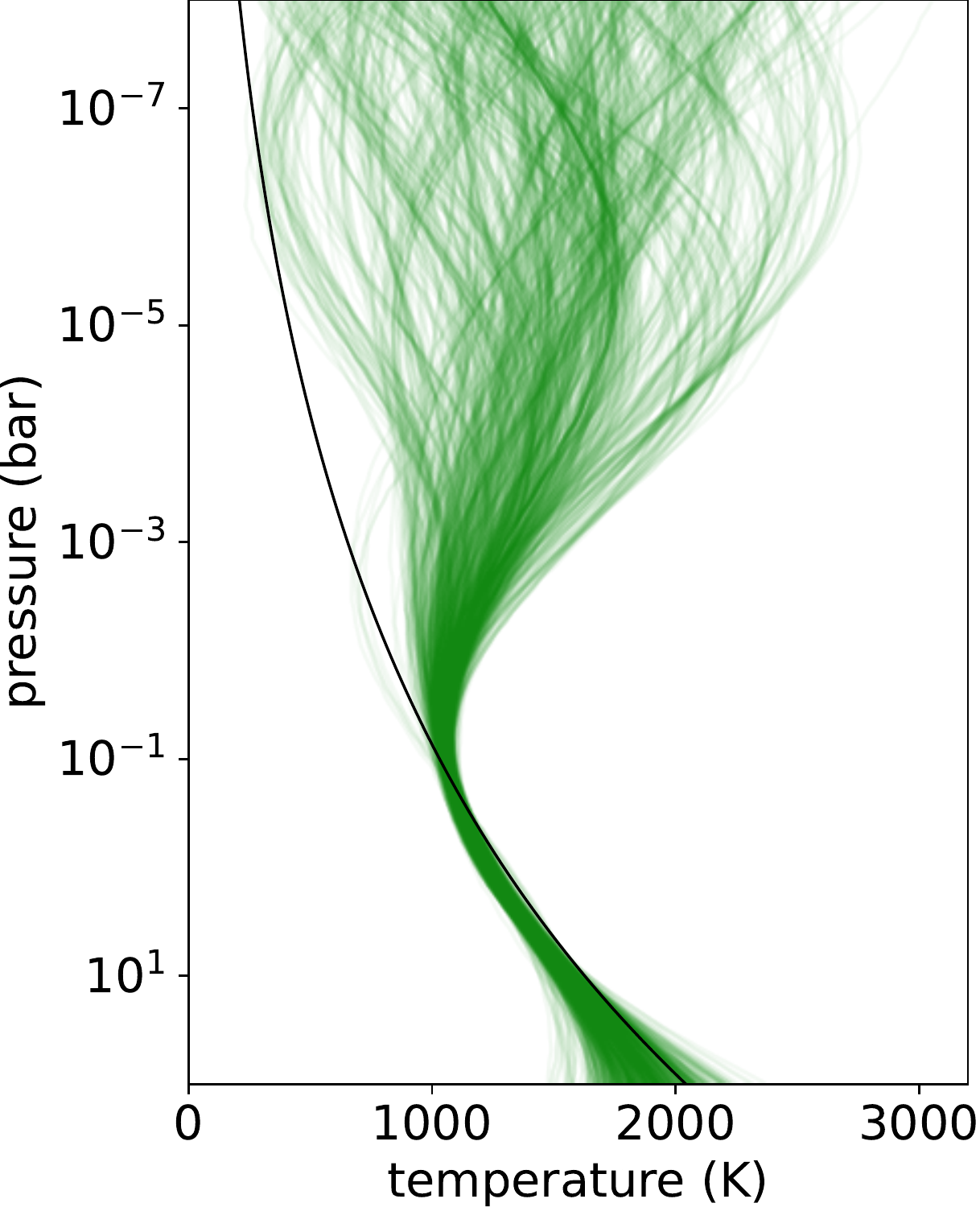}
\caption{ Prior (left) and posterior (right) sampling of the layer-by-layer temeprature model (green lines). The solid black lines indicate the best-fit power law model in Section 5.  \label{fig:hund}}
\end{center}
\end{figure}

}

\subsection{Applicability to High-Dispersion Coronagraphy}
One of the potential applications of our model is an analysis of companion exoplanets/brown dwarfs as observed by a high-dipsersion spectrograph 
\citep{2014Natur.509...63S,2016A&A...593A..74S} and/or HDC, a recently developed technology. 
Motivated by a reasonable agreement between the $K$-band mid-resolution spectrum of Luhman--16A and that of HR8799e \citep{2019A&A...623L..11G}, we simulate the high-dispersion spectrum of a self-luminous exoplanet as taken by HDC by injecting artificial noises to the real spectrum of Luhman 16 A.
We assume that the system is located at 40 pc and is observed for an hour using a spectrograph with a 5\% efficiency on a 8.2 m telescope. We also assume that the HDC suppresses speckle lights at a planet position up to 10 times the planet signal. Based on these assumptions, we injected the Gaussian noise with a standard deviation of 16 \% into the spectrum. 

\begin{figure}[htb]
\begin{center}
\includegraphics[width=\linewidth]{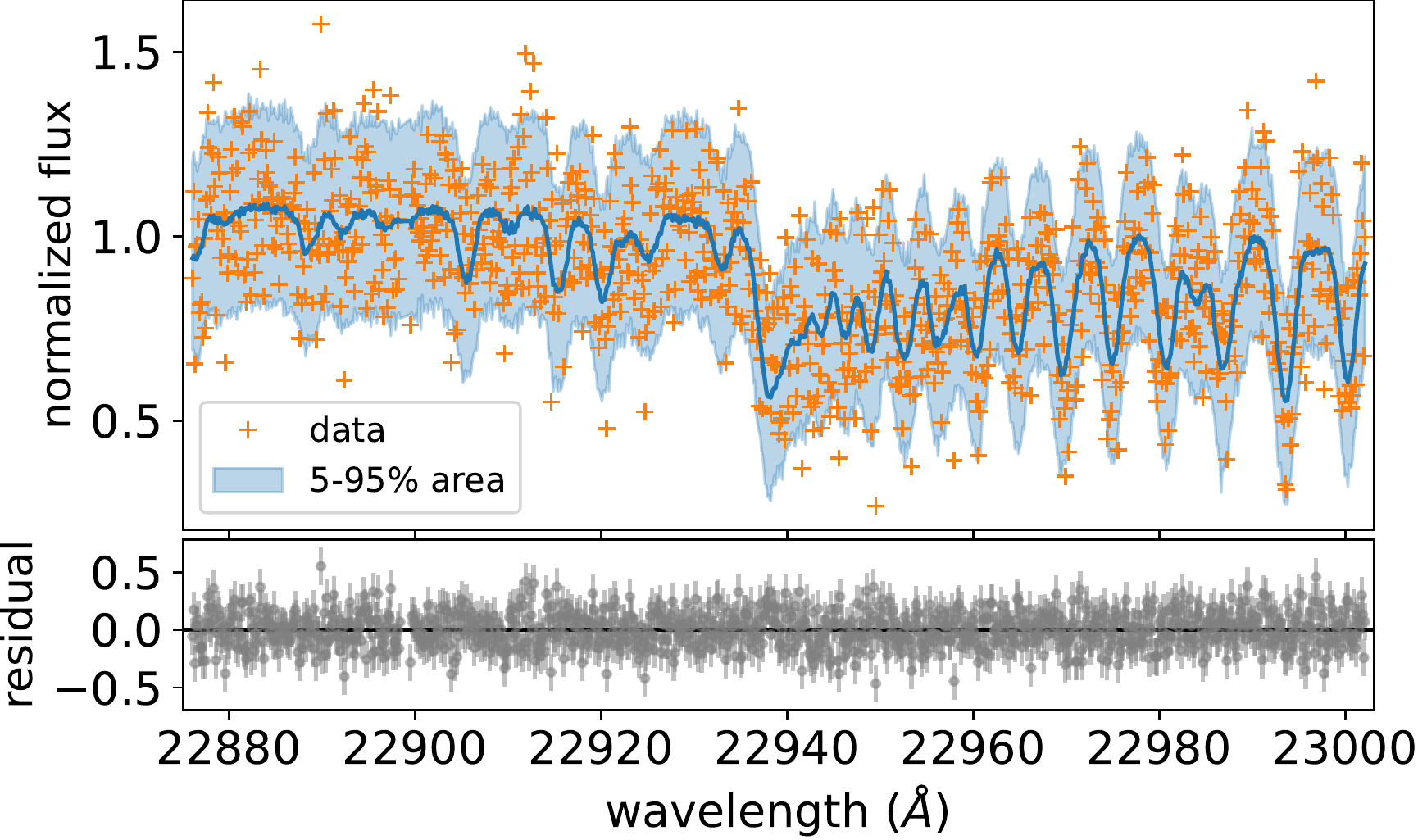}
\caption{Mock spectrum of a self-luminous planet as observed by HDC (top) and the mean and 90\% credible area
by {\sf exojax} (middle). The bottom panel shows the residuals of the median of the predictions from the data.  \label{fig:res2}}
\end{center}
\end{figure}

We did a similar analysis to those in Section \ref{ss:tp} (Case I). As shown in Figure \ref{fig:res2}, {\sf exojax} was able to fit the emission model even to this noisy spectrum. In addition, we obtained the 5--95\% credible intervals of $T_0 =1470_{-140}^{+120}$K,  C/O=$0.59_{-0.08}^{+0.09}$, and $V \sin{i}=15.6^{+0.09}_{-0.08}$ km/s. This result shows that even for such noisy data, our method can be used to constrain the C/O ratio and a planet rotation.

\subsection{Clouds/Dust}

Further developments are required to apply {\sf exojax} to different wavelength ranges and various types of brown dwarfs, exoplanets, and stars. For brown dwarfs, dust opacity becomes important at temperatures between the critical temperature for dust segregation ($T_\mathrm{cr}$) and the condensation temperature ($T_\mathrm{cond}$), i.e. for the region of $T_\mathrm{cr} \lesssim T \lesssim T_\mathrm{cond}$ \citep{2002ApJ...575..264T}.  For observations of L dwarfs/exoplanets hotter than Luhman 16A or in the bluer observing bands, the effect of dust clouds should be included to analyze the spectra in which the dust opacity is appreciably higher than the CIA opacity. In addition, the Rayleigh scattering by $\mathrm{H}_2$ and the line wings of alkali metals should be considered as sources of continuous opacity at the bluer side. 

%In the analysis in Section 5, we were able to exclude most of the water lines because the opacity of CIA is much stronger than most of the H$_2$O line opacity. In general, such a strong continuum is not always present. In the case where weak lines determine the continuum level, we need to include more molecular lines in the fitting. In this case, the computational cost will significantly increases. The next step therefore requires a method to include these weak lines effectively to overcome the computational cost. %{ A novel method for rapid opacity computation using a discrete integral transform \citep{van2021discrete}, which is already implemented in {\sf radis} package \citep{pannier2019radis}, is one of the potential future extensions of {\sf exojax} for massive molecular line lists.}

\subsection{Transmission Spectroscopy}

In principle, the automatic differentiation model of the planetary transmission spectrum is also possible. The current model supports only an emission spectrum model with pure absorption. Support for scattering is required for transmission spectroscopy. { One of the drawbacks of the current formulation of radiative transfer is that it cannot be applied directly to the scattering atmosphere. This is because two recurrence equations need to be solved for the scattering atmosphere, as described in Appendix \ref{ap:rt}. We postpone these issues until forthcoming work. } 

\acknowledgements

{ We greatly appreciate the constructive comments from an anonymous reviewer. We also thank Stevanus Nugroho, Takayuki Kotani, Masayuki Kuzuhara, Hiroyuki Ishikawa, Michael Gully, Brett M. Morris, Daniel Kitzmann, and the REACH collaboration for fruitful discussions. } This study was supported by JSPS KAKENHI grant nos. JP18H04577, JP18H01247, JP20H00170, and 21H04998 (H.K.). In addition, this study was supported by the JSPS Core-to-Core Program Planet2 and SATELLITE Research from the Astrobiology center (AB022006). Y.K. is supported by Special Postdoctoral Researcher Program at RIKEN. 

%Software: 
\software{
{\sf JAX} \citep{jax2018github}, {\sf NumPyro} \citep{2019arXiv191211554P}, {\sf arviz} \citep{arviz_2019}, {\sf HAPI} \citep{2016isms.confETG12K}, {\sf matplotlib} \citep{2007CSE.....9...90H}, {\sf seaborn} \citep{waskom2021seaborn}, {\sf scipy} \citep{virtanen2020scipy}, {\sf numpy} \citep{van2011numpy}.
}

\vspace{\baselineskip}

\appendix

\section{{Comparison {\sf exojax} implementaions of special functions with {\sf scipy}}}\label{ap:scipy}

In this Appendix, we compare the functions used in {\sf exojax}, {\sf hjert}, $\mathcal{T} (x)$ ({\sf trans2E3}), and {\sf erfcx}, with the corresponding functions in {\sf scipy}. The differences not only come from the algorithm and/or approximation of the functions, but also from the fact that {\sf JAX} uses a single--floating point format (F32) as a nominal type. 

\begin{figure}[htb]
\begin{center}
\includegraphics[width=0.5\linewidth]{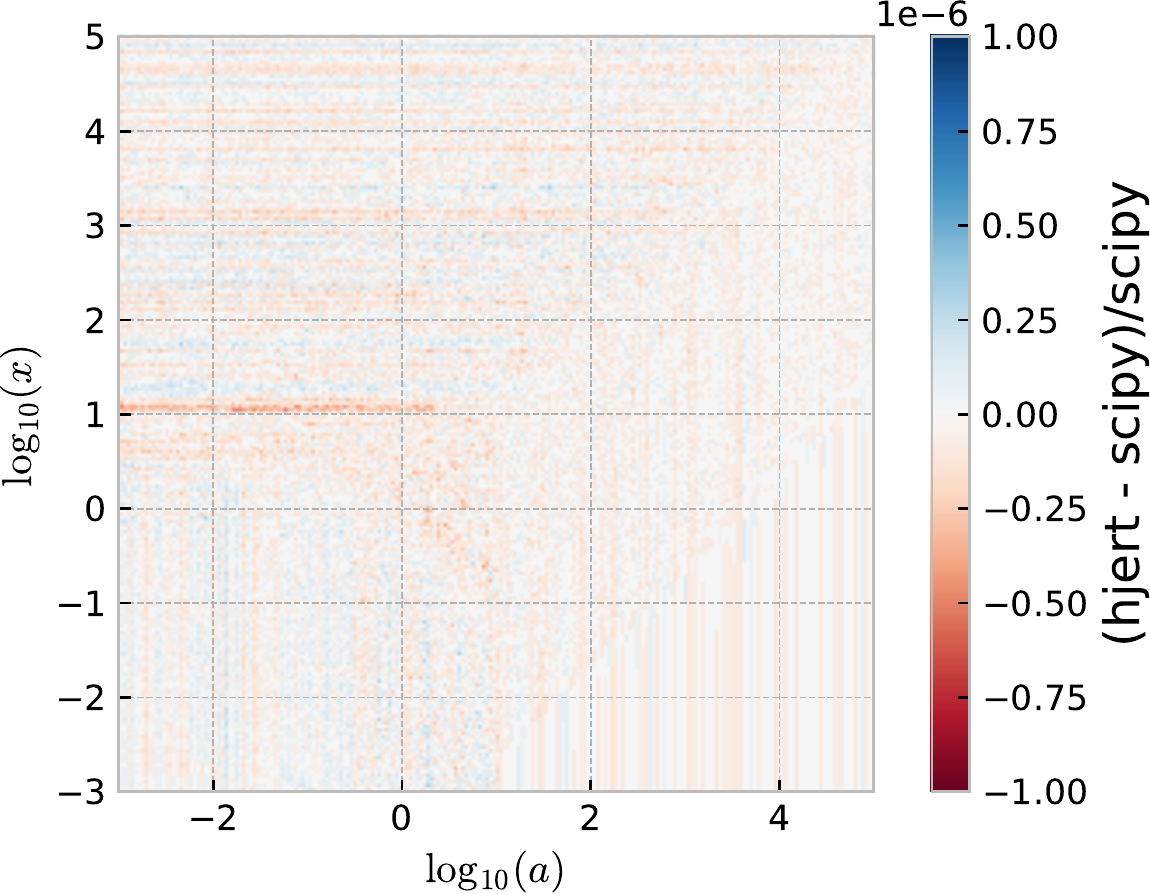}
\caption{Difference between {\sf hjert} (this paper) and {\sf scipy.special.wofz}, following \cite{2021arXiv210102005G}. \label{fig:wofz_ej2}}
\end{center}
\end{figure}

\begin{figure}[]
\begin{center}
\includegraphics[width=\linewidth]{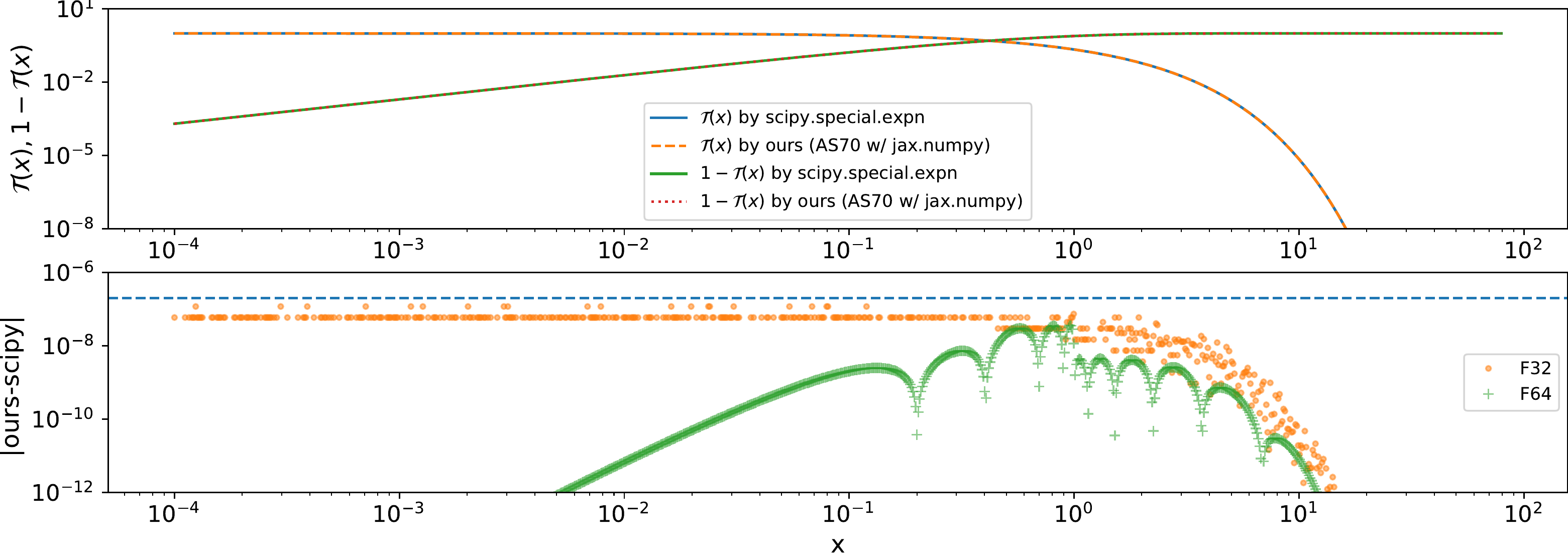}
\caption{Comparison of $\mathcal{T}(x)$ and $1-\mathcal{T}(x)$  using AS70 \citep{1970hmfw.book.....A,2014ApJS..215....4H,2017AJ....154...91L}+{\sf JAX} (ours,  {\sf exojax.rtransfer.trans2E3} ) with {\sf scipy.special.expn}. The bottom panel shows the difference ($|\mathcal{T}(x) \mathrm{(ours)} - \mathcal{T}(x) \mathrm{(scipy)} |$). { The dot and cross symbols correspond to F32 and F64.} The horizontal dashed line indicates $2 \times 10^{-7}$.  \label{fig:E3}}
\end{center}
\end{figure}

\begin{figure}[htb]
\begin{center}
\includegraphics[width=0.7\linewidth]{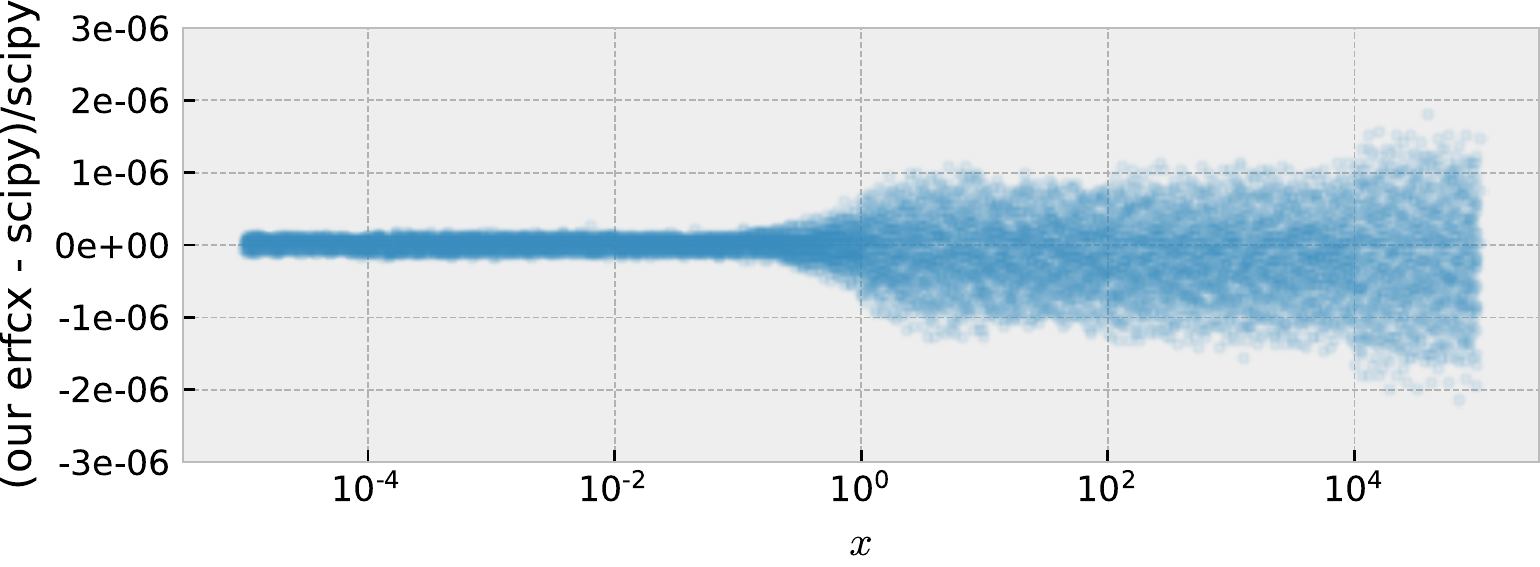}
\caption{Comparison of {\sf exojax.special.erfcx} (ours) with {\sf scipy.erfcx} for $x>0.0$. \label{fig:erfcx}}
\end{center}
\end{figure}

Figure \ref{fig:wofz_ej2} shows the residual of this function ({\sf hjert}) from the real part of {\sf scipy.special.wofz}. The residual is within $10^{-6}$ at least in the range of $10^{-3} < x < 10^5$ and $10^{-3} < a < 10^5$. Figure \ref{fig:E3} compares the transmission function between our implementation of AS70 and those using a {sf scipy} special function, {\sf expn}. The difference is within $2 \times 10^{-7}$ in $10^{-4} < x < 10^2$. Figure \ref{fig:erfcx} shows the relative error between our implementation and {\sf scipy.erfcx}. The difference is smaller than $2 \times 10^{-6}$ in the range we use. 

%exojax/tests/xs/f64

{

The residuals described above were computed using a single--precision floating--point format (F32) because {\sf JAX} uses F32 as a nominal precision. {\sf JAX} can use the double--precision floating--point format (F64), although XLA is not available for F64. Table \ref{tab:f64} shows the maximum differences between {\sf erfcx}, {\sf hjert}, and {\sf trans2E3} from {\sf scipy} using F32 and F64. The residual of {\sf erfcx} using F64 is two orders of magnitude smaller than that obtained using F32. The residual of {\sf hjert} using F64 is at most one order of magnitude smaller than that using F32 in most parameter ranges. However, F32 and F64 for {\sf trans2E3} have a similar level of residuals at $x>0.1$, as shown in Figure \ref{fig:E3}. 

\begin{table}[]
  \caption{Parameters for molecular lines \label{tab:f64}}
 \begin{center}
  \begin{tabular}{lccc} 
  \hline\hline
function & F32 & F64 \\
\hline
{\sf erfcx} & $< 2 \times 10^{-6}$  & $< 2 \times 10^{-8}$  \\
{\sf hjert} & $< 1 \times 10^{-6}$  & $< 1 \times 10^{-6}$  \\
{\sf trans2E3} & $< 2 \times 10^{-7}$  & $< 5 \times 10^{-8}$  \\
  \end{tabular}
  \end{center}
  \tablecomments{}
\end{table}

\section{Difference in F32 and F64 for MODIT}\label{sec:moditf64}

Because MODIT uses FFT in its algorithm, F32 has a relatively large impact on the accuracy of profiles with a wide dynamic range. In particular, F32 affects the precision of the bottom of the lines in the wide--dynamic case. Figure \ref{fig:comp_modit} shows the difference between the direct LPF and MODIT for F32 and F64. In this case, the dynamic range of opacity was approximately six orders of magnitude. At the bottom of the line profile, the residual becomes larger than 1 \% for F32, while it remains below 1 \% for F64. However, in many cases where MODIT is necessary, such a wide dynamic range of opacity is unlikely to occur because a large number of lines exist in a narrow wavelength range.

\begin{figure*}[htb]
\begin{center}
\includegraphics[width=\linewidth]{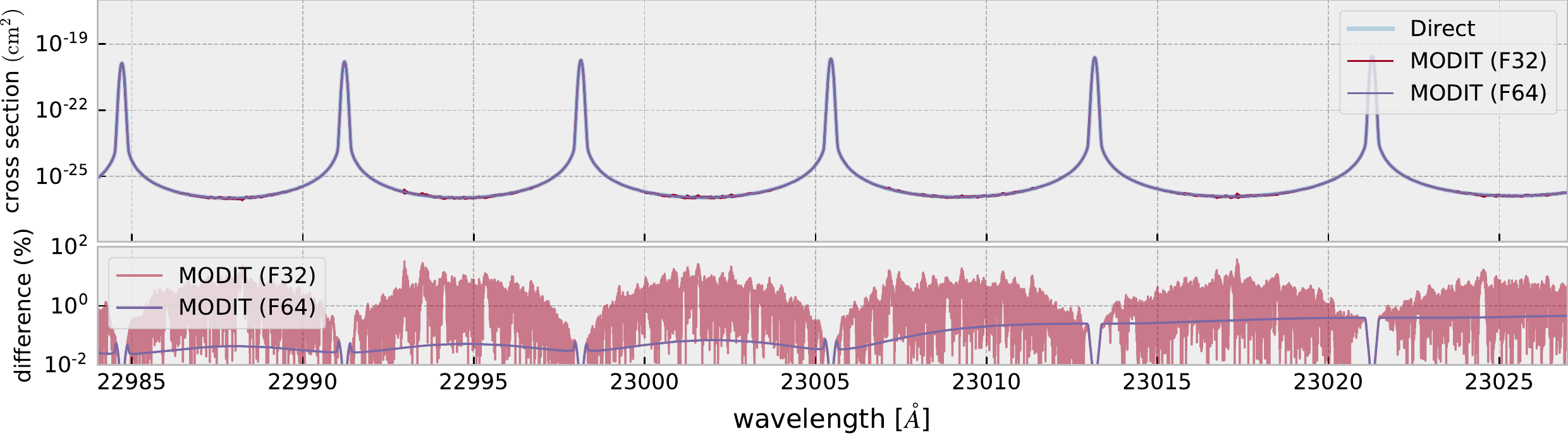}
\caption{Comparison of MODIT F32 and F64 with the direct LPF. \label{fig:comp_modit}}
\end{center}
\end{figure*}
}

\section{Imaginary part of the Faddeeva function}\label{ap:Lxa}

Here, we consider the imaginary part of the wofz function $w(x+i a)$ as 
\begin{eqnarray}
L(x,a) \equiv \mathrm{Im}[w(x +i a)].
\end{eqnarray}
The expansion of $L(x,a)$ was given by \cite{2011arXiv1106.0151Z} as
\begin{eqnarray}
\label{eq:algo916L}
\hat{L}_{\Ncut}(x,a) &=& - e^{-x^2} \mathrm{erfcx} (a) \sin{(2 x a)} \nonumber \\
&+& \frac{2 \eta x \sin{(a x)}}{\pi}  e^{-x^2} \mathrm{sinc} (2 a x/\pi) \nonumber \\
&+& \frac{2 \eta}{\pi} \left[ a \sin{(2 a x)} \Sigma_1 - \frac{a}{2} \Sigma_4 + \frac{a}{2} \Sigma_5 \right],
\end{eqnarray}
where
\begin{eqnarray}
\Sigma_4 &=& \sum_{n=1}^{\Ncut} \left( \frac{\eta n}{\eta^2 n^2 + a^2}\right) e^{-(\eta n + x)^2}  \\
\Sigma_5 &=& \sum_{n=1}^{\Ncut} \left( \frac{\eta n}{\eta^2 n^2 + a^2}\right) e^{-(\eta n - x)^2} 
\end{eqnarray}
Similar to $H(x,a)$, we implemented {\sf ljert} by combining $\hat{L}_{\Ncut}(x,a)$ and the asymptotic form of the wofz function (\ref{eq:asywofz}) as
\begin{eqnarray}
L(x,a) = 
  \begin{cases}
    \hat{L}_{27}(x,a) & \mbox{for $x^2 + a^2 < 111$} \\
    \mathrm{Im} [ \,w_2^\mathrm{asy}(x+i a)\,] & \mbox{otherwise.} 
  \end{cases}
\end{eqnarray}

\section{Jacobian Vector Product}\label{ap:JVP}

Here, we consider $\fv(\xv)$ where $\xv \in \mathbb{R}^N$,$\fv \in \mathbb{R}^M$. 
Jacobian 
\begin{eqnarray}
J_{\fv,\xv} \equiv \frac{\partial \fv}{\partial \xv} 
= \displaystyle{\left(
    \begin{array}{cccc}
      \frac{\partial f_1}{\partial x_1} &\frac{\partial f_1}{\partial x_2} & \ldots & \frac{\partial f_1}{\partial x_N} \\
            \frac{\partial f_2}{\partial x_1} &\frac{\partial f_2}{\partial x_2} & \ldots & \frac{\partial f_2}{\partial x_N} \\
      \vdots & \vdots & \ddots & \vdots \\
      \frac{\partial f_M}{\partial x_1} &\frac{\partial f_M}{\partial x_2} & \ldots & \frac{\partial f_M}{\partial x_N} \\
    \end{array}
  \right)}
\in \mathbb{R}^{M \times N} 
\end{eqnarray}
is essential for automatic differentiation. Let us consider $\fv(\xv) = \fv_2(\fv_1(\xv))$. The Jacobian chain rule is expressed as
\begin{eqnarray}
J_{\fv,\xv} &=& \frac{\partial \fv}{\partial \xv} = \left( \frac{\partial \fv_2 (\xv_1)}{\partial \xv_1}  \right)  \left( \frac{\partial \xv_1}{\partial \xv}  \right) = \left( \frac{\partial \fv_2 (\xv_1)}{\partial \xv_1}  \right)  \left( \frac{\partial \fv_1(\xv)}{\partial \xv}  \right) \\
&=& J_{\fv_2,\xv_1} J_{\fv_1,\xv}.
\end{eqnarray}
In general, a chain rule for $\fv(\xv) = \fv_L(\fv_{L-1}( \cdots \fv_2(\fv_1(\xv))))$, is given by 
\begin{eqnarray}
\label{eq:chain}
J_{\fv,\xv}  =  J_{\fv_L,\xv_{L-1}} \cdots J_{\fv_2,\xv_1} J_{\fv_1,\xv}. 
\end{eqnarray}
Automatic differentiation has two types for computing Jacovian for each $\fv_i$. The Jacobian vector product (JVP) is defined as 
\begin{eqnarray}
\mathrm{JVP}(\fv,\uv) = J_{\fv,\xv} \uv = \left(\frac{\partial f_1}{\partial \xv} \uv,\ldots,\frac{\partial f_M}{\partial \xv} \uv \right)^\top.
\end{eqnarray}
Once JVP is given, the $(i,j)$-th component of the Jacobian of $\partial f_i/\partial x_j$ can be computed by the $i$-th component of 
\begin{eqnarray}
\label{eq:extJac}
\mathrm{JVP}(\fv,\ev_j) &=& \frac{\partial \fv}{\partial x_j},
\end{eqnarray}
where $\ev_j$ is the $j$-th unit vector. This type of auto-differentiation is called the forward differentiation because the JVP for each function is sequentially applied from the right side as
\begin{eqnarray}
\label{eq:chain_c}
\mathrm{JVP}(\fv,\uv) = J_{\fv,\xv} \uv  =  J_{\fv_L,\xv_{L-1}} \cdots J_{\fv_2,\xv_1} J_{\fv_1,\xv} \uv = \mathrm{JVP}(\fv_L, \ldots \mathrm{JVP}(\fv_{2}, \mathrm{JVP}(\fv_{1}, \uv)) \ldots ). 
\end{eqnarray}

{
\section{Two-stream approximation with pure absorption}\label{ap:rt}
We derive Equation (\ref{eq:rerr}) from a general expression for the two-stream approximation. The two-stream expression of the 1 dimensional parallel plane atmosphere considers the diffusive flux in the upward and downward directions. The former is the upward flux $F$ (we omitted the superscript of $\uparrow$ here) considered in Equation (\ref{eq:rerr}) and the latter is denoted by $F^{\downarrow}$. A general form of the two-stream approximation, including scattering, can be written as follows \citep{1989JGR....9416287T}:
\begin{eqnarray}
\label{eq:twostream}
\frac{\partial F}{\partial \tau} &=& \gamma_1 F - \gamma_2 F^{\downarrow} - \mathcal{S} \nonumber \\
\frac{\partial F^{\downarrow}}{\partial \tau} &=& - \gamma_1 F^{\downarrow} + \gamma_2 F + \mathcal{S} 
\end{eqnarray}
where the factors $\gamma_1$ and $\gamma_2$ depend on a particular form such as Eddington, quadrature, hemispheric mean approximations, and the source term is 
\begin{eqnarray}
\label{eq:twostream_S}
 \mathcal{S} = 2 \pi \overline{\mu} (1 - \omega_0) B_\nu (T)
\end{eqnarray}
and $\omega_0$ is the single scattering albedo and $\overline{\mu}$ is the characteristic cosine of the co--latitude. For a general solution of (\ref{eq:twostream}), one needs to solve a second-order differential equation for $F_n$ using iterative methods such as a tridiagonal approach. 

\begin{figure*}[htb]
\begin{center}
\includegraphics[width=0.55\linewidth]{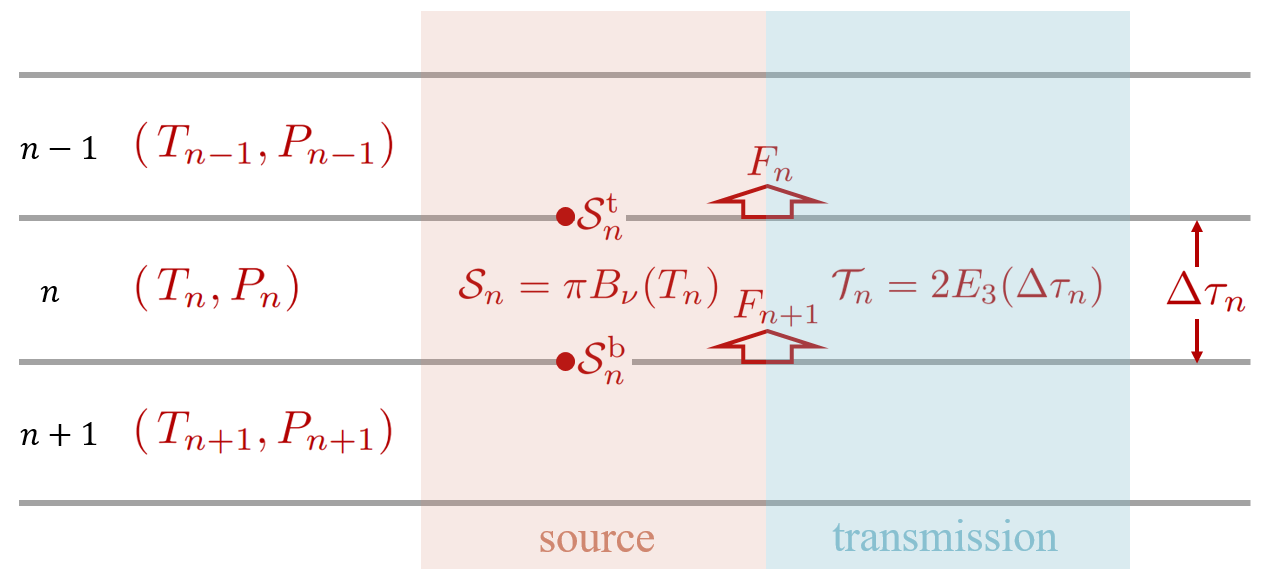}
\caption{Definitions of physical quantities in the atmospheric layers. \label{fig:layer}}
\end{center}
\end{figure*}

\cite{heng2017exoplanetary} derived the expression for a pair of representative neighboring points in atmospheric layers from Equation (\ref{eq:twostream_S}) using a Taylor expansion of the source function.
The definitions of the indexing of the atmospheric layers are shown in Figure \ref{fig:layer}. The source functions at the upper and lower boundaries of the $n$-th layer are denoted by $\mathcal{S}^t_n$ and $\mathcal{S}^b_n$.  
Rewriting the expression in  \cite{heng2017exoplanetary}  to those for the upper and lower boundaries in the $n$-th layer,\footnote{
The conversions of (3.34) in \cite{heng2017exoplanetary} to our notations are as follows:\\
$\pi B_{1+} = \mathcal{S}^b_n - \mathcal{S}_n^\prime \Delta \tau + \mathcal{S}_n^\prime/2 = \mathcal{S}^t_n  + \mathcal{S}_n^\prime/2 $,\\
$\pi B_{1-} = \mathcal{S}^t_n  - \mathcal{S}_n^\prime/2 $,\\
$\pi B_{2+} = \mathcal{S}^b_n + \mathcal{S}_n^\prime/2$,\\
$\pi B_{2-} = \mathcal{S}^t_n + \mathcal{S}_n^\prime \Delta \tau - \mathcal{S}_n^\prime/2 = \mathcal{S}^b_n  - \mathcal{S}_n^\prime/2$,
}
 we obtain
\begin{eqnarray}
\label{eq:twostream_Stwo}
%F_{n} &=& \Gamma_{1,n} \mathcal{T}_n F_{n+1} - \Gamma_{2,n} F^\downarrow_{n} + (\mathcal{S}^b_n - \mathcal{S}_n^\prime \Delta \tau + \mathcal{S}_n^\prime/2) - \Gamma_{1,n} \mathcal{T}_n (\mathcal{S}^b_{n} + \mathcal{S}_n^\prime/2) + \Gamma_{2,n} (\mathcal{S}^t_{n} - \mathcal{S}_n^\prime/2)   \\
F_{n} &=& \Gamma_{1,n} \mathcal{T}_n F_{n+1} - \Gamma_{2,n} F^\downarrow_{n} + (\mathcal{S}^t_n  + \mathcal{S}_n^\prime/2) - \Gamma_{1,n} \mathcal{T}_n (\mathcal{S}^b_{n} + \mathcal{S}_n^\prime/2) + \Gamma_{2,n} (\mathcal{S}^t_{n} - \mathcal{S}_n^\prime/2)   \\
%F^\downarrow_{n} &=& \Gamma_{1,n-1} \mathcal{T}_{n-1} F^\downarrow_{n-1} - \Gamma_{2,n-1} F_{n} + (\mathcal{S}^t_{n-1} + \mathcal{S}_{n-1}^\prime \Delta \tau - \mathcal{S}_{n-1}^\prime/2)  - \Gamma_{1,n-1} \mathcal{T}_{n-1} (\mathcal{S}^t_{n-1} - \mathcal{S}_{n-1}^\prime/2) + \Gamma_{2,n-1} (\mathcal{S}^b_{n-1} + \mathcal{S}_{n-1}^\prime/2) \nonumber \\
F^\downarrow_{n} &=& \Gamma_{1,n-1} \mathcal{T}_{n-1} F^\downarrow_{n-1} - \Gamma_{2,n-1} F_{n} + (\mathcal{S}^b_{n-1}  - \mathcal{S}_{n-1}^\prime/2)  - \Gamma_{1,n-1} \mathcal{T}_{n-1} (\mathcal{S}^t_{n-1} - \mathcal{S}_{n-1}^\prime/2) + \Gamma_{2,n-1} (\mathcal{S}^b_{n-1} + \mathcal{S}_{n-1}^\prime/2) \nonumber \\
\end{eqnarray}
where 
\begin{eqnarray}
\Gamma_{1,n} &\equiv& \frac{\zeta^2_- - \zeta_+^2}{( \mathcal{T}_n \zeta_-)^2 - \zeta_+^2} \\
\Gamma_{2,n} &\equiv& \frac{(1-\mathcal{T}_n^2) \zeta_- \zeta_+ }{( \mathcal{T}_n \zeta_-)^2 - \zeta_+^2} \\
\zeta_\pm &\equiv& \frac{1}{2} (1 \pm \sqrt{1 - \omega_0}) \\
\mathcal{S}_n^\prime &=& \left. \frac{\partial \mathcal{S} (\tau)}{\partial \tau} \right|_{n-\mathrm{th \,layer}} \approx \frac{\mathcal{S}_n^{\mathrm{b}} - \mathcal{S}_n^{\mathrm{t}}}{\Delta \tau_n}
\end{eqnarray}

A pure absorption assumption (no scattering, i.e., $\omega_0=0$, $\zeta_+=1$, $\zeta_-=0$, $\Gamma_{1,n}=1$, $\Gamma_{2,n}=0$ ) results in Equation (\ref{eq:twostream_Stwo}) to a closed form for $F_{n}$ as 
\begin{eqnarray}
\label{eq:rerr_ex}
F_{n} = \mathcal{T}_n F_{n+1} + \mathcal{S}_n^{\mathrm{t}} - \mathcal{T}_n \mathcal{S}_n^{\mathrm{b}} + (1 - \mathcal{T}_n) \frac{\mathcal{S}_n^\prime}{2}.
\end{eqnarray}
For a pure absorption atmosphere, the transmission can be written as \citep{2014ApJS..215....4H}
\begin{eqnarray}
\mathcal{T}_n = 2 E_3 (\Delta \tau_n). 
\end{eqnarray}
We denote a representative value of the source function of the $n$--th layer by
\begin{eqnarray}
\mathcal{S}_n = \pi B_\nu(T_n).
\end{eqnarray}
Then, we can write 
\begin{eqnarray}
\label{eq:rerr_exade}
\mathcal{S}_n^{\mathrm{t}} &=& \mathcal{S}_n - (\delta \mathcal{S})^{\mathrm{t}} \\ 
\mathcal{S}_n^{\mathrm{b}} &=& \mathcal{S}_n + (\delta \mathcal{S})^{\mathrm{b}} 
\end{eqnarray}
where $(\delta \mathcal{S})^{\mathrm{t}} $ and $(\delta \mathcal{S})^{\mathrm{b}} $ take a (small) positive value for a no--temperature--inversion atmosphere. Using these quantities, Equation (\ref{eq:rerr_ex}) can be rewritten as:
\begin{eqnarray}
\label{eq:rerr_exb}
F_{n} &=& \mathcal{T}_n F_{n+1} +  (1-\mathcal{T}_n) \, \mathcal{S}_n - \Delta^{(-)} + \Delta^{(+)} \\
\Delta^{(-)} &\equiv& (\delta \mathcal{S})^{\mathrm{t}} + (\delta \mathcal{S})^{\mathrm{b}} \mathcal{T}_n \approx \frac{1}{2} (\mathcal{S}_n^{\mathrm{b}} - \mathcal{S}_n^{\mathrm{t}}) (1-\mathcal{T}_n ) =  \Delta \tau_n (1-\mathcal{T}_n ) \frac{\mathcal{S}_n^\prime}{2}\\
\Delta^{(+)} &\equiv&  (1 - \mathcal{T}_n) \frac{\mathcal{S}_n^\prime}{2}
\end{eqnarray}
Comparing with Equation (\ref{eq:rerr}), we find that the present formulation of the radiative transfer ignores $\Delta^{(+)}$ and  $\Delta^{(-)}$. 

One can estimate the error of this approximation by considering 
\begin{eqnarray}
\label{eq:rerr_exbx}
\Delta^{(+)}/\mathcal{Q}_n &=& \frac{\Delta^{(+)}}{(1-\mathcal{T}_n) \, \mathcal{S}_n} = \frac{1}{2} \frac{\partial_\tau B(T_n)}{B(T_n)} \approx \frac{1}{2}  \frac{\partial_T B (T_n)}{B(T_n)} \frac{d T}{d \tau} = G(\chi) \frac{\Delta T/\Delta \tau}{T} \\
\Delta^{(-)}/\mathcal{Q}_n &=& \frac{\Delta^{(-)}}{(1-\mathcal{T}_n) \, \mathcal{S}_n} \approx \frac{1}{2}  \frac{\Delta B}{B (T_n)} \approx \frac{1}{2}  \frac{\partial_T B (T_n)}{B(T_n)} \Delta T = G(\chi) \frac{\Delta T}{T}
\end{eqnarray}
where
$ G(\chi) \equiv \chi e^\chi/ (2 e^\chi - 2) $
and $\chi \equiv h c \nu/k_B T$, respectively. For a small wavenumber limit, $\chi \ll 1$ and $G(\chi) \to \chi/2$. For a large wavenumber limit of $\chi \gg 1$, $G(\chi) \to 1/2$. Because most of the emission comes from the layers near $\tau = 1$, the error is approximately of the order of $\Delta T/T$ except for much longer wavelength compared to $k_B T/h c$.

}{
\section{Credible interval of the model+GP prediction}\label{ap:ci}
The probability of the prediction $\dv^\ast$ for an arbitrary wavenumber vector $\nuv^\ast$ conditioned on the given data $\dv$ is expressed as 
\begin{eqnarray}
\label{eq:predgp_model}
p(\dv^\ast|\dv) =  \Ng &(&\Fv(\nuv^\ast) + K_{\times}^\top K_\sigma^{-1} (\dv - \Fv(\nuv)) \nonumber \\
&,& K_{\ast,\sigma} - K_\times^\top K_\sigma^{-1} K_\times) 
\end{eqnarray}
where
\begin{eqnarray}
(K_{\times})_{ij} &=& a \,  \kGP(|\nu_i-\nu^\ast_j|;\tau) \\
(K_{\sigma})_{ij} &=& a \,  \kGP(|\nu_i-\nu_j|;\tau)  + \sigma_{e,i}^2 \delta_{ij}\\
(K_{\ast,\sigma})_{ij} &=& a \,  \kGP(|\nu^\ast_i-\nu^\ast_j|;\tau) + (\sigma_{e,i}^\ast)^2 \delta_{ij},
\end{eqnarray}
where $\delta_{ij}$ is the Kronecker delta. 

An HMC simulation provides a sampling of the other parameters $\thetav^\dagger$ than the GP parameters. Then, the prediction can be sampled by
\begin{eqnarray}
\label{eq:predgp2_}
\dv^\ast_k \sim  \Ng &(&\Fv(\nuv^\ast; \thetav^\dagger_k) + K_{\times}^\top K_\sigma^{-1} (\dv - \Fv(\nuv; \thetav^\dagger_k))\nonumber \\
&,& K_{\ast,\sigma} - K_\times^\top K_\sigma^{-1} K_\times),
\end{eqnarray}
where $\thetav^\dagger_k$ is the $k$-th sampling of $\thetav^\dagger$. The credible interval can be computed using the sampling given by Equation (\ref{eq:predgp2_}). This prediction includes independent Gaussian noise $\sigma_{e,i}^\ast$. When we adopt $\sigma_{e,i}^\ast = 0$, Equation (\ref{eq:predgp_model}) simply provides the prediction of the spectral model + trend.
}

\section{Choosing molecular lines that contribute to the emission spectrum for the direct LPF}\label{ap:weakline}
\begin{figure}[htb]
\begin{center}
\includegraphics[width=0.45\linewidth]{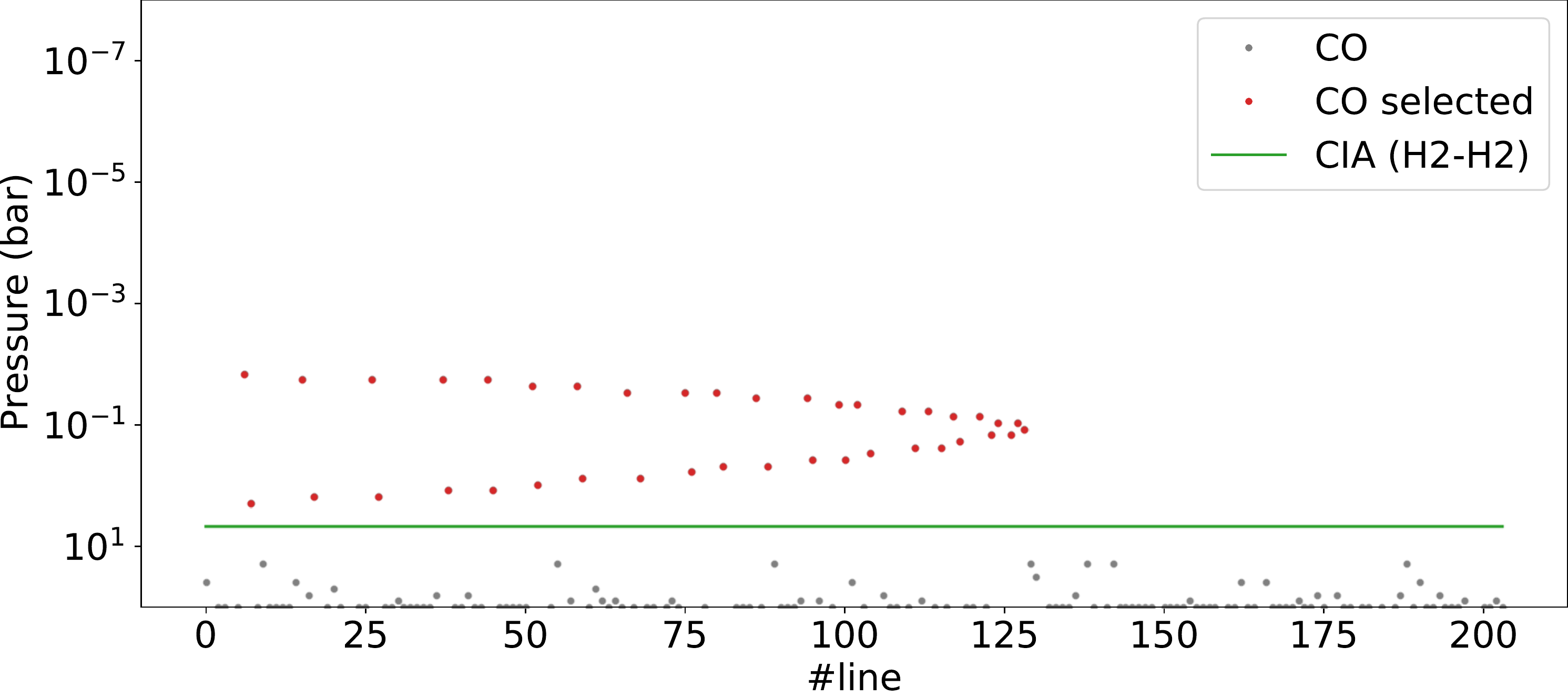}
\includegraphics[width=0.45\linewidth]{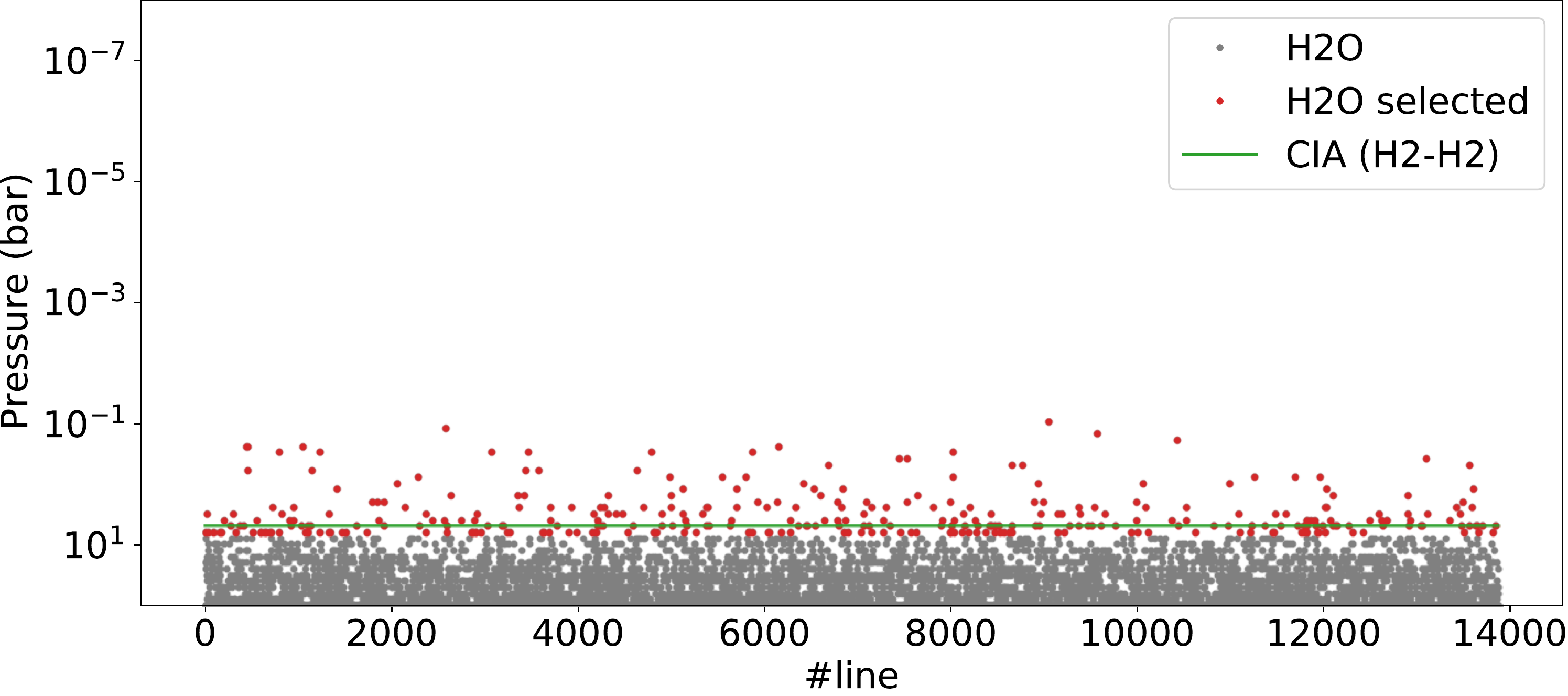}
\caption{Pressure points with the largest contribution to the radiation at the line centers of CO (top) and $\mathrm{H_2O}$ (bottom). We assumed an isothermal atmosphere with $T=$1700 \,K and logg=5. The green line corresponds to the largest contribution of the CIA ($\mathrm{H_2}$-- $\mathrm{H_2}$). \label{fig:maxpoint}}
\end{center}
\end{figure}

The layer that contributes the most to the radiation can be evaluated using the contribution function \citep{[e.g.][]2009ApJ...690..822K}, expressed as 
\begin{eqnarray}
 &B_\nu& (T) \left| \frac{d e^{-\tau}}{d \log{P}} \right| \propto \frac{\nu^3}{e^{h c \nu/k_B T(P)} -1} \frac{e^{-\tau_{x}(\nu;P)} \Delta \tau_x(\nu;P)}{\Delta P/P}  \nonumber \\ &\equiv& \psi_\nu^{x}(P),
\end{eqnarray}
where $x$ is the label of the absorber (e.g. $m,l$ is the $l$-th line of the $m$-th molecule, or CIA($\mathrm{H_2}$,$\mathrm{H_2}$)), and $\tau_x(\nu;P)$ is $\tau(\nu)$ at a pressure of $P$. Because it is time consuming to evaluate the contribution in all of the wavenumber bins for each line, we evaluate the contribution only for the line center. 
The layer at the maximum contribution function for the $l$-th line of the $m$-th molecule ($x = (m,l)$) is expressed as
\begin{eqnarray}
\hat{n}_{l,m} (\nulc_{m,l})  = \mathrm{argmax}_{n} [\psi_{\nulc_{m,l}}^{m,l}(P_n)].
\end{eqnarray}
To evaluate $\hat{n}_{l,m} (\nulc_{m,l}) $, we simply need to compute
\begin{eqnarray}
\Delta \tau_{m,l} (\nulc_{m,l}; P) 
%&=& \frac{\Delta P }{ g} \frac{\mmr_{m}}{\molmass_m} \sigma_{m,l} (0) \\ 
&=& \frac{\Delta P }{ g} \frac{\mmr_{m}}{\molmass_m} \frac{S_{m,l}}{\sqrt{2 \pi} \beta} \mathrm{erfcx}\left(\frac{\gamma_L}{\sqrt{2}\beta}\right),
\end{eqnarray}
for each line. We used an isothermal layer with a maximum mass mixing ratio in the fitting parameter ranges. 
Because the contribution function depends on the atmospheric structure and molecular abundance, we need to specify the T--P profile and mass mixing ratio. 
Figure \ref{fig:maxpoint} shows the layer at the maximum contribution function as a function of the label of the lines $l$. While some of the lines have their maximum contribution in the middle layers of the atmosphere, most of them 
contribute only in the lowest layers below 10~bar.
In our case, the most significant continuum opacity is CIA($\mathrm{H_2}$,$\mathrm{H_2}$). The green line shows the contribution function of CIA$(\mathrm{H_2},\mathrm{H_2})$, expressed as
\begin{eqnarray}
\hat{n}_{\mathrm{CIA}(\mathrm{H_2},\mathrm{H_2})} (\nulc_{m,l}) = \mathrm{argmax}_{n} [\psi_{\nulc_{m,l}}^{\mathrm{CIA(\mathrm{H_2},\mathrm{H_2})}}(P_n)].
\end{eqnarray}
In practice, this means that the lines that appear below the green line are shielded and have little effect on the resultant spectrum. To be safe, we adopted the condition that the $\hat{n}_{l,m}$-th layer exists above the pressure of the CIA line minus two layers, as marked by red dots in Figure \ref{fig:maxpoint}.

The line strength depends on the temperature of each line. For CO, the temperature at the maximum line strength $T_M$ is  $\gtrsim 1700$K. Therefore, we used $T= 1700 K $ for the isothermal temperature. However, for water, $T_M$ is much lower than 1700 K for parts of the lines. Therefore, we evaluated the maximum line strength for each line by changing the isothermal temperature from 500 K to 1700 K in 100 K intervals. The above procedure reduces the number of CO lines from 204 to 40 and the water lines from 13874\footnote{Pre-selected lines that satisfies $S_0 > 10^{-46} \mathrm{cm^5}$ from POKAZATEL.} to 334. Hence, we used 374 molecular lines for the spectral modeling\footnote{ Adding 468 water lines weaker than the 334 lines to the model, we performed the same analysis and found that the posteriors were not changed. The weak lines below the CIA photosphere did not affect the results.}. 

\bibliography{sample63}{}

\begin{thebibliography}{}
\expandafter\ifx\csname natexlab\endcsname\relax\def\natexlab#1{#1}\fi
\providecommand{\url}[1]{\href{#1}{#1}}
\providecommand{\dodoi}[1]{doi:~\href{http://doi.org/#1}{\nolinkurl{#1}}}
\providecommand{\doeprint}[1]{\href{http://ascl.net/#1}{\nolinkurl{http://ascl.net/#1}}}
\providecommand{\doarXiv}[1]{\href{https://arxiv.org/abs/#1}{\nolinkurl{https://arxiv.org/abs/#1}}}

\bibitem[{{Abramowitz} \& {Stegun}(1970)}]{1970hmfw.book.....A}
{Abramowitz}, M., \& {Stegun}, I.~A. 1970, {Handbook of mathematical functions
  : with formulas, graphs, and mathematical tables}

\bibitem[{{Agol} {et~al.}(2020){Agol}, {Dorn}, {Grimm}, {Turbet}, {Ducrot},
  {Delrez}, {Gillon}, {Demory}, {Burdanov}, {Barkaoui}, {Benkhaldoun},
  {Bolmont}, {Burgasser}, {Carey}, {de Wit}, {Fabrycky}, {Foreman-Mackey},
  {Haldemann}, {Hernandez}, {Ingalls}, {Jehin}, {Langford}, {Leconte},
  {Lederer}, {Luger}, {Malhotra}, {Meadows}, {Morris}, {Pozuelos}, {Queloz},
  {Raymond}, {Selsis}, {Sestovic}, {Triaud}, \& {Van
  Grootel}}]{2020arXiv201001074A}
{Agol}, E., {Dorn}, C., {Grimm}, S.~L., {et~al.} 2020, arXiv e-prints,
  arXiv:2010.01074.
\newblock \doarXiv{2010.01074}

\bibitem[{{Albert}(2020)}]{2020arXiv201215286A}
{Albert}, J.~G. 2020, arXiv e-prints, arXiv:2012.15286.
\newblock \doarXiv{2012.15286}

\bibitem[{{Allard} {et~al.}(2013){Allard}, {Homeier}, {Freytag},
  {Schaffenberger}, {}, \& {Rajpurohit}}]{2013MSAIS..24..128A}
{Allard}, F., {Homeier}, D., {Freytag}, B., {et~al.} 2013, Memorie della
  Societa Astronomica Italiana Supplementi, 24, 128.
\newblock \doarXiv{1302.6559}

\bibitem[{{Asplund} {et~al.}(2021){Asplund}, {Amarsi}, \&
  {Grevesse}}]{2021arXiv210501661A}
{Asplund}, M., {Amarsi}, A.~M., \& {Grevesse}, N. 2021, arXiv e-prints,
  arXiv:2105.01661.
\newblock \doarXiv{2105.01661}

\bibitem[{{Asplund} {et~al.}(2009){Asplund}, {Grevesse}, {Sauval}, \&
  {Scott}}]{2009ARA&A..47..481A}
{Asplund}, M., {Grevesse}, N., {Sauval}, A.~J., \& {Scott}, P. 2009, \araa, 47,
  481, \dodoi{10.1146/annurev.astro.46.060407.145222}

\bibitem[{{Bartoli{\'c}} {et~al.}(2021){Bartoli{\'c}}, {Luger},
  {Foreman-Mackey}, {Howell}, \& {Rathbun}}]{2021arXiv210303758B}
{Bartoli{\'c}}, F., {Luger}, R., {Foreman-Mackey}, D., {Howell}, R.~R., \&
  {Rathbun}, J.~A. 2021, arXiv e-prints, arXiv:2103.03758.
\newblock \doarXiv{2103.03758}

\bibitem[{{Barton} {et~al.}(2017){Barton}, {Hill}, {Czurylo}, {Li}, {Hyslop},
  {Yurchenko}, \& {Tennyson}}]{2017JQSRT.203..490B}
{Barton}, E.~J., {Hill}, C., {Czurylo}, M., {et~al.} 2017, \jqsrt, 203, 490,
  \dodoi{10.1016/j.jqsrt.2017.01.028}

\bibitem[{{Ben-Yami} {et~al.}(2020){Ben-Yami}, {Madhusudhan}, {Cabot},
  {Constantinou}, {Piette}, {Gandhi}, \& {Welbanks}}]{2020ApJ...897L...5B}
{Ben-Yami}, M., {Madhusudhan}, N., {Cabot}, S. H.~C., {et~al.} 2020, \apjl,
  897, L5, \dodoi{10.3847/2041-8213/ab94aa}

\bibitem[{{Betancourt}(2017)}]{2017arXiv170102434B}
{Betancourt}, M. 2017, arXiv e-prints, arXiv:1701.02434.
\newblock \doarXiv{1701.02434}

\bibitem[{{Birkby} {et~al.}(2013){Birkby}, {de Kok}, {Brogi}, {de Mooij},
  {Schwarz}, {Albrecht}, \& {Snellen}}]{2013MNRAS.436L..35B}
{Birkby}, J.~L., {de Kok}, R.~J., {Brogi}, M., {et~al.} 2013, \mnras, 436, L35,
  \dodoi{10.1093/mnrasl/slt107}

\bibitem[{{Birkby} {et~al.}(2017){Birkby}, {de Kok}, {Brogi}, {Schwarz}, \&
  {Snellen}}]{2017AJ....153..138B}
{Birkby}, J.~L., {de Kok}, R.~J., {Brogi}, M., {Schwarz}, H., \& {Snellen},
  I.~A.~G. 2017, \aj, 153, 138, \dodoi{10.3847/1538-3881/aa5c87}

\bibitem[{Bradbury {et~al.}(2018)Bradbury, Frostig, Hawkins, Johnson, Leary,
  Maclaurin, Necula, Paszke, Vander{P}las, Wanderman-{M}ilne, \&
  Zhang}]{jax2018github}
Bradbury, J., Frostig, R., Hawkins, P., {et~al.} 2018, {JAX}: composable
  transformations of {P}ython+{N}um{P}y programs, 0.2.5.
\newblock \url{http://github.com/google/jax}

\bibitem[{{Brogi} {et~al.}(2017){Brogi}, {Line}, {Bean}, {D{\'e}sert}, \&
  {Schwarz}}]{2017ApJ...839L...2B}
{Brogi}, M., {Line}, M., {Bean}, J., {D{\'e}sert}, J.~M., \& {Schwarz}, H.
  2017, \apjl, 839, L2, \dodoi{10.3847/2041-8213/aa6933}

\bibitem[{{Brogi} \& {Line}(2019)}]{2019AJ....157..114B}
{Brogi}, M., \& {Line}, M.~R. 2019, \aj, 157, 114,
  \dodoi{10.3847/1538-3881/aaffd3}

\bibitem[{{Brogi} {et~al.}(2012){Brogi}, {Snellen}, {de Kok}, {Albrecht},
  {Birkby}, \& {de Mooij}}]{2012Natur.486..502B}
{Brogi}, M., {Snellen}, I. A.~G., {de Kok}, R.~J., {et~al.} 2012, \nat, 486,
  502, \dodoi{10.1038/nature11161}

\bibitem[{{Buenzli} {et~al.}(2015){Buenzli}, {Saumon}, {Marley}, {Apai},
  {Radigan}, {Bedin}, {Reid}, \& {Morley}}]{2015ApJ...798..127B}
{Buenzli}, E., {Saumon}, D., {Marley}, M.~S., {et~al.} 2015, \apj, 798, 127,
  \dodoi{10.1088/0004-637X/798/2/127}

\bibitem[{{Burgasser} {et~al.}(2013){Burgasser}, {Sheppard}, \&
  {Luhman}}]{2013ApJ...772..129B}
{Burgasser}, A.~J., {Sheppard}, S.~S., \& {Luhman}, K.~L. 2013, \apj, 772, 129,
  \dodoi{10.1088/0004-637X/772/2/129}

\bibitem[{{Cabot} {et~al.}(2020){Cabot}, {Madhusudhan}, {Welbanks}, {Piette},
  \& {Gandhi}}]{2020MNRAS.494..363C}
{Cabot}, S. H.~C., {Madhusudhan}, N., {Welbanks}, L., {Piette}, A., \&
  {Gandhi}, S. 2020, \mnras, 494, 363, \dodoi{10.1093/mnras/staa748}

\bibitem[{{Casasayas-Barris} {et~al.}(2019){Casasayas-Barris}, {Pall{\'e}},
  {Yan}, {Chen}, {Kohl}, {Stangret}, {Parviainen}, {Helling}, {Watanabe},
  {Czesla}, {Fukui}, {Monta{\~n}{\'e}s-Rodr{\'\i}guez}, {Nagel}, {Narita},
  {Nortmann}, {Nowak}, {Schmitt}, \& {Zapatero Osorio}}]{2019A&A...628A...9C}
{Casasayas-Barris}, N., {Pall{\'e}}, E., {Yan}, F., {et~al.} 2019, \aap, 628,
  A9, \dodoi{10.1051/0004-6361/201935623}

\bibitem[{{Cont} {et~al.}(2021){Cont}, {Yan}, {Reiners}, {Casasayas-Barris},
  {Molli{\`e}re}, {Pall{\'e}}, {Henning}, {Nortmann}, {Stangret}, {Czesla},
  {L{\'o}pez-Puertas}, {S{\'a}nchez-L{\'o}pez}, {Rodler}, {Ribas},
  {Quirrenbach}, {Caballero}, {Amado}, {Carone}, {Khaimova}, {Kreidberg},
  {Molaverdikhani}, {Montes}, {Morello}, {Nagel}, {Oshagh}, \&
  {Zechmeister}}]{2021arXiv210510230C}
{Cont}, D., {Yan}, F., {Reiners}, A., {et~al.} 2021, arXiv e-prints,
  arXiv:2105.10230.
\newblock \doarXiv{2105.10230}

\bibitem[{{Crossfield} {et~al.}(2014){Crossfield}, {Biller}, {Schlieder},
  {Deacon}, {Bonnefoy}, {Homeier}, {Allard}, {Buenzli}, {Henning}, {Brandner},
  {Goldman}, \& {Kopytova}}]{2014Natur.505..654C}
{Crossfield}, I.~J.~M., {Biller}, B., {Schlieder}, J.~E., {et~al.} 2014, \nat,
  505, 654, \dodoi{10.1038/nature12955}

\bibitem[{{Dopita} \& {Sutherland}(2003)}]{2003adu..book.....D}
{Dopita}, M.~A., \& {Sutherland}, R.~S. 2003, {Astrophysics of the diffuse
  universe}

\bibitem[{{Duane} {et~al.}(1987){Duane}, {Kennedy}, {Pendleton}, \&
  {Roweth}}]{1987PhLB..195..216D}
{Duane}, S., {Kennedy}, A.~D., {Pendleton}, B.~J., \& {Roweth}, D. 1987,
  Physics Letters B, 195, 216, \dodoi{10.1016/0370-2693(87)91197-X}

\bibitem[{{Ehrenreich} {et~al.}(2020){Ehrenreich}, {Lovis}, {Allart}, {Zapatero
  Osorio}, {Pepe}, {Cristiani}, {Rebolo}, {Santos}, {Borsa}, {Demangeon},
  {Dumusque}, {Gonz{\'a}lez Hern{\'a}ndez}, {Casasayas-Barris},
  {S{\'e}gransan}, {Sousa}, {Abreu}, {Adibekyan}, {Affolter}, {Allende Prieto},
  {Alibert}, {Aliverti}, {Alves}, {Amate}, {Avila}, {Baldini}, {Bandy}, {Benz},
  {Bianco}, {Bolmont}, {Bouchy}, {Bourrier}, {Broeg}, {Cabral}, {Calderone},
  {Pall{\'e}}, {Cegla}, {Cirami}, {Coelho}, {Conconi}, {Coretti}, {Cumani},
  {Cupani}, {Dekker}, {Delabre}, {Deiries}, {D'Odorico}, {Di Marcantonio},
  {Figueira}, {Fragoso}, {Genolet}, {Genoni}, {G{\'e}nova Santos}, {Hara},
  {Hughes}, {Iwert}, {Kerber}, {Knudstrup}, {Landoni}, {Lavie}, {Lizon},
  {Lendl}, {Lo Curto}, {Maire}, {Manescau}, {Martins}, {M{\'e}gevand},
  {Mehner}, {Micela}, {Modigliani}, {Molaro}, {Monteiro}, {Monteiro},
  {Moschetti}, {M{\"u}ller}, {Nunes}, {Oggioni}, {Oliveira}, {Pariani},
  {Pasquini}, {Poretti}, {Rasilla}, {Redaelli}, {Riva}, {Santana Tschudi},
  {Santin}, {Santos}, {Segovia Milla}, {Seidel}, {Sosnowska}, {Sozzetti},
  {Span{\`o}}, {Su{\'a}rez Mascare{\~n}o}, {Tabernero}, {Tenegi}, {Udry},
  {Zanutta}, \& {Zerbi}}]{2020Natur.580..597E}
{Ehrenreich}, D., {Lovis}, C., {Allart}, R., {et~al.} 2020, \nat, 580, 597,
  \dodoi{10.1038/s41586-020-2107-1}

\bibitem[{{Flagg} {et~al.}(2019){Flagg}, {Johns-Krull}, {Nofi}, {Llama},
  {Prato}, {Sullivan}, {Jaffe}, \& {Mace}}]{2019ApJ...878L..37F}
{Flagg}, L., {Johns-Krull}, C.~M., {Nofi}, L., {et~al.} 2019, \apjl, 878, L37,
  \dodoi{10.3847/2041-8213/ab276d}

\bibitem[{{Follert} {et~al.}(2014){Follert}, {Dorn}, {Oliva}, {Lizon},
  {Hatzes}, {Piskunov}, {Reiners}, {Seemann}, {Stempels}, {Heiter}, {Marquart},
  {Lockhart}, {Anglada-Escude}, {L{\"o}winger}, {Baade}, {Grunhut}, {Bristow},
  {Klein}, {Jung}, {Ives}, {Kerber}, {Pozna}, {Paufique}, {Kaeufl}, {Origlia},
  {Valenti}, {Gojak}, {Hilker}, {Pasquini}, {Smette}, \&
  {Smoker}}]{2014SPIE.9147E..19F}
{Follert}, R., {Dorn}, R.~J., {Oliva}, E., {et~al.} 2014, in Society of
  Photo-Optical Instrumentation Engineers (SPIE) Conference Series, Vol. 9147,
  Ground-based and Airborne Instrumentation for Astronomy V, ed. S.~K.
  {Ramsay}, I.~S. {McLean}, \& H.~{Takami}, 914719, \dodoi{10.1117/12.2054197}

\bibitem[{{Foreman-Mackey} {et~al.}(2021){Foreman-Mackey}, {Luger}, {Agol},
  {Barclay}, {Bouma}, {Brandt}, {Czekala}, {David}, {Dong}, {Gilbert},
  {Gordon}, {Hedges}, {Hey}, {Morris}, {Price-Whelan}, \&
  {Savel}}]{2021arXiv210501994F}
{Foreman-Mackey}, D., {Luger}, R., {Agol}, E., {et~al.} 2021, arXiv e-prints,
  arXiv:2105.01994.
\newblock \doarXiv{2105.01994}

\bibitem[{{Gandhi} {et~al.}(2019){Gandhi}, {Madhusudhan}, {Hawker}, \&
  {Piette}}]{2019AJ....158..228G}
{Gandhi}, S., {Madhusudhan}, N., {Hawker}, G., \& {Piette}, A. 2019, \aj, 158,
  228, \dodoi{10.3847/1538-3881/ab4efc}

\bibitem[{{Garcia} {et~al.}(2017){Garcia}, {Ammons}, {Salama}, {Crossfield},
  {Bendek}, {Chilcote}, {Garrel}, {Graham}, {Kalas}, {Konopacky}, {Lu},
  {Macintosh}, {Marin}, {Marois}, {Nielsen}, {Neichel}, {Pham}, {De Rosa},
  {Ryan}, {Service}, \& {Sivo}}]{2017ApJ...846...97G}
{Garcia}, E.~V., {Ammons}, S.~M., {Salama}, M., {et~al.} 2017, \apj, 846, 97,
  \dodoi{10.3847/1538-4357/aa844f}

\bibitem[{{Gharib-Nezhad} \& {Line}(2019)}]{2019ApJ...872...27G}
{Gharib-Nezhad}, E., \& {Line}, M.~R. 2019, \apj, 872, 27,
  \dodoi{10.3847/1538-4357/aafb7b}

\bibitem[{{Giacobbe} {et~al.}(2021){Giacobbe}, {Brogi}, {Gandhi}, {Cubillos},
  {Bonomo}, {Sozzetti}, {Fossati}, {Guilluy}, {Carleo}, {Rainer},
  {Harutyunyan}, {Borsa}, {Pino}, {Nascimbeni}, {Benatti}, {Biazzo},
  {Bignamini}, {Chubb}, {Claudi}, {Cosentino}, {Covino}, {Damasso}, {Desidera},
  {Fiorenzano}, {Ghedina}, {Lanza}, {Leto}, {Maggio}, {Malavolta}, {Maldonado},
  {Micela}, {Molinari}, {Pagano}, {Pedani}, {Piotto}, {Poretti}, {Scandariato},
  {Yurchenko}, {Fantinel}, {Galli}, {Lodi}, {Sanna}, \&
  {Tozzi}}]{2021Natur.592..205G}
{Giacobbe}, P., {Brogi}, M., {Gandhi}, S., {et~al.} 2021, \nat, 592, 205,
  \dodoi{10.1038/s41586-021-03381-x}

\bibitem[{{Gibson} {et~al.}(2020){Gibson}, {Merritt}, {Nugroho}, {Cubillos},
  {de Mooij}, {Mikal-Evans}, {Fossati}, {Lothringer}, {Nikolov}, {Sing},
  {Spake}, {Watson}, \& {Wilson}}]{2020MNRAS.493.2215G}
{Gibson}, N.~P., {Merritt}, S., {Nugroho}, S.~K., {et~al.} 2020, \mnras, 493,
  2215, \dodoi{10.1093/mnras/staa228}

\bibitem[{{Gravity Collaboration} {et~al.}(2019){Gravity Collaboration},
  {Lacour}, {Nowak}, {Wang}, {Pfuhl}, {Eisenhauer}, {Abuter}, {Amorim},
  {Anugu}, {Benisty}, {Berger}, {Beust}, {Blind}, {Bonnefoy}, {Bonnet},
  {Bourget}, {Brandner}, {Buron}, {Collin}, {Charnay}, {Chapron}, {Cl{\'e}net},
  {Coud{\'e} Du Foresto}, {de Zeeuw}, {Deen}, {Dembet}, {Dexter}, {Duvert},
  {Eckart}, {F{\"o}rster Schreiber}, {F{\'e}dou}, {Garcia}, {Garcia Lopez},
  {Gao}, {Gendron}, {Genzel}, {Gillessen}, {Gordo}, {Greenbaum}, {Habibi},
  {Haubois}, {Hau{\ss}mann}, {Henning}, {Hippler}, {Horrobin}, {Hubert},
  {Jimenez Rosales}, {Jocou}, {Kendrew}, {Kervella}, {Kolb}, {Lagrange},
  {Lapeyr{\`e}re}, {Le Bouquin}, {L{\'e}na}, {Lippa}, {Lenzen}, {Maire},
  {Molli{\`e}re}, {Ott}, {Paumard}, {Perraut}, {Perrin}, {Pueyo}, {Rabien},
  {Ram{\'\i}rez}, {Rau}, {Rodr{\'\i}guez-Coira}, {Rousset}, {Sanchez-Bermudez},
  {Scheithauer}, {Schuhler}, {Straub}, {Straubmeier}, {Sturm}, {Tacconi},
  {Vincent}, {van Dishoeck}, {von Fellenberg}, {Wank}, {Waisberg}, {Widmann},
  {Wieprecht}, {Wiest}, {Wiezorrek}, {Woillez}, {Yazici}, {Ziegler}, \&
  {Zins}}]{2019A&A...623L..11G}
{Gravity Collaboration}, {Lacour}, S., {Nowak}, M., {et~al.} 2019, \aap, 623,
  L11, \dodoi{10.1051/0004-6361/201935253}

\bibitem[{{Gray}(2005)}]{Gray}
{Gray}, D.~F. 2005, {The Observation and Analysis of Stellar Photospheres}

\bibitem[{{Grimm} {et~al.}(2021){Grimm}, {Malik}, {Kitzmann},
  {Guzm{\'a}n-Mesa}, {Hoeijmakers}, {Fisher}, {Mendon{\c{c}}a}, {Yurchenko},
  {Tennyson}, {Alesina}, {Buchschacher}, {Burnier}, {Segransan}, {Kurucz}, \&
  {Heng}}]{2021arXiv210102005G}
{Grimm}, S.~L., {Malik}, M., {Kitzmann}, D., {et~al.} 2021, arXiv e-prints,
  arXiv:2101.02005.
\newblock \doarXiv{2101.02005}

\bibitem[{{Guilluy} {et~al.}(2019){Guilluy}, {Sozzetti}, {Brogi}, {Bonomo},
  {Giacobbe}, {Claudi}, \& {Benatti}}]{2019A&A...625A.107G}
{Guilluy}, G., {Sozzetti}, A., {Brogi}, M., {et~al.} 2019, \aap, 625, A107,
  \dodoi{10.1051/0004-6361/201834615}

\bibitem[{{Gunes Baydin} {et~al.}(2015){Gunes Baydin}, {Pearlmutter},
  {Andreyevich Radul}, \& {Siskind}}]{2015arXiv150205767G}
{Gunes Baydin}, A., {Pearlmutter}, B.~A., {Andreyevich Radul}, A., \&
  {Siskind}, J.~M. 2015, arXiv e-prints, arXiv:1502.05767.
\newblock \doarXiv{1502.05767}

\bibitem[{{Hawker} {et~al.}(2018){Hawker}, {Madhusudhan}, {Cabot}, \&
  {Gandhi}}]{2018ApJ...863L..11H}
{Hawker}, G.~A., {Madhusudhan}, N., {Cabot}, S. H.~C., \& {Gandhi}, S. 2018,
  \apjl, 863, L11, \dodoi{10.3847/2041-8213/aac49d}

\bibitem[{Heng(2017)}]{heng2017exoplanetary}
Heng, K. 2017, Exoplanetary Atmospheres (Princeton University Press)

\bibitem[{{Heng} {et~al.}(2014){Heng}, {Mendon{\c{c}}a}, \&
  {Lee}}]{2014ApJS..215....4H}
{Heng}, K., {Mendon{\c{c}}a}, J.~M., \& {Lee}, J.-M. 2014, \apjs, 215, 4,
  \dodoi{10.1088/0067-0049/215/1/4}

\bibitem[{Hirsch \& Smale(1974)}]{hirsch1974differential}
Hirsch, M., \& Smale, S. 1974

\bibitem[{{Hjerting}(1938)}]{1938ApJ....88..508H}
{Hjerting}, F. 1938, \apj, 88, 508, \dodoi{10.1086/144000}

\bibitem[{{Hoeijmakers} {et~al.}(2018{\natexlab{a}}){Hoeijmakers}, {Schwarz},
  {Snellen}, {de Kok}, {Bonnefoy}, {Chauvin}, {Lagrange}, \&
  {Girard}}]{2018A&A...617A.144H}
{Hoeijmakers}, H.~J., {Schwarz}, H., {Snellen}, I.~A.~G., {et~al.}
  2018{\natexlab{a}}, \aap, 617, A144, \dodoi{10.1051/0004-6361/201832902}

\bibitem[{{Hoeijmakers} {et~al.}(2018{\natexlab{b}}){Hoeijmakers},
  {Ehrenreich}, {Heng}, {Kitzmann}, {Grimm}, {Allart}, {Deitrick},
  {Wyttenbach}, {Oreshenko}, {Pino}, {Rimmer}, {Molinari}, \& {Di
  Fabrizio}}]{2018Natur.560..453H}
{Hoeijmakers}, H.~J., {Ehrenreich}, D., {Heng}, K., {et~al.}
  2018{\natexlab{b}}, \nat, 560, 453, \dodoi{10.1038/s41586-018-0401-y}

\bibitem[{{Hoeijmakers} {et~al.}(2019){Hoeijmakers}, {Ehrenreich}, {Kitzmann},
  {Allart}, {Grimm}, {Seidel}, {Wyttenbach}, {Pino}, {Nielsen}, {Fisher},
  {Rimmer}, {Bourrier}, {Cegla}, {Lavie}, {Lovis}, {Patzer}, {Stock}, {Pepe},
  \& {Heng}}]{2019A&A...627A.165H}
{Hoeijmakers}, H.~J., {Ehrenreich}, D., {Kitzmann}, D., {et~al.} 2019, \aap,
  627, A165, \dodoi{10.1051/0004-6361/201935089}

\bibitem[{{Hoeijmakers} {et~al.}(2020){Hoeijmakers}, {Cabot}, {Zhao},
  {Buchhave}, {Tronsgaard}, {Davis}, {Kitzmann}, {Grimm}, {Cegla}, {Bourrier},
  {Ehrenreich}, {Heng}, {Lovis}, \& {Fischer}}]{2020A&A...641A.120H}
{Hoeijmakers}, H.~J., {Cabot}, S. H.~C., {Zhao}, L., {et~al.} 2020, \aap, 641,
  A120, \dodoi{10.1051/0004-6361/202037437}

\bibitem[{{Hoffman} \& {Gelman}(2011)}]{2011arXiv1111.4246H}
{Hoffman}, M.~D., \& {Gelman}, A. 2011, arXiv e-prints, arXiv:1111.4246.
\newblock \doarXiv{1111.4246}

\bibitem[{{Hunter}(2007)}]{2007CSE.....9...90H}
{Hunter}, J.~D. 2007, Computing in Science and Engineering, 9, 90,
  \dodoi{10.1109/MCSE.2007.55}

\bibitem[{{Ishizuka} {et~al.}(2021){Ishizuka}, {Kawahara}, {Nugroho},
  {Kawashima}, {Hirano}, \& {Tamura}}]{2021AJ....161..153I}
{Ishizuka}, M., {Kawahara}, H., {Nugroho}, S.~K., {et~al.} 2021, \aj, 161, 153,
  \dodoi{10.3847/1538-3881/abdb25}

\bibitem[{{Jovanovic} {et~al.}(2019){Jovanovic}, {Delorme}, {Bond}, {Cetre},
  {Mawet}, {Echeverri}, {Wallace}, {Bartos}, {Lilley}, {Ragland}, {Ruane},
  {Wizinowich}, {Chun}, {Wang}, {Wang}, {Fitzgerald}, {Matthews}, {Pezzato},
  {Calvin}, {Millar-Blanchaer}, {Martin}, {Wetherell}, {Wang}, {Jacobson},
  {Warmbier}, {Lockhart}, {Hall}, {Jensen-Clem}, \&
  {McEwen}}]{2019arXiv190904541J}
{Jovanovic}, N., {Delorme}, J.~R., {Bond}, C.~Z., {et~al.} 2019, arXiv
  e-prints, arXiv:1909.04541.
\newblock \doarXiv{1909.04541}

\bibitem[{{Karman} {et~al.}(2019){Karman}, {Gordon}, {van der Avoird},
  {Baranov}, {Boulet}, {Drouin}, {Groenenboom}, {Gustafsson}, {Hartmann},
  {Kurucz}, {Rothman}, {Sun}, {Sung}, {Thalman}, {Tran}, {Wishnow},
  {Wordsworth}, {Vigasin}, {Volkamer}, \& {van der
  Zande}}]{2019Icar..328..160K}
{Karman}, T., {Gordon}, I.~E., {van der Avoird}, A., {et~al.} 2019, \icarus,
  328, 160, \dodoi{10.1016/j.icarus.2019.02.034}

\bibitem[{{Kawahara}(2012)}]{2012ApJ...760L..13K}
{Kawahara}, H. 2012, \apjl, 760, L13, \dodoi{10.1088/2041-8205/760/1/L13}

\bibitem[{{Kawahara} \& {Hirano}(2014)}]{2014arXiv1409.5740K}
{Kawahara}, H., \& {Hirano}, T. 2014, arXiv e-prints, arXiv:1409.5740.
\newblock \doarXiv{1409.5740}

\bibitem[{{Kawahara} {et~al.}(2014){Kawahara}, {Murakami}, {Matsuo}, \&
  {Kotani}}]{2014ApJS..212...27K}
{Kawahara}, H., {Murakami}, N., {Matsuo}, T., \& {Kotani}, T. 2014, \apjs, 212,
  27, \dodoi{10.1088/0067-0049/212/2/27}

\bibitem[{{Kawashima} \& {Ikoma}(2018)}]{2018ApJ...853....7K}
{Kawashima}, Y., \& {Ikoma}, M. 2018, \apj, 853, 7,
  \dodoi{10.3847/1538-4357/aaa0c5}

\bibitem[{{Kipping}(2013)}]{2013MNRAS.435.2152K}
{Kipping}, D.~M. 2013, \mnras, 435, 2152, \dodoi{10.1093/mnras/stt1435}

\bibitem[{{Knutson} {et~al.}(2009){Knutson}, {Charbonneau}, {Cowan}, {Fortney},
  {Showman}, {Agol}, {Henry}, {Everett}, \& {Allen}}]{2009ApJ...690..822K}
{Knutson}, H.~A., {Charbonneau}, D., {Cowan}, N.~B., {et~al.} 2009, \apj, 690,
  822, \dodoi{10.1088/0004-637X/690/1/822}

\bibitem[{{Kochanov} {et~al.}(2016){Kochanov}, {Gordon}, {Rothman}, {Wcislo},
  {Hill}, \& {Wilzewski}}]{2016isms.confETG12K}
{Kochanov}, R.~V., {Gordon}, I.~E., {Rothman}, L.~S., {et~al.} 2016, in 71st
  International Symposium on Molecular Spectroscopy, TG12,
  \dodoi{10.15278/isms.2016.TG12}

\bibitem[{{Kopal}(1950)}]{1950HarCi.454....1K}
{Kopal}, Z. 1950, Harvard College Observatory Circular, 454, 1

\bibitem[{{Kotani} {et~al.}(2018){Kotani}, {Tamura}, {Nishikawa}, {Ueda},
  {Kuzuhara}, {Omiya}, {Hashimoto}, {Ishizuka}, {Hirano}, {Suto}, {Kurokawa},
  {Kokubo}, {Mori}, {Tanaka}, {Kashiwagi}, {Konishi}, {Kudo}, {Sato},
  {Jacobson}, {Hodapp}, {Hall}, {Aoki}, {Usuda}, {Nishiyama}, {Nakajima},
  {Ikeda}, {Yamamuro}, {Morino}, {Baba}, {Hosokawa}, {Ishikawa}, {Narita},
  {Kokubo}, {Hayano}, {Izumiura}, {Kambe}, {Kusakabe}, {Kwon}, {Ikoma}, {Hori},
  {Genda}, {Fukui}, {Fujii}, {Kawahara}, {Olivier}, {Jovanovic}, {Harakawa},
  {Hayashi}, {Hidai}, {Machida}, {Matsuo}, {Nagata}, {Ogihara}, {Takami},
  {Takato}, {Terada}, \& {Oh}}]{2018SPIE10702E..11K}
{Kotani}, T., {Tamura}, M., {Nishikawa}, J., {et~al.} 2018, in Society of
  Photo-Optical Instrumentation Engineers (SPIE) Conference Series, Vol. 10702,
  Ground-based and Airborne Instrumentation for Astronomy VII, 1070211,
  \dodoi{10.1117/12.2311836}

\bibitem[{{Kotani} {et~al.}(2020){Kotani}, {Kawahara}, {Ishizuka}, {Jovanovic},
  {Vievard}, {Lozi}, {Sahoo}, {Guyon}, {Yoneta}, \&
  {Tamura}}]{2020SPIE11448E..78K}
{Kotani}, T., {Kawahara}, H., {Ishizuka}, M., {et~al.} 2020, in Society of
  Photo-Optical Instrumentation Engineers (SPIE) Conference Series, Vol. 11448,
  Society of Photo-Optical Instrumentation Engineers (SPIE) Conference Series,
  1144878, \dodoi{10.1117/12.2561755}

\bibitem[{Kumar {et~al.}(2019)Kumar, Carroll, Hartikainen, \&
  Martin}]{arviz_2019}
Kumar, R., Carroll, C., Hartikainen, A., \& Martin, O. 2019, Journal of Open
  Source Software, 4, 1143, \dodoi{10.21105/joss.01143}

\bibitem[{{Kuntz}(1997)}]{1997JQSRT..57..819K}
{Kuntz}, M. 1997, \jqsrt, 57, 819, \dodoi{10.1016/S0022-4073(96)00162-8}

\bibitem[{{Laraia} {et~al.}(2011){Laraia}, {Gamache}, {Lamouroux}, {Gordon}, \&
  {Rothman}}]{2011Icar..215..391L}
{Laraia}, A.~L., {Gamache}, R.~R., {Lamouroux}, J., {Gordon}, I.~E., \&
  {Rothman}, L.~S. 2011, \icarus, 215, 391,
  \dodoi{10.1016/j.icarus.2011.06.004}

\bibitem[{{Lavie} {et~al.}(2017){Lavie}, {Mendon{\c{c}}a}, {Mordasini},
  {Malik}, {Bonnefoy}, {Demory}, {Oreshenko}, {Grimm}, {Ehrenreich}, \&
  {Heng}}]{2017AJ....154...91L}
{Lavie}, B., {Mendon{\c{c}}a}, J.~M., {Mordasini}, C., {et~al.} 2017, \aj, 154,
  91, \dodoi{10.3847/1538-3881/aa7ed8}

\bibitem[{{Lazorenko} \& {Sahlmann}(2018)}]{2018A&A...618A.111L}
{Lazorenko}, P.~F., \& {Sahlmann}, J. 2018, \aap, 618, A111,
  \dodoi{10.1051/0004-6361/201833626}

\bibitem[{{Li} {et~al.}(2015){Li}, {Gordon}, {Rothman}, {Tan}, {Hu}, {Kassi},
  {Campargue}, \& {Medvedev}}]{2015ApJS..216...15L}
{Li}, G., {Gordon}, I.~E., {Rothman}, L.~S., {et~al.} 2015, \apjs, 216, 15,
  \dodoi{10.1088/0067-0049/216/1/15}

\bibitem[{{Line} {et~al.}(2013){Line}, {Wolf}, {Zhang}, {Knutson}, {Kammer},
  {Ellison}, {Deroo}, {Crisp}, \& {Yung}}]{2013ApJ...775..137L}
{Line}, M.~R., {Wolf}, A.~S., {Zhang}, X., {et~al.} 2013, \apj, 775, 137,
  \dodoi{10.1088/0004-637X/775/2/137}

\bibitem[{{Luhman}(2013)}]{2013ApJ...767L...1L}
{Luhman}, K.~L. 2013, \apjl, 767, L1, \dodoi{10.1088/2041-8205/767/1/L1}

\bibitem[{{Masuda} \& {Hirano}(2021)}]{2021ApJ...910L..17M}
{Masuda}, K., \& {Hirano}, T. 2021, \apjl, 910, L17,
  \dodoi{10.3847/2041-8213/abecdc}

\bibitem[{{Mawet} {et~al.}(2017){Mawet}, {Ruane}, {Xuan}, {Echeverri},
  {Klimovich}, {Randolph}, {Fucik}, {Wallace}, {Wang}, {Vasisht}, {Dekany},
  {Mennesson}, {Choquet}, {Delorme}, \& {Serabyn}}]{2017ApJ...838...92M}
{Mawet}, D., {Ruane}, G., {Xuan}, W., {et~al.} 2017, \apj, 838, 92,
  \dodoi{10.3847/1538-4357/aa647f}

\bibitem[{{Molli{\`e}re} {et~al.}(2019){Molli{\`e}re}, {Wardenier}, {van
  Boekel}, {Henning}, {Molaverdikhani}, \& {Snellen}}]{2019A&A...627A..67M}
{Molli{\`e}re}, P., {Wardenier}, J.~P., {van Boekel}, R., {et~al.} 2019, \aap,
  627, A67, \dodoi{10.1051/0004-6361/201935470}

\bibitem[{{Neal}(2012)}]{2012arXiv1206.1901N}
{Neal}, R.~M. 2012, arXiv e-prints, arXiv:1206.1901.
\newblock \doarXiv{1206.1901}

\bibitem[{{Nugroho} {et~al.}(2020){Nugroho}, {Gibson}, {de Mooij}, {Herman},
  {Watson}, {Kawahara}, \& {Merritt}}]{2020ApJ...898L..31N}
{Nugroho}, S.~K., {Gibson}, N.~P., {de Mooij}, E. J.~W., {et~al.} 2020, \apjl,
  898, L31, \dodoi{10.3847/2041-8213/aba4b6}

\bibitem[{{Nugroho} {et~al.}(2017){Nugroho}, {Kawahara}, {Masuda}, {Hirano},
  {Kotani}, \& {Tajitsu}}]{2017AJ....154..221N}
{Nugroho}, S.~K., {Kawahara}, H., {Masuda}, K., {et~al.} 2017, \aj, 154, 221,
  \dodoi{10.3847/1538-3881/aa9433}

\bibitem[{{Nugroho} {et~al.}(2021){Nugroho}, {Kawahara}, {Gibson}, {de Mooij},
  {Hirano}, {Kotani}, {Kawashima}, {Masuda}, {Brogi}, {Birkby}, {Watson},
  {Tamura}, {Zwintz}, {Harakawa}, {Kudo}, {Kuzuhara}, {Hodapp}, {Ishizuka},
  {Jacobson}, {Konishi}, {Kurokawa}, {Nishikawa}, {Omiya}, {Serizawa}, {Ueda},
  \& {Vievard}}]{2021ApJ...910L...9N}
{Nugroho}, S.~K., {Kawahara}, H., {Gibson}, N.~P., {et~al.} 2021, \apjl, 910,
  L9, \dodoi{10.3847/2041-8213/abec71}

\bibitem[{{Otten} {et~al.}(2020){Otten}, {Vigan}, {Muslimov}, {N'Diaye},
  {Choquet}, {Seemann}, {Dohlen}, {Houll{\'e}}, {Cristofari}, {Phillips},
  {Charles}, {Baraffe}, {Beuzit}, {Costille}, {Dorn}, {El Morsy}, {Kasper},
  {Lopez}, {Mordasini}, {Pourcelot}, {Reiners}, \&
  {Sauvage}}]{2020arXiv200901841O}
{Otten}, G.~P.~P.~L., {Vigan}, A., {Muslimov}, E., {et~al.} 2020, arXiv
  e-prints, arXiv:2009.01841.
\newblock \doarXiv{2009.01841}

\bibitem[{{Pelletier} {et~al.}(2021){Pelletier}, {Benneke}, {Darveau-Bernier},
  {Boucher}, {Cook}, {Piaulet}, {Coulombe}, {Artigau}, {Lafreni{\`e}re},
  {Deslile}, {Allart}, {Doyon}, {Donati}, {Fouqu{\'e}}, {Moutou}, {Cadieux},
  {Delfosse}, {H{\'e}brard}, {Martins}, {Martioli}, \&
  {Vandal}}]{2021arXiv210510513P}
{Pelletier}, S., {Benneke}, B., {Darveau-Bernier}, A., {et~al.} 2021, arXiv
  e-prints, arXiv:2105.10513.
\newblock \doarXiv{2105.10513}

\bibitem[{Phan {et~al.}(2019)Phan, Pradhan, \& Jankowiak}]{phan2019composable}
Phan, D., Pradhan, N., \& Jankowiak, M. 2019, arXiv preprint arXiv:1912.11554

\bibitem[{{Phan} {et~al.}(2019){Phan}, {Pradhan}, \&
  {Jankowiak}}]{2019arXiv191211554P}
{Phan}, D., {Pradhan}, N., \& {Jankowiak}, M. 2019, arXiv e-prints,
  arXiv:1912.11554.
\newblock \doarXiv{1912.11554}

\bibitem[{{Polyansky} {et~al.}(2018){Polyansky}, {Kyuberis}, {Zobov},
  {Tennyson}, {Yurchenko}, \& {Lodi}}]{2018MNRAS.480.2597P}
{Polyansky}, O.~L., {Kyuberis}, A.~A., {Zobov}, N.~F., {et~al.} 2018, \mnras,
  480, 2597, \dodoi{10.1093/mnras/sty1877}

\bibitem[{{Pope} {et~al.}(2020){Pope}, {Pueyo}, {Xin}, \&
  {Tuthill}}]{2020arXiv201109780P}
{Pope}, B. J.~S., {Pueyo}, L., {Xin}, Y., \& {Tuthill}, P.~G. 2020, arXiv
  e-prints, arXiv:2011.09780.
\newblock \doarXiv{2011.09780}

\bibitem[{Poppe \& Wijers(1990)}]{poppe1990algorithm}
Poppe, G., \& Wijers, C.~M. 1990, ACM Transactions on Mathematical Software
  (TOMS), 16, 47

\bibitem[{{Richard} {et~al.}(2012){Richard}, {Gordon}, {Rothman}, {Abel},
  {Frommhold}, {Gustafsson}, {Hartmann}, {Hermans}, {Lafferty}, {Orton},
  {Smith}, \& {Tran}}]{2012JQSRT.113.1276R}
{Richard}, C., {Gordon}, I.~E., {Rothman}, L.~S., {et~al.} 2012, \jqsrt, 113,
  1276, \dodoi{10.1016/j.jqsrt.2011.11.004}

\bibitem[{{Rodler} {et~al.}(2012){Rodler}, {Lopez-Morales}, \&
  {Ribas}}]{2012ApJ...753L..25R}
{Rodler}, F., {Lopez-Morales}, M., \& {Ribas}, I. 2012, \apjl, 753, L25,
  \dodoi{10.1088/2041-8205/753/1/L25}

\bibitem[{{Rothman} {et~al.}(2009){Rothman}, {Gordon}, {Barbe}, {Benner},
  {Bernath}, {Birk}, {Boudon}, {Brown}, {Campargue}, {Champion}, {Chance},
  {Coudert}, {Dana}, {Devi}, {Fally}, {Flaud}, {Gamache}, {Goldman},
  {Jacquemart}, {Kleiner}, {Lacome}, {Lafferty}, {Mandin}, {Massie},
  {Mikhailenko}, {Miller}, {Moazzen-Ahmadi}, {Naumenko}, {Nikitin}, {Orphal},
  {Perevalov}, {Perrin}, {Predoi-Cross}, {Rinsland}, {Rotger},
  {{\v{S}}ime{\v{c}}kov{\'a}}, {Smith}, {Sung}, {Tashkun}, {Tennyson}, {Toth},
  {Vandaele}, \& {Vander Auwera}}]{2009JQSRT.110..533R}
{Rothman}, L.~S., {Gordon}, I.~E., {Barbe}, A., {et~al.} 2009, \jqsrt, 110,
  533, \dodoi{10.1016/j.jqsrt.2009.02.013}

\bibitem[{{Ruyten}(2004)}]{2004JQSRT..86..231R}
{Ruyten}, W. 2004, \jqsrt, 86, 231, \dodoi{10.1016/j.jqsrt.2003.12.027}

\bibitem[{Salvatier {et~al.}(2016)Salvatier, Wiecki, \&
  Fonnesbeck}]{salvatier2016probabilistic}
Salvatier, J., Wiecki, T.~V., \& Fonnesbeck, C. 2016, PeerJ Computer Science,
  2, e55

\bibitem[{{Schwarz} {et~al.}(2015){Schwarz}, {Brogi}, {de Kok}, {Birkby}, \&
  {Snellen}}]{2015A&A...576A.111S}
{Schwarz}, H., {Brogi}, M., {de Kok}, R., {Birkby}, J., \& {Snellen}, I. 2015,
  \aap, 576, A111, \dodoi{10.1051/0004-6361/201425170}

\bibitem[{{Schwarz} {et~al.}(2016){Schwarz}, {Ginski}, {de Kok}, {Snellen},
  {Brogi}, \& {Birkby}}]{2016A&A...593A..74S}
{Schwarz}, H., {Ginski}, C., {de Kok}, R.~J., {et~al.} 2016, \aap, 593, A74,
  \dodoi{10.1051/0004-6361/201628908}

\bibitem[{{Shepherd} \& {Laframboise}(1981)}]{shepherd1981chebyshev}
{Shepherd}, M., \& {Laframboise}, J. 1981, Mathematics of Computation, 36, 249

\bibitem[{{Smith} {et~al.}(2003){Smith}, {Tsuji}, {Hinkle}, {Cunha}, {Blum},
  {Valenti}, {Ridgway}, {Joyce}, \& {Bernath}}]{2003ApJ...599L.107S}
{Smith}, V.~V., {Tsuji}, T., {Hinkle}, K.~H., {et~al.} 2003, \apjl, 599, L107,
  \dodoi{10.1086/381248}

\bibitem[{{Snellen} {et~al.}(2015){Snellen}, {de Kok}, {Birkby}, {Brandl},
  {Brogi}, {Keller}, {Kenworthy}, {Schwarz}, \& {Stuik}}]{2015A&A...576A..59S}
{Snellen}, I., {de Kok}, R., {Birkby}, J.~L., {et~al.} 2015, \aap, 576, A59,
  \dodoi{10.1051/0004-6361/201425018}

\bibitem[{{Snellen} {et~al.}(2014){Snellen}, {Brandl}, {de Kok}, {Brogi},
  {Birkby}, \& {Schwarz}}]{2014Natur.509...63S}
{Snellen}, I. A.~G., {Brandl}, B.~R., {de Kok}, R.~J., {et~al.} 2014, \nat,
  509, 63, \dodoi{10.1038/nature13253}

\bibitem[{{Snellen} {et~al.}(2010){Snellen}, {de Kok}, {de Mooij}, \&
  {Albrecht}}]{2010Natur.465.1049S}
{Snellen}, I. A.~G., {de Kok}, R.~J., {de Mooij}, E. J.~W., \& {Albrecht}, S.
  2010, \nat, 465, 1049, \dodoi{10.1038/nature09111}

\bibitem[{{Stangret} {et~al.}(2020){Stangret}, {Casasayas-Barris}, {Pall{\'e}},
  {Yan}, {S{\'a}nchez-L{\'o}pez}, \& {L{\'o}pez-Puertas}}]{2020A&A...638A..26S}
{Stangret}, M., {Casasayas-Barris}, N., {Pall{\'e}}, E., {et~al.} 2020, \aap,
  638, A26, \dodoi{10.1051/0004-6361/202037541}

\bibitem[{{Tennyson} {et~al.}(2016){Tennyson}, {Yurchenko}, {Al-Refaie},
  {Barton}, {Chubb}, {Coles}, {Diamantopoulou}, {Gorman}, {Hill}, {Lam},
  {Lodi}, {McKemmish}, {Na}, {Owens}, {Polyansky}, {Rivlin}, {Sousa-Silva},
  {Underwood}, {Yachmenev}, \& {Zak}}]{2016JMoSp.327...73T}
{Tennyson}, J., {Yurchenko}, S.~N., {Al-Refaie}, A.~F., {et~al.} 2016, Journal
  of Molecular Spectroscopy, 327, 73, \dodoi{10.1016/j.jms.2016.05.002}

\bibitem[{{Toon} {et~al.}(1989){Toon}, {McKay}, {Ackerman}, \&
  {Santhanam}}]{1989JGR....9416287T}
{Toon}, O.~B., {McKay}, C.~P., {Ackerman}, T.~P., \& {Santhanam}, K. 1989,
  \jgr, 94, 16287, \dodoi{10.1029/JD094iD13p16287}

\bibitem[{{Tremblay} {et~al.}(2020){Tremblay}, {Line}, {Stevenson}, {Kataria},
  {Zellem}, {Fortney}, \& {Morley}}]{2020AJ....159..117T}
{Tremblay}, L., {Line}, M.~R., {Stevenson}, K., {et~al.} 2020, \aj, 159, 117,
  \dodoi{10.3847/1538-3881/ab64dd}

\bibitem[{{Tsuji}(2002)}]{2002ApJ...575..264T}
{Tsuji}, T. 2002, \apj, 575, 264, \dodoi{10.1086/341262}

\bibitem[{{Turner} {et~al.}(2020){Turner}, {de Mooij}, {Jayawardhana}, {Young},
  {Fossati}, {Koskinen}, {Lothringer}, {Karjalainen}, \&
  {Karjalainen}}]{2020ApJ...888L..13T}
{Turner}, J.~D., {de Mooij}, E. J.~W., {Jayawardhana}, R., {et~al.} 2020,
  \apjl, 888, L13, \dodoi{10.3847/2041-8213/ab60a9}

\bibitem[{van~den Bekerom \& Pannier(2021)}]{van2021discrete}
van~den Bekerom, D., \& Pannier, E. 2021, Journal of Quantitative Spectroscopy
  and Radiative Transfer, 261, 107476

\bibitem[{Van Der~Walt {et~al.}(2011)Van Der~Walt, Colbert, \&
  Varoquaux}]{van2011numpy}
Van Der~Walt, S., Colbert, S.~C., \& Varoquaux, G. 2011, Computing in science
  \& engineering, 13, 22

\bibitem[{{Vanderburg} {et~al.}(2018){Vanderburg}, {Rappaport}, \&
  {Mayo}}]{2018AJ....156..184V}
{Vanderburg}, A., {Rappaport}, S.~A., \& {Mayo}, A.~W. 2018, \aj, 156, 184,
  \dodoi{10.3847/1538-3881/aae0fc}

\bibitem[{Virtanen {et~al.}(2020)Virtanen, Gommers, Oliphant, Haberland, Reddy,
  Cournapeau, Burovski, Peterson, Weckesser, Bright,
  {et~al.}}]{virtanen2020scipy}
Virtanen, P., Gommers, R., Oliphant, T.~E., {et~al.} 2020, Nature methods, 17,
  261

\bibitem[{{Wakeford} \& {Sing}(2015)}]{2015A&A...573A.122W}
{Wakeford}, H.~R., \& {Sing}, D.~K. 2015, \aap, 573, A122,
  \dodoi{10.1051/0004-6361/201424207}

\bibitem[{{Wang} {et~al.}(2017){Wang}, {Mawet}, {Ruane}, {Hu}, \&
  {Benneke}}]{2017AJ....153..183W}
{Wang}, J., {Mawet}, D., {Ruane}, G., {Hu}, R., \& {Benneke}, B. 2017, \aj,
  153, 183, \dodoi{10.3847/1538-3881/aa6474}

\bibitem[{Waskom(2021)}]{waskom2021seaborn}
Waskom, M.~L. 2021, Journal of Open Source Software, 6, 3021

\bibitem[{Wengert(1964)}]{wengert1964simple}
Wengert, R.~E. 1964, Communications of the ACM, 7, 463

\bibitem[{{Wilzewski} {et~al.}(2016){Wilzewski}, {Gordon}, {Kochanov}, {Hill},
  \& {Rothman}}]{2016JQSRT.168..193W}
{Wilzewski}, J.~S., {Gordon}, I.~E., {Kochanov}, R.~V., {Hill}, C., \&
  {Rothman}, L.~S. 2016, \jqsrt, 168, 193, \dodoi{10.1016/j.jqsrt.2015.09.003}

\bibitem[{{Yan} {et~al.}(2019){Yan}, {Casasayas-Barris}, {Molaverdikhani},
  {Alonso-Floriano}, {Reiners}, {Pall{\'e}}, {Henning}, {Molli{\`e}re}, {Chen},
  {Nortmann}, {Snellen}, {Ribas}, {Quirrenbach}, {Caballero}, {Amado},
  {Azzaro}, {Bauer}, {Cort{\'e}s Contreras}, {Czesla}, {Khalafinejad}, {Lara},
  {L{\'o}pez-Puertas}, {Montes}, {Nagel}, {Oshagh}, {S{\'a}nchez-L{\'o}pez},
  {Stangret}, \& {Zechmeister}}]{2019A&A...632A..69Y}
{Yan}, F., {Casasayas-Barris}, N., {Molaverdikhani}, K., {et~al.} 2019, \aap,
  632, A69, \dodoi{10.1051/0004-6361/201936396}

\bibitem[{{Yan} {et~al.}(2020){Yan}, {Pall{\'e}}, {Reiners}, {Molaverdikhani},
  {Casasayas-Barris}, {Nortmann}, {Chen}, {Molli{\`e}re}, \&
  {Stangret}}]{2020A&A...640L...5Y}
{Yan}, F., {Pall{\'e}}, E., {Reiners}, A., {et~al.} 2020, \aap, 640, L5,
  \dodoi{10.1051/0004-6361/202038294}

\bibitem[{{Zaghloul}(2018)}]{2018arXiv180601656Z}
{Zaghloul}, M.~R. 2018, arXiv e-prints, arXiv:1806.01656.
\newblock \doarXiv{1806.01656}

\bibitem[{{Zaghloul} \& {Ali}(2011)}]{2011arXiv1106.0151Z}
{Zaghloul}, M.~R., \& {Ali}, A.~N. 2011, arXiv e-prints, arXiv:1106.0151.
\newblock \doarXiv{1106.0151}

\end{thebibliography}
\bibliographystyle{aasjournal}
\end{document}